\documentclass[aps]{revtex4}
\usepackage{rotate}
\usepackage{epsfig}
\usepackage{psfrag}

\newcommand \be  {\begin{equation}}
\newcommand \bea {\begin{eqnarray} \nonumber }
\newcommand \ee  {\end{equation}}
\newcommand \eea {\end{eqnarray}}

\newcommand \N {{\cal N}}

\begin{document}

\title{The Random First-Order Transition Theory of Glasses: a critical assessment}

\author{Giulio Biroli \& Jean-Philippe Bouchaud}
\affiliation{Institut de Physique Th\'eorique, CEA, IPhT, F-91191 Gif-sur-Yvette, France and CNRS, URA 2306}
\affiliation{Science \& Finance, Capital Fund Management, 6 Bd Haussmann, 75009 Paris France}

\begin{abstract}
The aim of this chapter is to summarise the basic arguments and the intuition bolstering the RFOT
picture for glasses, based on a finite dimensional extension of mean-field models with an exponentially
large number of metastable states. We review the pros and cons that support or undermine the
theory, and the directions, both theoretical and experimental, where progress is needed to ascertain
the status of RFOT. We elaborate in particular on the notions of mosaic state and point-to-set
correlations, and insist on the importance of fluctuations in finite dimensions, that significantly blur
the expected cross-over between a Mode-Coupling like regime and the mosaic, activated regime. 
We discuss in detail the fundamental predictions of RFOT, in particular 
the possibility to force a small enough system into an ideal glass state, and present several new ones, concerning
aging properties or non-linear rheology. Finally, we compare RFOT to other recent theories, including elastic models, Frustration
Limited Domains or Kinetically Constrained models.
\end{abstract}

\maketitle

{\it Details that could throw doubt on your interpretation must be given, if you know them. 
You must do the best you can -- if you know anything at all wrong, or possibly wrong -- to explain it. 
If you make a theory, for example, and advertise it, or put it out, then you must 
also put down all the facts that disagree with it, as well as those that agree with it.}

R. P. Feynman

\tableofcontents 

\section*{List of Acronyms}

AG: Adam-Gibbs

FLD: Frustration Limited Domains

IID: Independent Identically Distributed

KCM: Kinetically Constrained Models

KTW: Kirkpatrick, Thirumalai and Wolynes

LPS: Locally preferred structures

MCT: Mode Coupling Theory

REM: Random Energy Model

RFOT: Random First Order Transition

ROM: Random Orthogonal Model

RSB: Replica Symmetry Breaking

1-RSB: 1-step Replica Symmetry Breaking

SER: Stokes-Einstein Relation

SK: Sherrington and Kirkpatrick

TAP: Thouless, Anderson, Palmer

TTS: Time-Temperature superposition

VF: Vogel-Fulcher

\section*{List of relevant temperatures (in decreasing order)}

$T_0$: onset temperature, at which activated effects first appear

$T_{GH}$: Ginzburg-Harris temperature, below which fluctuations deeply modify the nature of the MCT transition.

$T_d$: dynamical temperature, where Mode-Coupling Theory predicts a dynamical arrests and below which phase space is fragmented into local minima 

$T^*$: Goldstein's crossover temperature separating a free-flow regime from an activated viscosity in supercooled liquids. Empirically, $T^* \approx T_d$

$T_g$: calorimetric glass transition

$T_K$: Kauzmann temperature, where the configurational entropy vanishes or appears to vanish

$T_{VF}$: Vogel-Fulcher temperature, where the relaxation time appears to diverge. Empirically, $T_{VF} \approx T_K$

\section*{List of relevant length scales}

$a$: average inter-particle distance

$u$: particle displacement around its average position

$\Lambda$: range of interactions

$R$: variable size of the cavity

$\ell_d$: extension of the unstable soft modes above $T_d$, that diverges when $T \to T_d^+$

$\ell^*$: mosaic length, above which a TAP state is unstable, that diverges when $T \to T_K^+$

$\ell_{AG}$: Adam-Gibbs length, such that the number of metastable states ${\cal N}$ in a sphere of size $\ell_{AG}$ 
is ${\cal N} \approx 2$.

$\ell_{GH}$: Ginzburg-Harris length, beyond which fluctuations deeply modify the nature of the MCT transition.

$\ell_\sigma$: effective mosaic length in the presence of shear

$\xi_d$: dynamical correlation length, that measures the range over which a local perturbation affects the
dynamics in its surroundings.

$R^*$: size of the critical nucleus or critical void

\section*{List of symbols with different meanings}

$\alpha$: generic name of a TAP state//name of the terminal relaxation regime in supercooled liquids

$\beta$: inverse temperature//exponent of the stretched exponential decay/name of the early 
relaxation regime in supercooled liquids

$\gamma$: generic name of a TAP state//relaxation time divergence exponent in MCT

$\Gamma$: surface tension//cooling rate

$\sigma$: configurational entropy as a function of the free-energy//shear stress

$\omega$: frequency//sub-leading fluctuation exponent for the interface energy

$m$: fragility parameter//number of replicas

$z$: dynamical exponent $t \sim \ell^z$//partial partition function

$A$: generic numerical constant

\newpage

\section{Introduction}

When we have to argue why one should be interested in the theory of glasses, it is customary to
quote Phil Anderson \cite{Anderson-Science}, who wrote that {\it the deepest and most interesting unsolved problem 
in solid state theory is probably the nature of glass and the glass transition.} He was of course himself
deeply involved in this endeavour and as always made several outstanding contributions. He was 
convinced early on that spin-glasses would be a kind of appetiser, the theory of which would open up new 
avenues to understand glasses. As we all know, the theory of spin-glasses turned up to be exquisitely
complex, offering one of the most beautiful surprise in theoretical physics in the last 30 years. 
Very soon after the solution of mean-field spin-glasses was worked out by Parisi, Derrida and others \cite{Parisi,Derrida}, 
Kirkpatrick, Thirumalai and Wolynes, in a remarkable series of papers, proposed the spin-glass inspired
`Random First Order Transition' (RFOT) theory of the glass transition \cite{KTW1,KTW2,KTW3,KTW4}. It took ten more years for the 
theoretical physics community to understand the depth and scope of RFOT, and to start actively criticising,
reformulating and expanding the original papers and the more recent contributions of Wolynes and associates \cite{XW1,XW2,EW,LW}. 
Whereas many aspects of the theory are quite compelling, others still appear wobbly. It is fair to say 
that we are still far from a consensus on whether the basic tenets of the theory are correct or not, 
and whether Anderson's suggestion is the right track to follow. 

The aim of this chapter is to summarise the basic arguments and the intuition bolstering the RFOT 
picture for glasses (for recent reviews with a significant overlap, see \cite{LW,Tarjus-review,Cavagna-review,MP}). 
We want to review the pros and cons that support or
undermine the theory, and the directions, both theoretical and experimental, where progress is needed to 
ascertain the status of RFOT. We have deliberately written this paper using a narrative style, insisting on
ideas and concepts and leaving technical details to original publications or some appendices when we
felt they were particularly important. We have also chosen to present several conjectures or half-baked 
arguments that may well be wrong, with the hope that these will stimulate more research and discussions. 
Although we are ourselves somewhat biased in favour of RFOT, we are in fact motivated by a strong desire 
to understand the physical mechanisms underlying the glass transition, and hope that the following material 
will help making headway. 

\subsection{Glasses: back to basics}
\label{I-A}
So what is so special about glasses, that requires tools unavailable to classical statistical mechanics and
solid state theory? Coming back to the basic phenomenology, a glass is an {\it amorphous solid}: below the 
glass temperature $T_g$, the static shear modulus $G_0$ is for all practical purposes non zero (like for a crystal)
but there is no apparent long range order: glasses are liquids that cannot flow. This is puzzling, because 
the shear modulus can in principle be computed as a purely {\it thermodynamical average} of microscopic 
quantities, since it measures the change of free energy to an infinitesimal change of the shape of the container. 
This is necessarily {\it zero} for an ergodic amorphous state (liquid) \cite{JSTAT,GG}. As Anderson puts it \cite{Anderson-Book}, 
acquiring rigidity is not a minor fact: {\it We are so accustomed to this rigidity property that we don't 
accept its almost miraculous nature, that is an ``emergent property" not contained in the simple laws of physics, 
although it is a consequence of them.}

In the case of a crystal, we understand what is going on: the liquid undergoes a first order phase transition towards a symmetry broken 
state characterised by a periodic arrangement of the particles. The fact that symmetry is broken means that ergodicity is broken as well: disordered (liquid) 
configurations are no longer accessible to the system.  Glasses, on the other hand, are in an ergodicity broken state but with -- apparently --
the same symmetry as the liquid. Below $T_g$, glasses remain stuck around a {\it mechanically stable, but amorphous configuration} that can sustain,
for a very long time, an external shear without flowing.  
When microscopic energy barriers $\Delta$ are high, as in so-called strong glasses like SiO$_2$, this is simply due to the fact 
that elementary moves themselves become extremely slow, and the viscosity $\eta$ follows an Arrhenius law,
$\eta = \eta_0 \exp(\Delta/T)$ which eventually exceeds $10^{13}$ Poise at low temperatures, and we conventionally 
call the system a glass. But in fragile molecular liquids, the energy barrier itself grows as temperature is 
decreased. This suggests that some kind of thermodynamic ``amorphous order'' propagates (at least over medium scales), 
such that any change of configuration attempting to restore ergodicity requires the collective rearrangement of many particles 
-- this is the only way to prevent it from happening very quickly. In other words, the energy barrier for the rearrangement of 
a cluster of particles must grow with the size of that cluster, at least up to a certain size. 

\begin{figure}
\begin{center}
\centerline{\epsfig{file=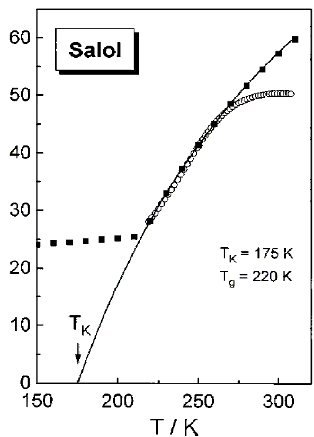,height=0.25\textheight}}
\end{center}
\caption{Excess entropy $S_{xs}(T)$ as a function of temperature for Salol (in $kJ/K/mol$). The solid line is a fit of the form
$S_{xs}(T)=A(1-T_K/T)$. Note (i) the abrupt change of slope at $T_g$, 
interpreted as the freezing of the configurational entropy contribution and (ii) the value of $S_{xs}(T_g) \approx 3 k_B$ per particle. 
From \cite{Angell-Richert}.}
\label{salol}
\end{figure}

From a theoretical point of view, it is convenient to think of an idealisation where the energy barrier and the
relaxation time are actually infinite, so the system is really in a new thermodynamic phase, and ask oneself how to
describe this new phase and the transition separating it from the ergodic liquid. After all, this is what happens when 
a system crystallises and one ignores defects, such as vacancies and dislocations. There are several crucial differences, though. 
First, as we alluded to above, there seems to be no simple static order parameter to landmark the transition; the
structure factor of the glass, for example, shows very little difference, if any, with that of the liquid. Second, there
is no latent heat at the glass transition, the glass freezes in whatever entropy was in the liquid at $T_g$, and this 
is large: the excess entropy $S_{xs}$ of the glass over the crystal at the same temperature (in order to remove at least 
part of the vibrational entropy) is substantial, of the order of $\sim 1\, k_B$ per molecule \cite{Angell-Richert} -- 
for example, $S_{xs}(T_g) \approx 3 k_B$ per particle for Salol -- see Fig. 1.
This means that the number of microscopic
configurations in which the glass can get stuck is exponentially large. There seems to be a very large degeneracy in the
ways molecules can arrange themselves such as to form mechanically stable, amorphous patterns around which they vibrate 
without exploring the other stable patterns. As a matter of fact, those thermal vibrations are small, the Lindemann
ratio that compares the root-mean square of the vibration amplitude $u$ to the intermolecular distance $a$ is around $10 \%$ at $T_g$
for most molecular glasses. \footnote{The Lindemann criterion appears to be somewhat less universal than for crystals, see \cite{Niss} for a 
recent discussion of this point.} Correspondingly, as soon as the glass transition is crossed, the Debye-Waller factor is close to unity.

This exponential degeneracy of the ``phases'' towards which the liquid can freeze is the feature that classical theories of
phase transitions cannot easily handle, and that requires new tools. At the same time, this feature seems to be 
the very essence of glassiness: in order to prevent fast crystallisation, the interaction between molecules must
be able to generate enough ``frustration'' to make the energy landscape rough and rocky and trap the system in a configuration
not very different from an arbitrary initial configuration of the liquid. Before explaining how the  solution of 
some spin glass mean-field models has afforded us the analytical tools needed to deal with exponentially degenerate amorphous 
system, we want to delve a little longer into the phenomenology of glass forming liquids, and identify several salient 
observations, common to some degree to all glass formers, and that ought to be explained by any viable theory.

\subsection{Glasses: more advanced phenomenology}
\label{I-B}
As recalled above, a distinctive properties of so-called fragile glass formers is the very fast increase of the relaxation time 
(or the viscosity) as temperature is decreased. 
The slowdown is said to be {\it super-Arrhenius} and cannot be explained by activation over {\it fixed} barriers. 
It is customary to define the fragility parameter $m$
as:
\be\label{m-def}
m = \left.\frac{\partial \log_{10} \tau_\alpha}{\partial \ln (1/T)} \right|_{T_g}, 
\ee
where the $\alpha$ relaxation time $\tau_\alpha$ is measured in microscopic time scale ($10^{-13}$ secs.). For a purely Arrhenius slowdown, the value of $m$ is $m_0=16$, by definition of the
glass temperature $T_g$. Large values of $m-m_0$ means that the effective barrier $\Delta(T)$ that one has to plug into an Arrhenius law strongly varies with temperature.
For the most fragile glass formers, the effective energy barrier at $T_g$ can be 5 times larger than its extrapolated high temperature value. This is a
very substantial increase pointing towards some collective mechanism. It is clear that many distinct numerical functions will do a good
job at fitting a five fold increase of an energy barrier over a $30 \%$ temperature variation (see the discussion in \cite{Stickel}). Well known examples of fits are (a) the Vogel-Fulcher form 
$\Delta_{VF} \propto TT_{VF}/(T-T_{VF})$, suggesting a divergence of the barriers at a finite temperature $T_{VF}$; (b) the B\"assler form $\Delta_{B} \propto T^{*2}/T$, for
which the barriers only diverge for $T=0$ \cite{Bassler}. $T^*$ is the temperature at which the liquid starts to be super-Arrhenius. 
More sophisticated fits, where the temperature dependent contribution 
to the effective barrier vanishes above $T^*$, have also been proposed in the recent literature \cite{Tarjus,Tarjus2,GCU}, see Fig. 2. The existence of such a crossover temperature $T^*$, 
where the slowdown mechanism appears to change from weakly activated to strongly activated has been advocated several times in the literature since early insights 
by Goldstein \cite{Goldstein}. Glasses lose their rigidity as temperature increases
not merely because the relaxation time scale decreases, but more radically because at some point {\it local stability is lost}. In the language of 
Mode-Coupling Theory \cite{Gotze1,Gotze2,Das}, that we will discuss below, local ``cages'' open when a certain critical temperature is reached, above which particles move quasi-freely around 
one another, and
below which thermal activation is necessary. We will see below how this crossover naturally appears in the context of RFOT.
Fig. 2 in fact advocates a cross-over from Arrhenius to super-Arrhenius behaviour at $T^*$, whereas the above MCT interpretation suggests that these barriers are actually absent above $T^*$.
We will not embark in a detailed discussion of the possible origin of this high temperature barrier, but as will be clear below (see in particular section \ref{V-C}), one actually 
expects activation processes to first appear at a higher ``onset'' temperature $T_0 > T^*$.

\begin{figure}
\begin{center}
\centerline{\epsfig{file=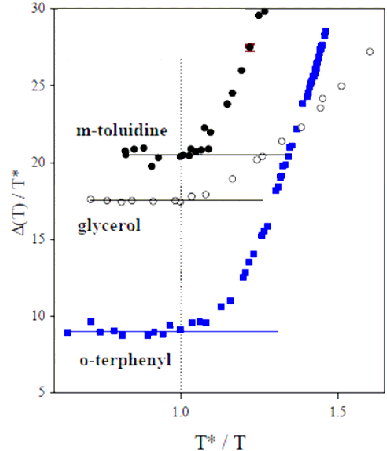,height=0.3\textheight}}
\end{center}
\caption{Energy barrier $\Delta \equiv k_B T \ln \tau_\alpha/\tau_0$ in units of the Goldstein crossover temperature $T^*$, as a function of $T^*/T$, for three fragile glass
formers (from \cite{Tarjus2}). One observes (i) the change of regime around $T^*$ and (ii) an energy barrier that increases by a factor up to five between $T^*$ and $T_g$.}
\label{tarjus_fig}
\end{figure}

Another striking empirical regularity of glass formers is remarkable correlations between thermodynamics and dynamics, for a large variety of materials (for introductory 
material, see \cite{dB1,dB2,Angell,Hodges}). These so-called 
Adam-Gibbs correlations \cite{AG1,AG2} hold both for single liquids as temperature is varied, and cross-sectionally for different liquids. For single liquids, it relates the relaxation 
time to the excess entropy mentioned above, and reads: $\ln \tau(T) \propto  [TS_{xs}(T)]^{-1}$ (see Fig. 3-a). This is the Adam-Gibbs relation. Since the excess entropy 
appears to extrapolate to zero at the so-called Kauzmann temperature $T_K$, the Adam-Gibbs relation is compatible with the Vogel-Fulcher description if $T_{VF} \approx T_K$. 
The coincidence 
between these two temperatures has been documented for many materials \cite{Angell}, 
but has also been disputed as an artefact induced by unwarranted fits (see e.g. \cite{Dyre-Review,GC-review}). 

Cross-sectionally, one observes substantial correlations between the fragility parameter $m$ and the 
jump of specific heat $\Delta C_p$ at $T_g$: $m-m_0 \propto \Delta C_p$ (see \cite{SW} and Fig. 3-b). 
More fragile glasses have a stronger variation of their excess entropy with temperature and a larger specific heat jump at $T_g$. 
Note that since super-cooled liquids are formed by very different microscopic molecules, it might be reasonable
to speak of a $\Delta C_p$ per ``bead'', the beads corresponding to the truly mobile elements inside one single molecule \cite{XW1}. 
This is an approximate way to compare diverse molecules in a more meaningful way. With this normalisation the correlation between $m$ and $\Delta C_p$ increases substantially \cite{SW}.

\begin{figure}
\begin{center}
\centerline{\epsfig{file=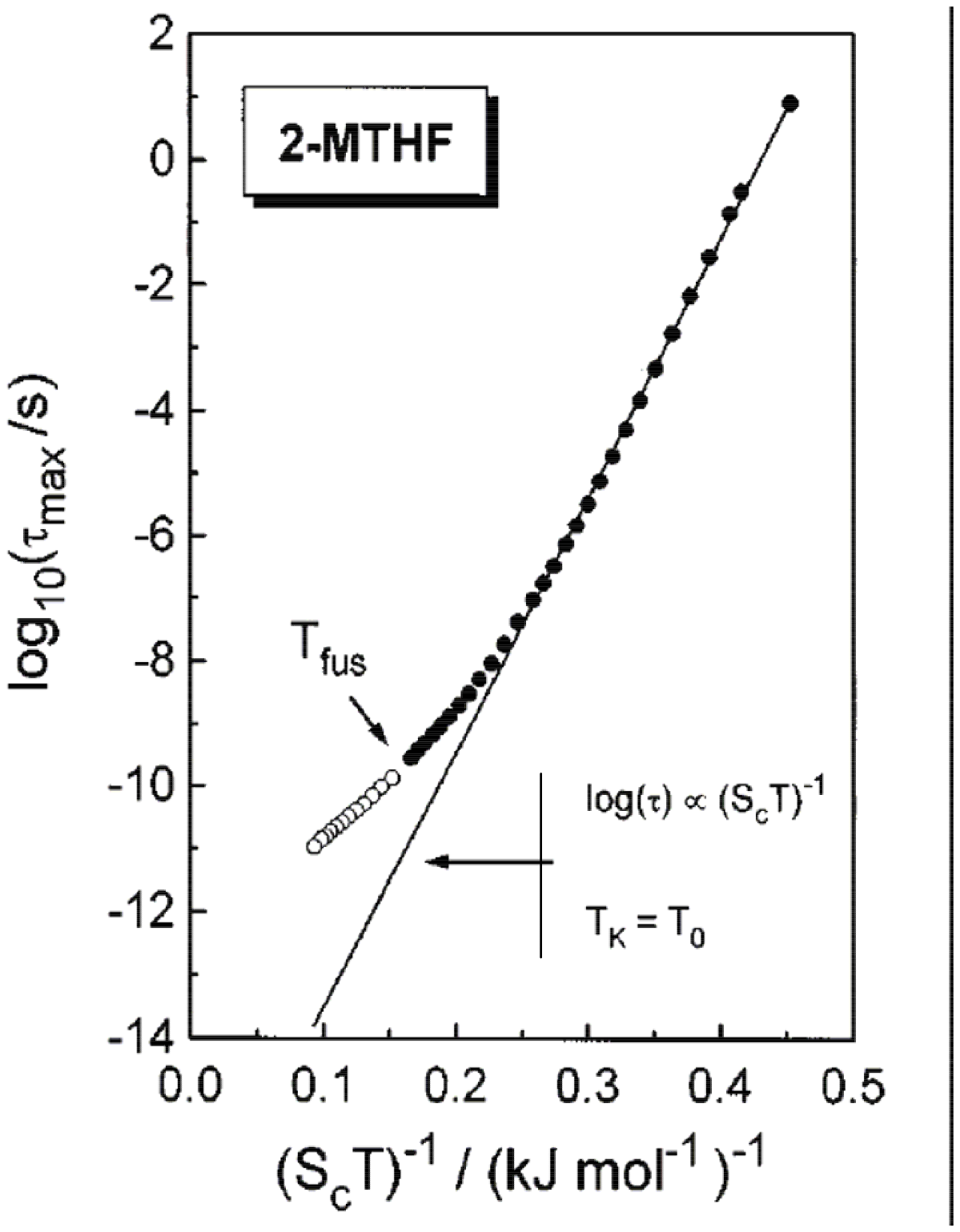,height=0.3\textheight} \epsfig{file=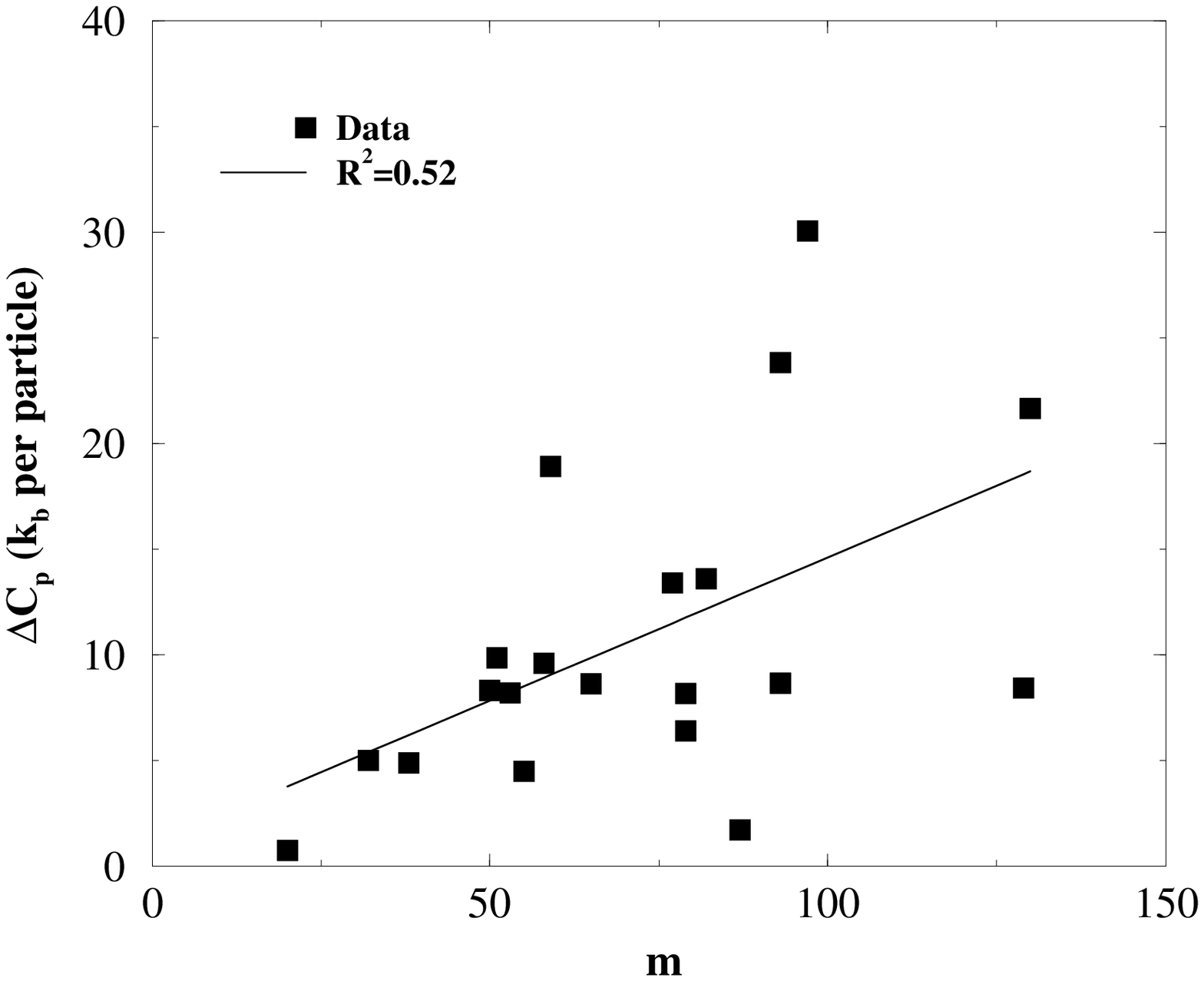,height=0.25\textheight}}
\end{center}
\caption{Left: Illustration of the Adam-Gibbs correlation: plot of $\log_{10} \tau_\alpha/\tau_0$ as a function of the inverse configurational entropy, revealing a reasonably linear region 
between $10^{-7}$ sec. (corresponding to $T = T^*$) and $10$ sec. ($T \approx T_g$). From \cite{Angell-Richert}. Right: correlation between the fragility parameter $m$ and the 
jump of specific heat at $T_g$, $\Delta C_p$, for a collection of 20 molecular glasses for which we were able to cross-check the results from different sources. The correlation is significant, 
albeit far from perfect ($R^2 = 0.52$). The regression gives $\Delta C_p \approx 0.15 m$. A better correlation, with much less scatter, is obtained when $\Delta C_p$ is counted per ``bead'' \cite{SW}.}
\label{angell}
\end{figure}

Although these correlations have been repeatedly reported in a very large body of experimental work, they are obviously not perfect (see Fig. 3-b) and can always be questioned, 
especially for  super-cooled polymeric melts \cite{McKenna}. \footnote{Polymers however add a new degree of complexity and we feel that one should first focus on the already very difficult problem 
of molecular glass-formers.} Furthermore, 
correlation does not mean causality -- in particular, it is not obvious that the {\it cause} for the super-Arrhenius slowdown of glass formers is the behaviour of the 
excess entropy, a purely thermodynamic quantity. Whereas RFOT essentially thrives on such a causal mechanism where thermodynamics drives dynamics, other scenarios have been put forth, 
where thermodynamics is completely irrelevant to understand the dynamics, and any correlation between the two is deemed fortuitous. 
We will come back to this point in sections \ref{VI-A},\ref{VI-B}.

Finally, as discovered more recently, dynamics is glasses is spatially heterogeneous and temporally intermittent \cite{Ediger,Richert}.  
A number of experimental, numerical and theoretical papers have established
that the slowdown of supercooled liquids as temperature is reduced is accompanied by the growth of a purely dynamical correlation length $\xi_d$, with no static counterpart. This length 
can be for example defined by measuring how a local perturbation of the system (for example a local density change) influences the dynamics at a distance $r$ from the perturbation, $\xi_d$ is the
decay length of this response function (see \cite{IMCT,JCP1}). 

The fact that dynamics in glasses is heterogeneous is thought to have observable consequences on macroscopic quantities \cite{Ediger,Richert}. For example, the stretched exponential nature 
$e^{-t^\beta}$ of the 
relaxation is often interpreted in terms of a mixture of exponential functions with different local relaxation time, reflecting the coexistence of slow and fast regions. Stronger 
heterogeneities should lead to smaller values of $\beta$. Similarly, violations of the Stokes-Einstein relation (SER) between the viscosity and the self-diffusion constant have also been
attributed to dynamical heterogeneities. Experimentally, the self-diffusion constant at $T_g$ can be $10^3$ times larger than expected from the value of the viscosity if the  
SER was valid. There seems to be a strong correlation between the violation of SER and the smallness of $\beta$, indicating that both phenomena should indeed have a common origin. 
Dynamical heterogeneities represent a relatively new facet of the glass transition problem. It is certainly a very important one
that contribured shifting the attention of theorists towards the real space, microscopic origin of glassy dynamics. Their existence demonstrate that glass formation 
cannot be thought of as a purely local process, due to the increase of a local energy barrier, or to the decrease of a local free volume, etc... 
Although dynamic heterogeneity may not be the cause of the slowing down of the dynamics, any theory of the glass transition should now account for the rich non-local 
space-time properties of the dynamics.  

Many other interesting phenomena and empirical regularities of glass forming liquids would be worth reviewing, but are beyond the scope of the present chapter. Some will however be
mentioned below, in relation with theoretical predictions. 

\section{Glass theory: where should one start?}
\subsection{General remarks and the Random Energy Model}
\label{II-A}
As we mentioned above, glasses below $T_g$ are for all purposes in a genuine {\it thermodynamic state}, characterised by a certain specific heat, a
shear modulus, a Debye-Waller factor, etc. that one wishes to predict from first principle. The fact that after a very long time the system actually
relaxes and flow, and is in fact technically speaking a liquid, is irrelevant except if one is prepared to wait a million years. After all, even a {\it bona fide} 
crystal subject to shear will eventually flow \cite{SBK}, and diamond eventually turns into graphite. Keeping in mind that the glass transition we are talking about
is actually waiting time dependent, a natural statistical mechanics approach is to find an appropriate mean-field theory that captures the physics qualitatively, 
even if quantitatively wrong -- for example, predicting a true transition while there is in fact only a sharp crossover, or predicting spurious metastability 
wiped out by nucleation effects, or inaccurate critical exponents renormalised by strong fluctuations. This is a very usual situation, and we know that 
hard work is usually required to understand and account for non-mean-field effects. The classical example is the Curie-Weiss theory of magnets that required 
more than sixty years and the Wilson-Fisher renormalisation group to account for real, three dimensional magnets.

In the case of glasses, we need a theory with a transition that has a mixed first/second order character. Empirically, there is indeed no latent heat nor volume 
jump at the glass transition, whereas the amplitude of frozen-in density fluctuations (again on a long but finite time scale) jumps discontinuously from zero in the 
liquid phase to a large value. As we noted above, as soon as the glass transition is crossed, the Debye-Waller factor is close to unity; in this sense the transition is
strongly first order. As intuitively expected, it seems impossible to maintain a classical system around an amorphous configuration if thermal fluctuations are too strong. 

As anticipated by Anderson, progress indeed came from spin-glasses, albeit in a rather unexpected way. The natural mean-field model for spin-glasses proposed by
Sherrington and Kirkpatrick (SK) was solved by Parisi, who invented along the way a framework (`replica symmetry breaking', RSB) to deal with systems with a large number of 
quasi-degenerate states \cite{Parisi}. Whether or not the mean-field limit of spin-glasses provides a reasonable starting point to understand real spin-glasses is an open issue which 
we do not want to even touch upon. What is interesting, in any case, is that the SK model is characterised by a genuine transition towards a state for which 
``long range amorphous order'' has a precise meaning. Spins freeze in a random configuration but the correlation length of the
fluctuations are infinite. As one approaches the transition, the correlation length diverges, and some observable
quantities, such as the non-linear (third order) susceptibility, are directly sensitive to this growing amorphous order \cite{BY,BBchi3}. However, the SK transition is second order, and the
spin-glass order parameter (i.e. the frozen-in magnetisation fluctuations) is infinitesimal just below the transition, in stark contrast with what happens in molecular glasses.

Because the mathematics of the Parisi solution is exquisitely baffling, Derrida \cite{Derrida} sought for a simpler model of spin-glasses, and came up with the Random Energy Model (REM), 
which is at first sight {\it too} simple for anything interesting to happen. In that model, the energies of microscopic configurations are independent and identically distributed (IID) random variables. 
Still, the thermodynamics is non trivial. There is a second order phase transition between a low temperature phase where the extensive part of the entropy is zero and the
system is trapped in a handful of low-lying energy states, and a high temperature phase where a very large number of configurations are relevant. Physically, this comes from
the fact that the entropy per spin $\sigma$ as a function of the energy per spin $e$ vanishes linearly at the edge of low-lying energy states:
\be\label{rem}
\sigma(e) \approx \beta_K (e-e_{\min}) - \frac{B}{2}(e-e_{\min})^2 + \dots \quad e \geq e_{\min}, \,\, B > 0.
\ee
This assumption is enough to get a transition temperature $T_K$, since from $(\partial \sigma/\partial e)=1/T$ one gets
for the entropy as a function of temperature $\Sigma(T)$:
\be
\Sigma(T \geq T_K) \approx \frac{\beta_K^2}{B} \left(1-\frac{T_K}{T}\right) + O((T-T_K)^2) \qquad {\mbox{with}} \, \, T_K = \beta_K^{-1}
\ee
and $\Sigma(T < T_K)=0$. Clearly, the specific heat has a jump at $T_K$, given by:
\be
\Delta C(T_K)=\frac{\beta_K^2}{B}.
\ee
The above shape of $\Sigma(T)$ for $T \geq T_K$ appears to be a good fit of the
excess entropy of a number of materials, see Fig. 1 and \cite{Angell-Richert}. 
This, together with the Adam-Gibbs relation $\ln \tau \propto (T \Sigma(T))^{-1}$, precisely leads to the Vogel-Fulcher law for the relaxation time. 
Note that in the analogy between the REM and real glasses, each REM configuration is an amorphous metastable state in which the liquid can be trapped in.
Energies of configurations are of course not IID in a real glass but one can argue, as we shall do in the following, that they behave as if they were so. 

The important step forward was to realize that instead of being just an abstract toy model, the scenario laid out by the REM is exactly realized for a large family of disordered mean-field models 
with well defined degrees of freedom, which allows one to make {\it bona fide} thermodynamical and dynamical first principle calculations \cite{Derrida,Gross-Mezard,KTW1,KTW2}. 
A particularly important model within that family is 
the 3-spin Hamiltonian:
\be\label{3spin}
{\cal H}  = \sqrt{3} \frac{J_0}{N} \, \sum_{i < j < k \leq N}  \epsilon_{ijk} S_i S_j S_k, \qquad S = \pm 1
\ee
where the sum runs over all triplets of spins and $\epsilon_{ijk}$ are IID random variables with zero mean and unit variance. In order to compute the
thermodynamical properties of this model at low temperatures, one has to use 
Parisi's replica symmetry breaking theory, but in this case the technicalities are different (and in fact simpler) than for the SK model -- these models are called `1-step RSB' (1-RSB). One
finds a REM type transition with a mixed first/second order character. The entropy and the specific heat of the low temperature phase are now non zero but the specific heat jumps at 
the transition temperature $T_K$. The amplitude of the frozen-in magnetisation fluctuations jumps discontinuously from $q=0$ for $T > T_K$ to $q^*=1/2$ for $T = T_K^-$. These similarities with 
the phenomenology of glasses suggest that 1-RSB models may indeed grasp some essential aspects of the glass formers. This observation motivated a series of papers by 
Kirkpatrick, Thirumalai and Wolynes (KTW), who established deeper still connections \cite{KTW1,KTW2,KTW3,KTW4}, as we detail below. 

The only certainty, at this stage, is that there exists well defined, but somewhat 
exotic mean-field models that exhibit a phase transition with a new phenomenology. That these fully connected disordered {\it spin} models have anything sensible to say about the
viscosity of supercooled liquids is obviously highly questionable, but it was the hunch of KTW that they do. 

\subsection{Physical properties of 1-RSB models}

\subsubsection{Thermodynamics}
\label{II-B.1}

1-RSB models are disordered models solved by the so-called one-step replica
symmetry breaking ansatz (see e.g. ~\cite{ABarrat,Cugliandolo,Cavagnapedestrians} and references therein).  
Examples are the 3-spin model defined above, or any $p$-spin model with
$p \geq 3$ (the REM corresponding mathematically to $p \to \infty$ \cite{Derrida}). A 2-spin model with 1-RSB properties
is the Random Orthogonal Model where the interaction matrix $J_{ij}$ between Ising spins is a random matrix
with all eigenvalues equal to $+1$ or $-1$ \cite{ROM1,ROM2,ROM3}. Other examples of direct physical interest will be given below.

The physical behaviour of these models is particularly transparent in terms of the Thouless-Anderson-Palmer 
(TAP)~\cite{TAP} approach that allows to scrutinise the free energy landscape of the model. Technically, one computes the
Legendre transform of the free energy as a function of all the local magnetisations. The minima of this TAP free energy correspond to the
thermodynamic states of the system, like the two minima of the Curie-Weiss free energy represent the two low temperature
ferromagnetic states in the Ising model. The configurational entropy, or complexity $\Sigma(T)$ is defined as $N^{-1} \ln \mathcal{N}$ 
where $\mathcal{N}$ is the number of solutions of the TAP equations that contribute to the free energy density. 
In the limit $p \to \infty$ the TAP states of the p-spin model become the REM
configurations.

For a 1-RSB system, the analysis of the complexity shows
that on top of the temperature $T_K$ below which $\Sigma(T)$ vanishes, there exists 
a second, higher temperature that we call $T_d > T_K$ (the subscript $d$ will become clear below) 
above which $\Sigma(T)$ drops discontinuously to zero again. However, the situation is 
markedly distinct below $T_K$ and above $T_d$. As these two temperatures are crossed, the properties of the
free-energy landscape change drastically, but in two very different ways.
At $T_K$ the number of minima is no more exponential in the system size,  $\Sigma(T < T_K)=0$, and this 
leads to a thermodynamic transition of the REM type, as explained above. The only difference with the REM
is that now the free energy minima are not single configurations; they also involve small vibrations around the minima
that contribute to the total entropy in this phase. At $T_d$, on the other
hand, there is no thermodynamical phase transition at all, but rather a ``fragmentation'' of phase space into disconnected minima.
For $T > T_d$ there is only one minimum of the TAP free energy with $f=f_p(T)$, where all local magnetisations are zero,
corresponding to a trivial paramagnet. For $T_K < T < T_d$, on the other hand, the partition function is built up 
from the superposition of an exponentially large number of disconnected minima, which all have the same free energy $f=f^*(T)$. 
As a consequence the free energy of the system, still given by $f_p(T)$, is smaller than $f^*(T)$ because of the additional entropic gain due to the complexity:
\be \label{fp}
f_p(T) = f^*(T) - T \Sigma(T),
\ee
However, this result implicitly assumes that all these minima are 
mutually accessible, so that the corresponding subtraction of $T \Sigma(T)$ is warranted. 
Quite surprisingly, one finds that 
$f_p(T)$ is not singular at $T_d$: the thermodynamics is completely blind to this fragmentation!
We will come back to this crucial point in the following subsection.

\begin{figure}
\begin{center}
\centerline{\epsfig{file=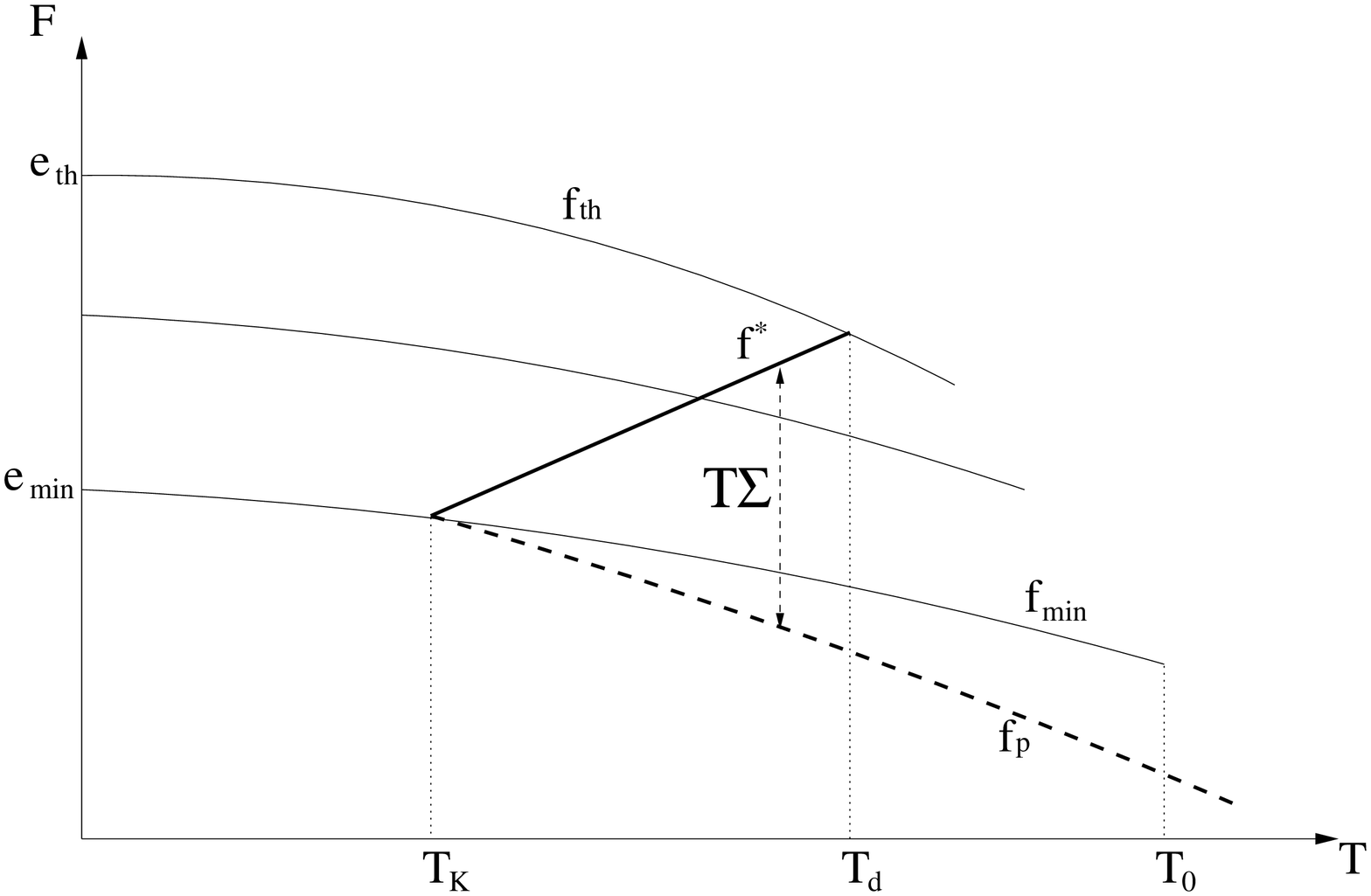,height=0.3\textheight}}
\end{center}
\caption{Summary of the mean-field 1-RSB scenario for $p$-spin models: free-energy of the different TAP states as a function of temperature, from the lowest (most stable) $f_{min}(T)$ 
to the highest (marginally stable) threshold states. We also show the free energy of the paramagnet $f_p(T)$ that crosses $f_{\min}(T)$ at the static transition $T_K$. We also show the free
energy $f^*(T)$ of the dominant TAP states in the temperature range $T \in [T_K,T_d]$, and the configurational entropy $\Sigma(T)$ that makes up the difference between 
$f^*$ and $f_p$. The most stable branch terminates at the onset temperature $T_0$.}
\label{tap}
\end{figure}

The abrupt vanishing of $\Sigma(T)$ above $T_d$ does {\it not} mean that there are no minima in the high temperature region, but rather that
these states are no longer numerous enough to compete with the trivial TAP minimum where all local magnetisations are zero. More formally, 
one can introduce a free-energy dependent complexity, $\sigma(f,T)$, that counts the number of TAP minima with free-energy density $f$ 
at temperature $T$. The partition function of the system then reads: \footnote{Here and below, we set $k_B=1$.}
 \footnote{One may wonder if one should also add the contribution from the paramagnetic (or liquid) 
state that has a free energy equal to $f_p(T)$. However, it has been understood that the paramagnetic state does not exist anymore below $T_d$. 
The clearest explanation comes from the study of dynamics and will be discussed in the next section.} 
\be
Z(T) = \int {\rm d}f \, \exp\left[-\frac{Nf}{T} + N\sigma(f,T)\right].
\ee
For large $N$, one can as usual perform a saddle-point
estimate of the integral, that fixes the dominant value of $f$, $f^*(T)$, such that:
\be
T \left.\frac{\partial \sigma(f,T)}{\partial f}\right|_{f=f^*(T)} = 1
\ee
The temperature dependent complexity is in fact defined by: $\Sigma(T) \equiv \sigma(f^*(T),T)$. From this point of view,
the fact that $f_p(T)$ is not singular at $T_d$ is actually far from obvious (see Fig. \ref{tap}). It is one of the most unexpected result emerging from the analytical solution of
1-RSB models. It suggests that at $T_d$ the paramagnetic (or liquid) state fractures in an exponential number of states
and that this transition is only a dynamic phenomenon with no consequences on the thermodynamics.
This physical scenario is key to the development of the RFOT theory of glasses. 
When $T > T_d$, there are still non-trivial TAP states but their contribution is sub-dominant with respect to the `trivial' paramagnetic
state, which exists in that region. The probability to sit in one of these TAP states is exponentially 
small in $N$ for $T > T_d$. 
Technically, this means that $\sigma(f^*(T),T)$ becomes too small to compensate the difference between $f^*$ and $f_p$.  
One has reach a third temperature $T_0 > T_d$ for the last TAP state to disappear, in the sense that $\sigma(f,T > T_0)\equiv 0$. 
A sketch of the rather intricate situation is provided in Fig. \ref{tap}. The temperature $T_0$ turns out to be important for the understanding of glasses, 
because the existence of locally stable states can slow down the dynamics of the system. It is often called the `onset' temperature, where activation effects first appear.

Another way to describe the above fragmentation scenario is to compute the free energy cost ${\cal V}(q)$
to keep an arbitrary configuration of the spins $\{S_i^1\}$, drawn from the canonical probability 
distribution at temperature $T$,  at a certain overlap $q = \frac1N \sum_i S_i^1S_i^2$ from another equilibrium 
configuration $\{S_i^2\}$ \cite{Franz-Parisi}. The form of ${\cal V}(q)$ for $T > T_d$ is such that the only minimum of ${\cal V}(q)$ is at $q=0$, for which ${\cal V}(q=0)=0$: spontaneously, 
two copies of the same system wander in phase space independently and the most probable is to find them in completely different states.
When $T_K < T < T_d$, a second minimum appears 
for a value of $q^* > 0$, but the absolute minimum is still at $q=0$, and ${\cal V}(q^*)>0$ is precisely equal to the complexity $\Sigma(T)$. This is perfectly in line with the above interpretation: 
the secondary minimum corresponds to both systems being in the same locally stable TAP state. However, the associated loss of configurational entropy makes this situation metastable, and the two
copies will end up in equilibrium in two different states with zero mutual overlap. Finally,
when $T = T_K$, ${\cal V}(q^*) = 0$ and there is a finite probability to find two independent copies of the system in the same state. 

\subsubsection{Energy landscape}
\label{II-B.2}

It is useful to rephrase, and make more precise, the above discussion of the {\it free-energy} of the
TAP states in terms of the properties of the {\it energy} landscape. The topological properties of the Hamiltonian of the system as a function
of the microscopic configurations are obviously temperature independent. Particularly important configurations are those corresponding to 
local minima. The TAP states can be seen as the ensemble of thermally populated configurations around a local minimum. The spread of 
the statistical weight around the minimum gives rise to magnetisation fluctuations and to a vibrational entropy contribution $s_v$, 
which depend on the local curvature of the Hamiltonian. \footnote{For clarity sake, we discuss here systems with continuous degrees of freedom.}
Each TAP state is therefore characterised by a `bottom-of-the-well' energy $e_0$ (per spin) and an entropy $s_v$, giving rise to a free energy 
$f = e_0 - Ts_v$. In the low temperature harmonic approximation, the free-energy per particle 
can be expressed in terms of the density of eigenvalues $\rho(\lambda)$ of the Hessian stability matrix as: \footnote{For Newtonian particles in $d=3$ dimensions, the
last term would read $-3T \ln (2 \pi T)$.}
\be\label{fTAP}
f(T) = e_0 + \frac{T}{2} \int d\lambda \rho(\lambda) \ln \lambda  - \frac{T}{2} \ln (2 \pi T).
\ee
It turns out that for p-spin models, the density of eigenvalues is a shifted Wigner semi-circle, 
\be
\rho(\lambda) = \frac{8}{\pi} \, \frac{\sqrt{(\lambda-\lambda_{\min})(\lambda_{\max}-\lambda)}}{(\lambda_{\max}-\lambda_{\min})^2},
\ee
with a lower edge of the spectrum $\lambda_{\min} > 0$ that only depends on the 
bottom of the well energy $e_0$ \cite{Kurchan-Laloux}. The remarkable feature is that $\lambda_{\min}$ decreases monotonously when $e_0$ increases, 
i.e. lower energy states are also {\it more stable} in the sense that 
the average curvature of the Hessian is larger, and thermal fluctuations weaker. 
There exists a particular `threshold' value $e_{th}$ of $e_0$ at which $\lambda_{\min}$ is exactly zero, i.e. states with the
threshold energy are marginally stable. There is an exponentially small number of minima with energy above $e_{th}$, stationary points of the Hamiltonian are mostly unstable
saddles with a negative $\lambda_{\min}$; $|\lambda_{\min}|$ in fact increases linearly with $e_0 - e_{th}$ close to the threshold. 

Interestingly, one finds that the bottom-of-the-well energy of the TAP states that are dominant right at $T_d$ is precisely $e_{th}$ (see Fig. \ref{tap}) and therefore these TAP states are also
marginally stable and have soft modes. This marginality has important physical consequences, in particular because it induces the divergence of a correlation length close to $T_d$ (see Eq. (\ref{elld})
below and the discussion there.) Note also that energy minima with $e_0=e_{th}$ are exponentially more numerous than any other minima. This means that the overwhelming majority of 
minima of 1-RSB systems are only marginally stable! 

On the other hand, the bottom-of-the-well energy of the TAP states that are dominant at $T_K$ is clearly the ground state energy $e_{\min}$, 
and these states are also the most stable ones, with the largest value of $\lambda_{\min}$. 

As the temperature increases, the harmonic approximation breaks down and at some temperature TAP states ``melt'' and disappear (see Fig. \ref{tap}). 
This melting temperature depends on $e_0$, and (as intuitively expected) increases as $e_0$ goes down, since the stability of the corresponding states increases. This is 
strongly reminiscent of the recent experiments on ultra-stable glasses by Ediger \cite{ediger-ultrastableglasses}.
Furthermore, as we shall discuss later, TAP states that are irrelevant for the thermodynamics within mean-field become important
in finite dimensions, especially for the dynamics that becomes activated above $T_d$. From this perspective
it seems natural to associate the onset temperature $T_0$ mentioned above to the melting of the most stable TAP states with $e_0=e_{\min}$ \cite{franz-tonset}. 

The interest of exactly solvable models in general is that they allow to characterise the physical properties of a system in a solid, rigorous way. These solvable models sometimes 
unveil effect that cannot be easily guessed using hand waving and plausibility arguments. In the case of p-spin models, analytic calculations indeed provide an extraordinary rich information 
on the structure of the energy landscape and the thermodynamical properties, which we tried to summarise in the above two sections. The most striking and unexpected result is the 
fragmentation transition at $T_d$, where the nature of paramagnetic phase changes without any thermodynamical signature. This phenomenon is bound to have some impact on the dynamics of
these 1-RSB models, to which we turn now.

\subsubsection{Dynamics} 
\label{II-B.3}

Below $T_d$, stable local minima are numerous enough to dominate the thermodynamics. These local minima have typically zero mutual overlap, which means that the local amorphous order 
is very different in the different TAP states. But since we are in mean-field, the energy barrier between these zero overlap states involve all spins and must diverge 
in the thermodynamic limit $N \to \infty$. Dynamically, the system must thus be trapped in a single TAP minimum when $T < T_d$; the thermodynamical average is only realized if the large time limit $t \to \infty$ is taken
{\it before} the limit $N \to \infty$. The temperature $T_d$ should therefore separate an ergodic high temperature region from a glassy phase disguised in a `non ergodic paramagnet' below $T_d$. 
These intuitive expectations are  confirmed by the exact treatment of the dynamics of these 1-RSB models, assuming Langevin dynamics and continuous spins. 
The dynamical equations can be cast in terms of two observables: the spin-spin correlation function, defined as:
\be
C(t) = \frac{1}{N} \sum_{i} \langle  S_i(0)S_i(t) \rangle,
\ee
and the corresponding response function $R(t)={\partial \langle S_i(t)\rangle }/{\partial h_i(0)}$, where $h_i(0)$
is the local magnetic field at site $i$ and time $0$. In equilibrium, the two functions are related by the Fluctuation-Dissipation Theorem: $TR(t)=-\partial_t C(t)$. In the large $N$ limit, and
for the 3-spin model, $C(t)$ is found to obey a closed non-linear integro-differential equation that reads (see \cite{KTW1,BCKM-review,Cugliandolo}):
\be\label{MCT}
\partial_t C(t) + \tau_0^{-1} C(t) = -\frac{3J_0}{2T} \int_0^t {\rm d}s C^2(s) \partial_{t-s} C(t-s).
\ee
Remarkably, this equation is {\it identical} to the so-called schematic Mode-Coupling approximation, developed since the mid-eighties by G\"otze and others as a self-consistent 
re-summation scheme for the density correlation function of strongly interacting liquids \cite{Gotze1,Gotze2,Charbonneau-Reichmann,Das}. 
The solution of the above Mode-Coupling equation is now very well understood mathematically. At high
temperature, interaction is unimportant and $C(t)$ relaxes exponentially. As the temperature is reduced towards $T_d=\sqrt{3/8}J_0$, the relaxation becomes slower and non-exponential.
There is a first rapid decay from $C(t=0)=1$ towards a plateau value $q^*$, then a slow evolution around it (the $\beta$ regime) and finally a very slow decay towards zero (the $\alpha$
relaxation). At a given temperature $T>T_d$, the dynamics in the $\beta$-regime is
described by power-laws. The approach to the plateau can be written as $C(t) \sim q^*+ct^{-a}$, and the later departure from it as $C(t)\sim q^*-c't^{b}$, where $c,c'$ are
constants, and $a$ and $b$ are exponents, found to be $a \approx 0.395$ and $b=1$ for the 3-spin model. In the $\alpha$-regime close to $T_d$, $C(t)$ verifies a scaling
law, called Time-Temperature superposition (TTS) in the structural glass literature:
\be
C(t) \simeq g(t/\tau_\alpha)\ ,
\label{tts}
\end{equation} 
where $g$ is a certain scaling function and $\tau_\alpha$ is the $\alpha$ relaxation time that diverges at the transition $T=T_d$ as:
\be\label{MCT-gamma}
\tau_\alpha \propto (T-T_d)^{-\gamma} \qquad \gamma=\frac{1}{2a} + \frac{1}{2b} \ .
\ee
When $T < T_d$, the asymptotic limit of $C(t)$ is non longer zero, which means that ergodicity is broken and the assumptions made to derive Eq. (\ref{MCT}) are not valid. 
For the 3-spin model, one can still write down the dynamical equations in terms of two-time (aging) quantities, $C(t,t_w)$ and $R(t,t_w)$ -- see \cite{CuKu,Cugliandolo} for details. If the
initial condition is taken instead with the Boltzmann measure at a temperature $T<T_d$ \cite{Barrat-Burioni-Mezard}
one finds a rapid decay of $C(t)$ towards a non zero value. This corresponds to the equilibrium 
relaxation inside one of the TAP states with free energy $f^*(T)$. No paramagnetic or liquid state is found below $T_d$ within this approach, in agreement with the fact that it does not exist
anymore since it is fractured into an exponential number of TAP states.

These findings are perfectly in line with the energy landscape picture brushed in the previous paragraph (see e.g. \cite{BCKM-review,Cavagna1,Cugliandolo}): for $T > T_d$, the system only explores 
with high probability unstable
stationary points of the Hamiltonian, and relaxation can proceed without any large barrier crossing: the correlation function decays fast and quasi-exponentially. As $T$ 
approaches $T_d$, both the number of unstable directions and the negative curvature of these unstable directions go to zero. One therefore expects a fast relaxation to a plateau, governed
by the stable directions, followed by a slow relaxation driven by the unstable directions that allow the system to relax towards zero (see \cite{Cavagna-toy} for a 
simple picture, and \cite{Thalmann} for more insights). Since the curvature of 
these unstable directions goes 
to zero, the relaxation time diverges continuously at $T_d$. Below $T_d$, all directions are stable and the system is trapped within a TAP state. 
The only relaxation mechanism is (collective) activation, but for mean-field models with $N \to \infty$ this relaxation channel is also forbidden, and $T_d$ is a genuine ergodicity breaking solution,
precisely of the same nature as found within the Mode-Coupling Theory (MCT) of glasses. Although it is very clear that non mean-field activated events will smear 
out this dynamical transition, the precise theoretical description of this mechanism is well beyond our current abilities. As we will detail below, the RFOT framework only brings partial 
answers to this crucial question. In any
case, the similarity between this smeared transition at $T_d$ and the dynamical crossover conjectured by Goldstein around $T^*$ is very striking.

It should be noted that the above discussion is not at all limited to the 3-spin model. All models that exhibit a 1-RSB transition and for which the dynamics can be studied analytically share the very
same features. Although the corresponding integro-differential equation for $C(t)$ can take more complicated forms, the two-step relaxation described above, with power-law
dynamics around the plateau and a diverging $\alpha$ relaxation time. Only the value of the plateau $q^*$ and of the exponents $a$ and $b$ are model dependent (although the relation 
between $a$ and $b$ is not). In fact, the MCT phenomenology is even more general since all the results can be derived based on general assumptions about the nature of the 
dynamical arrest without relying on any particular model \cite{ABB}.  

\subsection{But are we on the right track?}

Let us summarise where we are: we have a mean-field model of spin glasses with random multi-spin interactions. This model has a static transition 
temperature $T_K$ with a vanishing excess entropy {\it \`a la
Kauzmann} and a dynamical MCT transition at a higher temperature $T^* \sim T_d$ below which thermal activation becomes dominant, {\it \`a la Goldstein}. 

The analogy may however look far-fetched or even dubious to many. According to 
Jim Langer, for example, it is hard to believe that long range interacting spin model with weird interactions can teach us anything about molecules that stop jittering around \cite{Langer}. 
There are indeed different issues that need clarification. First, in what sense a model with spin degrees of freedom on a lattice can be used to describe positional degrees of freedom of molecules?
Second, spin-glass models assume from the start the existence of quenched disorder and encode it as random interactions, whereas molecules in a glass interact via rather simple deterministic potentials. 
Disorder in the latter case is self-induced, in the sense that the system freezes in an amorphous state, so that the potential seen by a given molecule can indeed be considered as random. But the 
freezing phenomenon is precisely what we want to model! The initial assumption of quenched randomness put by hand may not be warranted. Third, non mean-field, finite dimensional effects must play an
important role. As we just discussed, the MCT transition must be smeared by activation events and is at best only a crossover in finite dimensions. Does the physics 
of the 1-RSB transition survive in finite dimension or is it totally wiped out by fluctuation effects? Even more fundamentally, we have to address Langer's strongest 
claim \cite{Langer}: {\it{I do not think these mean-field models can be used to predict the divergence of the viscosity. 
On the contrary, I think the mechanisms that produce molecular rearrangements in glasses must be localised and that long-range models inevitably fail to describe such mechanisms properly}}. In the
next subsections, we try to tackle these three issues. 

\subsubsection{Mean-field models with spins and disorder?}
\label{II-C.1}

The first two points are actually deeply intertwined, and the answer seems to be that the 1-RSB scenario established within some specific spin-glass models is in fact far more general. 
The situation is analogous to the standard Curie-Weiss scenario for the ferromagnetic transition, or the Van der Waals theory of the liquid-gas transition, which both end up being very generic. 
There are many complementary ways to show this result. First, although historically researchers have started
to focus on mean-field disordered spin systems, it is now clear that the mean field theory of {\it bona fide} schematic models of glass-forming liquids leads naturally to the 1-RSB universality class. 
An example is provided by the so called lattice glass models \cite{birolimezard,conigliotarzia,dawsonfranz}. These are hard particles lattice models (devoid
of any quenched disorder) that do reproduce the phenomenology of glasses \cite{reichmanbiroli} and, when solved within the
Bethe approximation, all display the 1-RSB physics described previously. 

Following a completely independent path, variational or numerical  treatments of the Density Functional Theory for liquids have been performed in order to 
study whether amorphous solutions, akin to the TAP states appearing below $T_d$, emerge at low temperature or high density \cite{StoesselW,Dasgupta}. The answer is, again, positive: 
there are indeed amorphous density modulations, which are solutions of the variational problem below a certain temperature. 
The variational parameter is the width $w$ of the localised density peaks around an amorphous packing; the liquid corresponds to this width going to infinity. 
The free-energy  as a function of the inverse width $F(w^{-1})$ looks exactly as the potential ${\cal V}(q)$ described in section \ref{II-B.1}, with the same evolution as a function of temperature. 
As a matter of fact, $q$ and $w^{-1}$ play the very same role. 

Another independent route that has been followed to show the connection between 1-RSB physics and glass-formers is a formulation {\`a la} Landau of the problem. It has been shown that 
natural approximation schemes for supercooled liquid dynamics lead to MCT equations, which mathematically describe the progressive fragmentation of phase space into local minima. As recently argued 
in \cite{ABB}, MCT can be indeed formulated as a Landau theory where the ``order parameter'' is the difference between the time dependent correlation function and its plateau value.

There is another, perhaps deeper interpretation of the success of 1-RSB theories, based on the idea of generic properties of large dimensional energy landscapes. The energy landscape 
is a `height' function that gives the energy as a function of the $N$ degrees of freedom of the system under study. It is made of valleys and peaks and saddles, and it is natural to 
ask about the statistical topography of such a landscape \cite{Isichenko}, such as the number of minima or maxima around a certain level, or the index distribution of the saddles (the index is the number of
unstable directions of the Hessian matrix), etc. It turns out that for Gaussian random landscapes in large dimensions, the replica theory becomes exact and allows one to classify all landscapes
into two distinct categories. One, corresponding to short-range correlated landscapes, is precisely described by the 1-RSB phenomenology detailed above. In particular, one observes a de-mixing
phenomenon where (nearly) all saddle points are minima below a certain level $e_{th}$, whereas (nearly) all saddles above $e_{th}$ are unstable \cite{Kurchan-Laloux,Cavagna-old,Cugliandolo}. 
The second class, corresponding to long-range correlated landscapes, requires a `full' RSB solution analogous to regular spin-glasses with a continuous transition and describes a 
hierarchical landscape with valleys within valleys, etc. \footnote{A very interesting marginal case, intermediate between the two previous ones, 
has been studied in details in \cite{CLD,FB}.}
If the energy landscape is not a random function, but a generic deterministic function of all its arguments (i.e. the position of all particles in the case of glasses), it may be 
reasonable to imagine that the above classification still holds, exactly as predictions from Random Matrix Theory accurately describe the spectral properties of generic deterministic matrices. 
Up to now, all reasonable approximations applied to models that are `complex enough' to be glassy have
indeed led to a 1-RSB transition. An interesting explicit case is the frustrated Coulomb model studied by Schmalian \& Wolynes \cite{Sch-Wol}, see also \cite{KTW4,Grousson,KacW,Lopatin} for other similar 
investigations. In fact, the energy landscape of many hard satisfaction problems is exactly of the 1-RSB nature: see \cite{MM-Book,MKZ} for more on this aspect. 

Finally, let us illustrate the idea of `self-induced' disorder on the example of a ferromagnetic 3-spin model on a tree, where the $J_{ijk}$
appearing in Eq.(\ref{3spin}) are actually all identical for triplets of spins that are connected. It turns out that a numerical simulation of the dynamics of the model starting from a random
configuration leads to very strong glassy effects, with the same phenomenology as the disordered 3-spin model \cite{3-spin}. This is also confirmed by a theoretical calculation, 
where a 1-RSB transition is again found. Intuitively, this comes from that 
fact that $J S_i S_j S_k$ can be seen as a two-spin interaction model between $j$ and $k$ with a configuration dependent, random interaction
$J \sum_i S_i$. It might be that this idea is much more general (see e.g. \cite{BM,ROM1,4spin,Enzo}), and that it is possible to coarse-grain models of interacting particles in such a way
to generate these higher-order, effectively disordered interactions.

\subsubsection{Replicas for molecules?}
\label{II-C.1b}

As it should be clear from the discussion above, if we remove all these historical scaffoldings, it is now indisputable that there are very deep connections between the 
1-RSB universality class and the glass transition, although
one should of course worry about how fluctuations effects not contained in mean field theory could change or even completely wipe out the mean-field physics. 
One of the most relevant theoretical breakthroughs allowing one to get quantitative predictions using these ideas is the successful transplant of the 
replica formalism to describe systems without disorder (see the contribution of M\'ezard and Parisi in this
book \cite{MP}). A very important aspect of the theory, is once again the emergence of a 1-RSB solution to describe systems with a large number of metastable states: we 
show this in detail in Appendix A. Some general results on the glass transition, free of any approximation, can in fact be obtained within this approach (see Appendix A). 
The development of replica theory opened the path to a series of new analytical approximations to calculate the phase diagram and thermodynamic properties of realistic 
models such as hard- and soft-sphere systems, binary Lennard-Jones mixtures, etc. in the glassy region. Several 
predictions based on the 1-RSB formalism turn out to be in remarkable quantitative agreement with numerical simulations, including the existence of two transition temperatures 
$T_d$ and $T_K$, the temperature dependence of the configurational entropy, and a jump of specific heat at $T_K$ (see \cite{MP}). The quality of 
the predictions are particularly impressive for the case of hard-spheres, and the replica formalism has lead to new insights on that problem, in particular regarding the value of 
the density at Random Close Packing and the existence of a unique Jamming density -- we refer to \cite{ZP,Kurchan-Krzakala,MP} for details. More recently, Yoshino and M\'ezard \cite{YM} 
have shown how to compute within the same formalism the value of the static shear modulus $G_0=G(\omega \to 0)$, that we have argued in the introduction to be the most distinctive feature 
of the glass state. As expected on general grounds, one finds the thermodynamical shear modulus to be zero above $T_K$ and non zero below $T_K$. \footnote{We also expect $G_0$ to 
be non zero in the region $[T_K,T_d]$ if the system is confined in a single state. The value of $G_0(T)$ corresponds in that case to the shear modulus for frequencies 
much larger than $\tau_\alpha^{-1}$, see section \ref{III-E}}  

The reasons why the replica formalism is tailored to capture the phenomenology of glasses are discussed in the Appendix A, and in full detail in \cite{MP}. 
In a nutshell, this comes from the fact that
there is no {\it a priori} way of introducing an external field that selects the amorphous low energy configurations relevant in the glass phase. 
The trick \cite{Monasson} is to introduce $m-1$ additional `clones' 
of the same system with a small attractive interaction between them, that plays the role of a self-adaptive field guiding the system into the `right' amorphous configuration 
(see Appendix A and \cite{MP} for more precise technical statements). If the $m$ clones have a 
non zero probability to end up in the same configuration even when the attraction goes to zero, the system is a glass; if the clones all choose different configurations, the system is still a
liquid. Quite remarkably, the computation of the `cloned' partition function allows one to access the configurational entropy of the system, see \cite{Monasson,MP} and Appendix A.

Several quantitative predictions can therefore be obtained both for the dynamics (within MCT) and for the statics (using replicas). 
Reviewing them all is beyond the scope of this chapter (see \cite{MP}). However, it is important to underline that the possibility of 
obtaining several quantitative predictions is certainly a very strong advantage of RFOT when compared to other theories that often only provide qualitative insights.

\subsubsection{Finite dimensional effects?}
\label{II-C.2}

How much of the 1-RSB phenomenology survive in non mean-field, finite dimensional situations? Some authors, like Mike Moore, have argued that 
1-RSB models are inherently unstable in finite dimensions, and fall in another universality class, that of $p=2$ spin-glasses in a field, for which the
transition in continuous or in fact a mere crossover \cite{Moore1,Moore2}. 
If this were the case, then all the interesting results quoted above would obviously be useless to account for the properties
of real glasses. This worry stemmed in part from the fact that all spin (or Potts) models with a 1-RSB transition in mean field behave totally
differently when simulated on a finite dimensional lattice \cite{Bragian,3spinFranz}. All MCT/1-RSB features (such as a two-step relaxation function) seem 
to disappear and leave way to a continuous, spin-glass like behaviour. 

Eastwood \& Wolynes \cite{EW} have claimed that this is because these models are not ``hard'' enough (in the sense of the height of the non ergodic parameter $q^*$ or the Lindemann ratio) 
to sustain a discontinuous transition in finite dimensions. It would be interesting to have a precise formulation of their heuristic argument, and decide from 
first principle whether or not a given model exhibits the 1-RSB phenomenology in finite dimensions. Fortunately, some models that are described by 1-RSB in mean field have recently been found 
to maintain the same behaviour in finite dimensions behaviour, at least above the dynamical/MCT 
temperature $T_d$ -- for example the Random Orthogonal Model \cite{SBBB} or several lattice glass models \cite{coniglio,reichmanbiroli}. These findings are important because they shows 
that there does not seem to be any general principle that rules out the relevance of the 1-RSB scenario in finite dimensions, at least over experimentally relevant time scales.

However, one should indeed expect drastic changes going from mean-field models to finite dimensional systems. We have already mentioned that activation effects must smear out the dynamical/MCT 
transition that takes place at $T_d$. As we will argue in the next paragraph, the multiplicity of states with infinite lifetime found in mean-field theory cannot persist in finite dimension. 
Understanding the mechanisms leading to relaxation requires therefore a correct treatment of the inter-state dynamics. Make no mistake, Langer is right: these mechanisms, 
which produce molecular rearrangements in glasses, are indeed localised in space. A real-space understanding of finite dimensional 1-RSB physics in mandatory.  
But this is precisely the motivation of RFOT! We now turn to this crucial development of the theory. 

\section{Finite dimensions: droplets, cavities \& RFOT}

\subsection{The original nucleation argument}
\label{III-A}

As we have seen above, the dominant TAP minima at temperature $T \in ]T_K,T_d]$ have a free-energy $f^*$ per particle that is strictly larger than the paramagnetic free-energy $f_p$ at the
same temperature. Such a situation also arises when a ferromagnet in a non zero magnetic field $h > 0$ is studied in mean-field. When $h$ is not too large, one finds that the free-energy per spin has
two minima, $f_+$ and $f_-$, corresponding to positive and negative magnetization per spin, with $f_- > f_+$. The negative magnetization state is metastable, but the favoured up state cannot 
nucleate in the mean-field limit. In any finite dimension $d$, the well-known argument giving the critical nucleation radius is to compare the free-energy gain of a droplet of up spin in a
sea of down spins. This balance reads:
\be
\Delta(R) = (f_+-f_-) \Omega_d R^d + \Gamma S_d R^{d-1},
\ee
where $S_d, \Omega_d=S_d/d$ are respectively the surface and the volume of a d-dimensional sphere with unit radius, and $\Gamma$ the surface tension between the up and down states. The maximum
of $\Delta(R)$ is reached when $R = R^*= (d-1) \Gamma/(f_--f_+)$. Droplets with size $R < R^*$ shrink back to zero, while droplets with $R > R^*$ are unstable and 
grow indefinitely (see e.g. \cite{Langer-nucleation} for a good introduction). The time to homogeneously nucleate a marginally unstable droplet of size $R^*$ is 
$\tau(R^*) \propto \exp( \Delta(R^*)/ T)$, which can be very long when $(f_--f_+) \sim h \to 0$. 
In the mean-field limit $d \to \infty$, $R^*(d)$ diverges and one indeed finds strict metastability in this case.

In their early paper \cite{KTW3}, KTW argued that a similar mechanism is at play for 1-RSB models in finite dimension, with $f_--f_+ = f^* - f_p \equiv T \Sigma(T)$. In this case, they insisted, the 
mechanism driving nucleation is of entropic origin, which is the difference between one single TAP state and 
the liquid state. The only difference with nucleation would come from an anomalous dependence of the interface energy on $R$, which KTW postulated to 
be of the form $\Upsilon_0(T) R^\theta$, where $\theta$ is a certain exponent $\leq d-1$ and $\Upsilon_0(T)$ a temperature dependent effective surface tension (that contains a $d$ dependent 
prefactor akin to $S_d$, which diverges when $d \to \infty$). Following the same train of
thoughts, one would conclude that a given TAP minimum is unstable against the nucleation of a paramagnetic droplet of size larger than a certain $\ell^*$, given by:
\be\label{ellstar}
\ell^* = \left(\frac{\theta \Upsilon_0(T)}{d \Omega_d T \Sigma(T)} \right)^{\frac{1}{d-\theta}}.
\ee
However, in the mean-field model, the paramagnet is nothing but the superposition of all relevant TAP states, so it is
unclear in which sense this whole superposition 
can ``nucleate'' as a standard droplet. Furthermore, if this were the case, it would mean that TAP states can only be relevant in a transient, non equilibrium regime -- 
once the paramagnet has nucleated and invaded the whole system, TAP states would disappear, end of story. And if it is one of the exponentially many other TAP states that nucleates, 
why is the configurational entropy a driving force, since one particular TAP state has been singled out?
To say the least, the original KTW argument left a many people flummoxed and unsatisfied for a long time. Still, their interpretation of the mean-field 
scenario in terms of a ``mosaic'' of local TAP states with a typical size equal to $\ell^*$ is, we believe, correct, and becomes quite compelling of one rephrases the argument as follows.

\subsection{Entropy driven cavity melting}
\label{III-B}

We have argued that the 1-RSB scenario is expected to hold not only for abstract spin models but also for realistic particle models, on which we
now focus. In order to interpret this scenario in 
finite dimensions, we consider the following {\it Gedanken} experiment \cite{BB}. Suppose we can identify one of the exponentially numerous TAP state relevant at a given temperature $T$, which 
we call $\alpha$, and characterise the average position of all the particles in that state. We will establish that there exists a length scale 
above which the assumption that this TAP state has a well defined meaning is  
inconsistent. In order to do this, we freeze the motion of all particles outside a spherical cavity of radius $R$ and focus on the  {\it thermodynamics} of the 
particles inside the sphere, ${\cal S}(R)$, that are free to move but are subject to the boundary conditions imposed by the frozen particles
outside the sphere. Because of the `pinning' field imposed by these frozen particles, some configurations inside ${\cal S}(R)$ are particularly favoured energetically. 
When $\Sigma(T) R^{d}$ is much larger than unity there are many metastable states accessible to the particles in the cavity. 
The boundary condition imposed by the external particles, frozen in state $\alpha$, act as a random boundary 
field for all other metastable states except $\alpha$ itself, for which these boundary conditions perfectly
match. Any other metastable state $\gamma$ has a positive mismatch energy, otherwise our assumption that state $\alpha$ is locally stable would be violated. We assume this 
interface energy can be written as $\Upsilon_{\alpha;\gamma} R^\theta$, where $\Upsilon_{\alpha;\gamma} \geq 0$ is distributed around  a typical value equal to 
$\Upsilon_0$ and $\theta \le d-1$.  For simplicity, we assume in this section that $\Upsilon_{\alpha;\gamma}$ does not fluctuate much and 
is equal to $\Upsilon_0$ (see next section for a discussion of the effect of fluctuations). We first imagine that we wait long
enough so that the cavity embedded in state $\alpha$ is fully equilibrated. The partition function $Z_\alpha$ can then be decomposed into two contributions:
\begin{eqnarray}
Z_\alpha(R,T)=\exp[-\Omega_d R^d\frac{f_{\alpha }}{T}]+ \sum_{\gamma \neq \alpha }\exp \left[-\Omega_d R^{d} \frac{f_{\gamma
}}{T} - \frac{\Upsilon_0 R^{\theta}}{T}\right]\approx
\nonumber
\\ 
\approx \exp[-\Omega_dR^d\frac{
f_{\alpha }}{T}]+\int_{f_{\min}}^{f_{\max}}{\rm d}f\exp\left[\frac{(T \sigma(f,T)-f)\Omega_d R^d-\Upsilon_0 R^{\theta}}{ T}\right]\label{z2}
\end{eqnarray}
where $f_{\gamma}$ is the excess free energy per unit volume of state $\gamma$. (Here and below, all lengths are in units of the inter-particle distance $a$). 
When $T \to T_{K}$, the relevant $R$ is large, allowing one to make clear-cut statements using saddle point arguments. 
We focus on a typical state $\alpha$ at that temperature, i.e. a state with the free energy $f^{*}$ that dominates the integral over
$f$ above, such that $T \partial \sigma/\partial f = 1$. The partition function of the cavity immersed in the $\alpha$ state becomes independent of $\alpha$ and reads:
\be\label{final}
Z(R,T) \approx \exp[-\Omega_d R^{d} \frac{f^*}{T}] \left(1 + \exp [\Omega_d R^{d} \Sigma(T)- \frac{\Upsilon_0 R^{\theta}}{ T}] \right).
\ee
The above expression is central to our argument. When $R$ is smaller than the length $\ell^*$ defined by Eq. (\ref{ellstar}) but still large, the second term
is exponentially small even if an exponentially large number of terms contribute. The mismatch energy dominates and the state $\alpha$ favoured by the boundary conditions 
is the most probable state, even if the particles inside the cavity are free to move. In this sense, the cavity is in a glass phase, where only one (or a few) amorphous 
configurations, selected by the boundary conditions, are relevant. When $R > \ell^*$, on the other hand, the second term becomes overwhelming. There 
are {\it so many} other states to explore that it becomes very improbable to observe the $\alpha$ state. There leads to ``entropic melting'' of the cavity.

In order to derive Eq. (\ref{final}), we have assumed that the free energy displays many minima, which is only true in mean-field. 
Although in finite dimension the free energy has to be convex, the above procedure is correct: as for usual nucleation theory,
one starts from an approximate (mean-field) expression for the free energy with many minima, then account for the fluctuations that 
make these states unstable, sum over these states and eventually obtain a convex `true' free energy. 
However, it is instructive to reformulate the previous results directly in terms of the true TAP free energy $F$. Since the cavity is a finite dimensional and finite size system, 
the free energy must be convex with a single minimum. The existence of the cross-over length $\ell^*$ means in this context that
for all lengths $R<\ell^*$ there is a set of very different boundary conditions indexed by $\alpha$ such that $F$ has {\it a single} minimum
which corresponds to a density profile which is very close to the $\alpha$ one, as obtained from the approximate  
TAP free energy. However, for  $R>\ell^*$, the density profile at the minimum becomes insensitive to the boundary conditions, whatever they are, and is 
a structure-less liquid-like profile. This teaches us that TAP states are, strictly speaking, only defined on scales $R<\ell^*$. 
On larger length scales the system must ``phase separate''. In the present case this phase separation is not macroscopic as in the liquid-gas problem, \footnote{The whole 
cavity scenario has been tested in details in the context of ``usual'' nucleation theory, see \cite{Cavagna-Cammarotta}.}
but microscopic. This is the essence of the mosaic state that we explain in detail below. 
Note that in practical applications, obtaining a TAP state with the procedure outlined above is impossible. A natural approximate
procedure, which is commonly used, consists in associating TAP states with inherent structures. Although this is quite reasonable, one should not forget that energy and 
free-energy minima are not the same concepts at all \cite{Biroli-Monasson,Biroli-Kurchan} otherwise paradoxes emerge (like $T_K$ being necessarily equal to zero \cite{Stillinger}).

\subsection{Properties of the mosaic state}
\label{III-C}

The argument above indicates that the notion of a TAP state $\alpha$ can only be self-consistent if one restricts to a small enough region of space, 
such that the state is stable against spontaneous fragmentation within its bulk. As we just discussed, if one considers true TAP states then 
there is only a single minimum: the liquid state, independently of the boundary conditions above $\ell^*$ and essentially the amorphous state selected by the
boundaries below $\ell^*$. Interestingly, cavities smaller than $\ell^*$ behave effectively as if in a true thermodynamic glass phase, even when $T > T_K$. 
If on the other hand one considers approximate (finite lifetime) TAP states, the TAP states above $\ell^*$ should be obtained as a ``product" of those obtained on scale $\ell^*$.  
As a consequence the free energy of the system should be computed using Eq. (\ref{final}) above with $R \gg \ell^*$. This is indeed precisely equal to that of the liquid, 
where the configurational entropy  is subtracted from $f^*$. Up to a sub-leading excess contribution from `domain-walls', one finds: 
\be
f_{liq}(R \gg \ell^*) \approx f^*(T) - T \Sigma(T) + O(\ell^{*\theta-d}),
\ee
in close analogy with the mean-field result, Eq. (\ref{fp}) above. 

The resulting state is called a {\it mosaic liquid}. How should one interpret this? Thermodynamically,
the mosaic state is a superposition of all possible TAP states with free-energy $f^*$, as we just discussed. But a reasonable 
physical interpretation  is that it is an ever evolving patchwork of ``glassites'' 
(by analogy with ``crystallites'' in a polycrystalline phase) of size $\sim \ell^*$. The bulk of the glassites have a free-energy density $f^*$
(up to fluctuations of order $\ell^{*-d/2}$, see section \ref{III-F} below), whereas some excess energy of order $\ell^{*\theta-d}$ is 
localised within the grain boundaries between these glassites. One expects the local mean-squared particle fluctuations $\langle u^2(\vec r)\rangle$,
computed over a time much longer than $\tau_0$ but much shorter than $\tau_\alpha$, to reveal spatial correlations on a scale $\ell^*$. Curiously enough, 
we are not aware of any numerical work systematically testing this idea. Appendix B develops and makes more
precise these ideas on the simpler, but exactly soluble case of the 1-d Ising model.

Several additional comments are in order:
\begin{itemize}

\item When $\ell^*$ becomes large, the time needed to equilibrate the cavity will also be large, and the above
thermodynamical computation may not be warranted. We will discuss these dynamical aspects in a separate section below,
but already note that below $T_g$, the system is out of equilibrium and the characteristic size of the mosaic is
basically stuck at $\ell^*(T_g)$. The free energy of the glass is therefore:
\be
f_{glass}(T) \approx f_g(T) - T \Sigma(T_g),
\ee
where $f_g(T)$ is the free-energy of the most probable states at $T_g$, i.e. $f^*(T_g)=f_g(T_g)$ (see Fig. \ref{tap})
There is no entropy discontinuity at $T_g$, but a jump of specific heat given by the sum of two contributions: a) since the 
configurational entropy is frozen, one contribution is $\Delta C_p= T_g d\Sigma/dT|_{T_g}$, and b) since the local TAP states below
$T_g$ are no longer the equilibrium ones, there is a priori a second contribution to $\Delta C_p$ coming from the temperature dependence of the bottom of the well energy
and of the vibrational properties of the TAP states for $T > T_g$. In the harmonic approximation leading to Eq. (\ref{fTAP}), these contributions come 
from the temperature evolution of $e_0$ and of:
\be
{\cal K}^* = \int d\lambda \rho^*(\lambda) \ln \lambda,
\ee
where $\rho^*(\lambda)$ is the density of eigenstates of the Hessian characterising the typical TAP states at temperature $T$. Since these TAP states 
are more and more stable as $T$ decreases, $\partial {\cal K}^*/\partial T <0$. In the following, we will neglect these extra contributions 
and write $\Delta C_p \approx  T_g d\Sigma/dT|_{T_g}$ as in the REM, but this particular point would need a more detailed discussion, since the change of ${\cal K}^*$ 
with temperature is in fact related to the temperature dependence of the high frequency shear modulus $G_\infty(T)$, see section \ref{III-E} and Fig. \ref{ginfty}.

\item The value of the exponent $\theta$ and of the effective surface tension $\Upsilon_0$ are not well known at this point. Based on exact calculations by 
Franz \& Montanari in the Kac limit \cite{FM}, one finds the naive result $\theta=d-1$, which leads to $\ell^* \propto \Upsilon_0/T \Sigma$. Recent numerical 
simulations are not incompatible with this result, albeit with a strong sub-leading correction to the surface energy \cite{Cavagna-surface,Cavagna-surface0}. Wolynes and
collaborators, one the other hand, argue that $\theta=d/2 \neq d-1$ and $\Upsilon_0 = \kappa T$ with $\kappa$ to a large extent independent
of all molecular details. State to state fluctuations of $\Upsilon_{\alpha;\gamma}$ must however play an important role, see below.

\item Because the surface to volume ratio goes to infinity when $d \to \infty$, we expect $\Upsilon_0/\Omega_d$ to diverge in that limit. Therefore $\ell^*$ also
diverges in large dimensions, and the notion of ``states'' that span the whole system does make perfect sense in that limit.

\item The Adam-Gibbs argument also predicts the existence of a characteristic length $\ell_{AG}$. The question they asked was: how large must a region be
such that at least two metastable states can fit in? \cite{AG2} In other words, we want $\Sigma(T) \ell_{AG}^d \approx \ln 2$, leading to:
\be
\ell_{AG} = \left(\frac{\ln 2}{\Omega_d \Sigma(T)}\right)^{1/d}
\ee
The above cavity melting argument only makes sense if $\Sigma(T)\ell^{*d} > \ln 2$, 
corresponding to $\ell^* > \ell_{AG}$.
Forgetting numerical constants (of order unity when $d$ is finite), this inequality reads:
\be
\Sigma(T) < \left(\frac{\Upsilon_0(T)}{T}\right)^{d/\theta},
\ee
which is surely satisfied close to $T_K$, since $\Sigma(T) \to 0$. Note that even when the number of possible TAP states inside the cavity is large, the 
{\it actual} entropy of the cavity (as computed from $Z(R,T)$) is small until $R$ reaches $\ell^*$, and only becomes $\Sigma(T)$ for $R \gg \ell^*$. Franz \& 
Semerjian \cite{FS} quote $\Sigma(R,T)=\Sigma(T)[1-\ell^*/R]$ in the Kac limit (see below).

\item The cavity argument is consistent only if the number of TAP states inside the cavity is much less than the number of possible boundary 
conditions, otherwise these boundary conditions would not be able to select one given state. For $R=\ell^*$, this inequality reads:
\be
e^{S_d \ell^{*d-1}} \geq e^{\Omega_d \ell^{*d}\Sigma(T)} \to \frac{\Upsilon_0(T)}{T} \Sigma(T)^{d-1-\theta} \leq C_d,
\ee
where $C_d$ is a $d$-dependent numerical constant. This inequality is always satisfied close to $T_K$ provided $\theta \leq  d-1$. 
So there is no contradiction in thinking that boundary conditions can fix the phase inside the cavity, even in short-ranged interacting models. 
\end{itemize}

\subsection{Growth of amorphous order: order parameter and point-to-set correlations length}
\label{III-D}

The sections above makes clear that RFOT is a theory for an ideal glass transition where ``amorphous long range order'' 
emerges \footnote{As we discussed already, it could be that the transition is actually avoided, for example because
of crystal nucleation.}. The glass phase should be characterised by an order parameter and a growing static correlation 
length as the transition is approached. What do these concepts mean in the present context? The question of an order parameter for the glass transition is old. Back in 1983, Anderson (again)
anticipated that there might be an hidden order parameter: {\it Some -- but not all -- transitions to rigid, glass-like states, may entail a hidden, 
microscopic order parameter which is not a microscopic variable in any usual sense, and describes the rigidity of the system. 
This is the fundamental difficulty of the order-parameter concept: at no point can one be totally certain that one can really exclude a priori the appearance of some new hidden order.} Insightful
indeed: we now understand that within RFOT the order parameter is the overlap between two equilibrated configurations with the same boundary condition, and the 
correlation length is related to the ``stiffness" of this overlap field. As outlined in Appendix A, the formal idea is to compute using replicas the thermodynamics of the cavity by 
constraining the overlap between the configuration of the system and the reference state $\alpha$ to be large outside the cavity. One can make analytical progress by 
considering a Kac model, where the range of interactions $\Lambda$ becomes very large \cite{Franz-Kac,FM}. An interesting quantity is the resulting average overlap $q(0;R)$ between the 
configuration at the centre of the cavity and the reference state $\alpha$, as a function of the radius of the cavity $R$. The quantity $q(0;R)$ is 
called a ``point-to-set'' correlation \cite{MM,MS,FM}. One expects on general ground that $q(0;R)$ is large for small $R$ and drops to zero at large $R$, since in that 
limit the system will explore configuration totally unrelated to $\alpha$. The point-to-set correlation length $\ell^*$ characterises 
the extension of amorphous order. Montanari \& Semerjian \cite{MS} proved for a very broad class of systems that if the relaxation timescale diverges 
at a finite temperature then the point to set correlation length 
must diverge too. 

In fact, from our previous 
hand waving argument, we expect that the scale over which $q(0;R)$ drops to zero is precisely $\ell^*$. This is exactly what Franz \& Montanari 
found in the Kac limit $\Lambda \to \infty$; they indeed find a length $\ell^*$ beyond which the thermodynamical solution corresponding to a small $q(0;R)$
has a smaller free-energy than the solution corresponding to a high value of the overlap. This length $\ell^*$ is found to behave as $\Lambda T_K/(T-T_K)$ 
when $T \to T_K$, corresponding to Eq. (\ref{ellstar}) with $\Sigma(T \to T_K) \propto T-T_K$, $\Upsilon_0(T_K) > 0$ and, in this Kac model, $\theta = d-1$.

Quite remarkably, another information naturally comes out of the calculation of Franz \& Montanari, now about the finite dimensional physics close to 
$T_d$ \cite{FM,FS}.
The high overlap branch appears at temperatures above the dynamical MCT transition $T_d$ and remains metastable up to a length $\ell_d$ larger than $\ell^*$, 
that diverges approaching $T_d$ from above as:
\be \label{elld}
\ell_d \propto \Lambda \left(\frac{T_d}{T-T_d}\right)^{\nu}, \qquad \nu=\frac14
\ee
This means that typical TAP states, which should be unstable above $T_d$, are {\it stabilised} by the presence of a frozen boundary condition, but only 
if the cavity is sufficiently small, i.e. $R < \ell_d$. For $T \to T_d$, the boundary condition is able to stabilise arbitrary large systems. One can
understand this result in terms of delocalized soft modes: typical TAP states are indeed unstable above $T_d$, but the corresponding soft modes have
an extension $\sim \ell_d(T)$. This phenomenon is very similar to the one occurring in hard spheres packings close to the isostatic point (see \cite{Wyart}, where the importance
of these soft modes is clearly discussed, and the analogy with the physics of glasses is underlined.)
The existence of such a diverging length scale was in fact predicted within the context of MCT, on the basis of a very different dynamical calculation but with the 
same final prediction, Eq. (\ref{elld}) \cite{FPMCT,BBMCT,IMCT,JCP1}. Within MCT, $\ell_d$ is precisely the dynamical correlation length $\xi_d$, 
that measures how far a local perturbation affects the relaxation dynamics. As was argued in \cite{IMCT,BBSER}, the value of $\nu$ is expected to change whenever $d < 8$. An Harris-like
argument on the fluctuations of the local free-energy suggests that if the transition survives, $\nu$ should be $\geq 2/d$ (see section \ref{V-C} for more about this).

Finally, note that a distinct but possibly related static correlation length has been introduced by Kurchan and Levine \cite{Kurchan-New}. 
This is defined in terms of pattern repetition in a given configuration and, as the point-to-set length, should be infinite if there is amorphous long range order.

\subsection{Stability of the TAP states and high frequency shear modulus} 
\label{III-E}

As the temperature is reduced, the typical TAP states not only have a smaller free-energy $f^*$ but are also 
more stable. The average curvature of the stability matrix is larger. Physically, this means that the  
fluctuations of the particles around their average position, $\langle u^2 \rangle \sim T/G_\infty(T)$, is smaller not 
only because the temperature is smaller, but also because the stiffness, proportional to the high frequency shear 
modulus $G_\infty(T)$, is larger. \footnote{The reason we call $G_\infty$ a high frequency shear modulus is that in finite dimensions, TAP states are 
unstable when $R > \ell^*$, and as a result the zero frequency modulus is zero (see discussion in section \ref{IV-A}). High frequency 
here means frequencies comparable to the inverse $\beta$-relaxation time, such that inter-state transitions are neglected.} 
Although TAP states become marginally unstable when $T \to T_d^-$, in the sense
that the smallest eigenvalue of the Hessian $\lambda_{\min}$ touches zero there, we expect that $G_\infty$ in 
fact does {\it not} vanish at $T_d$, but has a square-root singularity inherited from the square-root singularity of the eigenvalue density:
\be\label{Ginfty}
G_\infty(T) \propto \left[\int d\lambda \frac{\rho(\lambda)}{\lambda} \right]^{-1} \approx G_\infty(T_d) \left(1+ A \sqrt{\frac{T_d - T}{T_d}}\right), \qquad A > 0.
\ee
This result can be established within MCT \cite{GinftyMCT}, and a replica calculation of $G_\infty$ {\it \`a la} Yoshino-M\'ezard (forbidding inter-state transitions)
should confirm this. Note however that there is no consensus on this point; for example, \cite{Berthier-Kurchan} claim that $G_\infty(T_d^-)=0$ in mean-field.

This shear modulus gives a natural energy scale for the effective surface tension $\Upsilon_0$, its singular behaviour around $T_d$ may turn out to be important for 
detailed comparison with experiments, see sections \ref{IV-B},\ref{VI-A}. The fate of the square-root
singularity in low dimensions and for realistic systems is however not known at this point. For elastic spheres close to the jamming point, for example, one finds 
that the shear modulus indeed vanishes \cite{Wyart}.

\begin{figure}
\begin{center}
\centerline{\epsfig{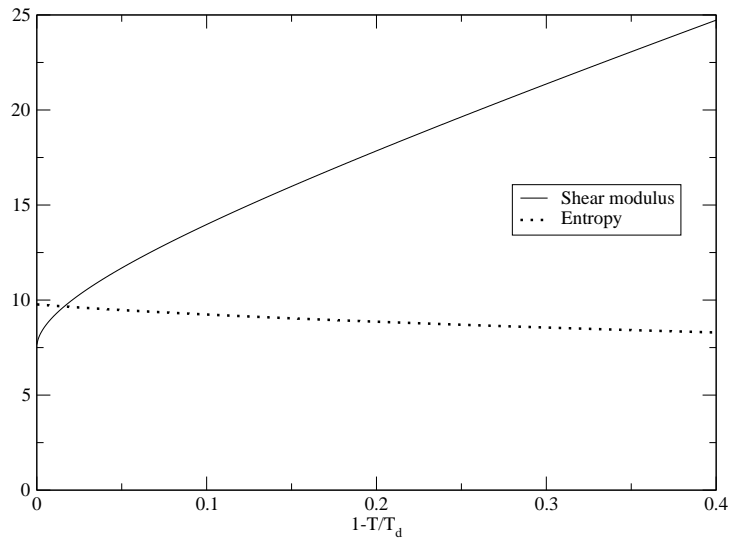}}
\end{center}
\caption{High frequency modulus $G_\infty$ (arbitrary units) as a function of $\epsilon=1-T/T_d$ for a semi-circle distribution of Hessian eigenvalues $\rho^*(\lambda)$ with 
$2$ shear modes per particles. Note that $G_\infty$ increases substantially below $T_d$: this is  a crucial ingredient in Dyre's shoving model (see section \ref{VI-A}). We show on the 
same graph the vibrational contribution to the entropy ${\cal K}^*$ (up to an arbitrary vertical shift).} 
\label{ginfty}
\end{figure}

\subsection{Entropy driven melting: numerical evidence and fluctuations}
\label{III-F}

As we just discussed, the above cavity {\it Gedanken} experiment can in fact be used to make analytical progress. It can also be faithfully implemented in numerical
simulations, for which the freezing of particles outside a spherical cavity is an easy operation. \footnote{This might even be feasible experimentally using colloids and 
optical tweezers.} One then monitors how the overlap $q(0;R)$ 
between the reference state that acts as a frozen 
boundary condition and the configuration at the centre of the cavity varies with $R$. The point-to-set 
correlation length $\ell^*$ can be defined, for example, such as $q(0;R > \ell^*)/q(0;a) < 0.5$. This length scale is
clearly found to increase substantially when temperature is reduced below $T_d$, at least in the temperature range where numerical 
simulations can be equilibrated \cite{Cavagna-PRL,Nphys} (see also \cite{Verrocchio} for a related effect). Quite interestingly, the shape of $q(0;R)$ as a function of $R$ also evolves
very significantly from a simple exponential decay at high temperatures to a ``compressed exponential'',
$\ln q(0;R) \propto -R^\zeta$ with $\zeta > 1$ at lower temperatures. As argued in \cite{Nphys}, the compressed
exponential shape may be a signature of fluctuations. One type of fluctuation that we already mentioned comes
from the effective surface tension between different TAP states, $\Upsilon_{\alpha;\gamma}$. For a fixed 
external state $\alpha$, it is indeed conceivable that for a given $R$ some $\gamma$ states are good matches and have a
particularly low $\Upsilon_{\alpha;\gamma}$.

\subsubsection{Surface energy fluctuations and renormalisation of $\theta$}

To understand how these fluctuations might strongly affect the physics, we now write the partition function for
the mobile cavity surrounded by the pinning state $\alpha$ as: \cite{Nphys}
\begin{equation}
  Z_\alpha(R,T) = e^{-\beta R^d f_\alpha} + \sum_{\gamma\neq\alpha} e^{-\beta R^d f_\gamma - \beta R^\theta \Upsilon_{\alpha;\gamma}}.
\end{equation}
The probability $p_{\overline{\alpha}}$ to leave state $\alpha$ is therefore:
\begin{equation}
  p_{\overline{\alpha}}(R) = \frac{\sum_{\gamma\neq\alpha} e^{-\beta R^d
    f_\gamma - \beta R^\theta \Upsilon_{\alpha;\gamma}}}{Z_\alpha(R,T)}.
\end{equation}
Introducing $\N_\alpha(f,\Upsilon) = \sum_\gamma \delta(f-f_\gamma)
\delta(\Upsilon - \Upsilon_{\alpha;\gamma})$, the sum can be written
\begin{eqnarray}
  \sum_{\gamma\neq\alpha} e^{-\beta R^d f_\gamma - \beta R^\theta
    \Upsilon_{\alpha\gamma}} &=& \int\!\!df\!\!\int\!\!d\Upsilon\,
  e^{-\beta R^d f - \beta R^\theta \Upsilon} \N_\alpha(f,\Upsilon), \\
  & =&
  \int\!\!df\!\!\int\!\!d\Upsilon\, 
  e^{-\beta R^d f - \beta R^\theta \Upsilon + R^d \sigma(f,T)}
  P_\alpha(\Upsilon|f),
\end{eqnarray}
where in the last equality we have defined $p_\alpha(\Upsilon|f) =
\N_\alpha(f,\Upsilon) / \N(f)$, and $\N(f)=\exp[\sigma(f,T)]$ is the
number of states with free energy $f$. $P_\alpha(\Upsilon|f)$ is the fraction of states 
$\gamma$ (inside the cavity) with free energy $f$ and effective interface tension
$\Upsilon$. 
The integrals above can be simplified using the saddle point
method, which is a very good approximation even for rather small
values of $R$. The integral over $f$ picks up the most probable value $f=f^*$, as
above. Because of the negative exponential term in $R^\theta$, the
integral over $\Upsilon$ is dominated by the lowest values of $\Upsilon$ allowed by
the distribution $P_{\alpha}(\Upsilon|f^*)$. There are two possible
cases: (a) If this function vanishes for $\Upsilon<\Upsilon_0(\alpha)$, one finds that 
up to sub-leading terms one can forget about the fluctuations of $\Upsilon_{\alpha,\gamma}$ 
and use $\Upsilon_0(\alpha) R^\theta$ in the cavity argument. However, if (b) there are
arbitrarily small effective tensions $\Upsilon$, the value of $\theta$ can be renormalised. For instance
in the case where $P_{\alpha}(\Upsilon|f^*)\simeq \exp[-(\Upsilon_0(\alpha)/\Upsilon)^y]$, where $y$ is a positive exponent, 
one that the effective surface energy is now proportional to $R^{\theta'}$, with 
$\theta'=y\theta/(y+1) < \theta$, and again an $\alpha$ dependent effective value of $\Upsilon_0$. 
This is quite important because the mosaic length $\ell^*$ is determined by this renormalised value of $\theta$.
Wolynes et al. have obtained $\theta=d/2 \leq d-1$ on the basis of a ``wetting'' mechanism that softens the
interface. It is unclear to us whether this wetting mechanism and the existence of rare, but arbitrary small
surface tensions, in fact describe the same physical effect (in which case, one should argue why $y=d/(d-2)$).

We have shown how the fluctuations of $\Upsilon_{\alpha;\gamma}$ for a fixed $\alpha$ can be accounted for. But
it may still be that the effective surface tension $\Upsilon_0(\alpha)$ strongly depends on $\alpha$ -- for 
example, $\alpha$ could be exceptionally stable locally, corresponding to a large $\Upsilon_0(\alpha)$. The
average overlap $q(0;R)$ receives contributions from samples that are locked in state $\alpha$, i.e.:
\be
q(0;R) \approx \int {\rm d} \Upsilon_0 P(\Upsilon_0) \frac{1}{1 + \exp(\Sigma(T)R^d - \Upsilon_0 R^{\theta'}/T)}
\ee
where $P(\Upsilon_0)$ is the probability (over $\alpha$) to obtain a renormalised surface tension $\Upsilon_0$,
which may depend both on $R$ and $T$. As shown in \cite{Nphys}, this assumption can be used rationalise the compressed exponential shape of $q(0;R)$ found numerically. 
In a recent work  $P(\Upsilon_0)$ has been directly measured using forced excitations \cite{Cavagna-surface0}. 
The results are compatible with the shape surmised in  \cite{Nphys} and underline the importance of the fluctuations of $\Upsilon_0$.

\subsubsection{Free-energy fluctuations and ``locally preferred structures''}

A second, very important source of fluctuations comes from the local free-energy of the selected TAP state. In the
limit where the size of the cavity $\ell^*$ goes to infinity (i.e. close to $T_K$ or in the mean-field or Kac limit),
only states with free-energy density $f^*$ have a significant probability of being observed. But when $\ell^*$ is
finite, there is some probability to observe exceptionally low free-energy states; when $f-f^*$ is
small, the distribution is Gaussian with a width $\propto \ell^{*-d/2}$. Let us assume for simplicity that the
complexity $\sigma(f,T)$ takes the following form (by analogy with Eq.(\ref{rem}) above):
\be\label{rem2}
\sigma(f,T) \approx \beta_K (f-f_{\min}(T)) - \frac{B}{2}(f-f_{\min}(T))^2 + \dots \quad f \geq f_{\min}, \,\, B > 0,
\ee
with $f_{\min}(T)=e_{\min} + f_{vib}(T)$, where $f_{vib}(T)$ is a certain function that describes the free-energy of the
(possibly an-harmonic) vibrations. This explicit choice leads to:
\be
f^* = f_{\min}(T) + \frac{\beta_K - \beta}{B} + \dots, \qquad \Sigma(T)=\frac{\beta_K^2 - \beta^2}{2B} + \dots
\ee
and $f-f^*=u/\sqrt{B\ell^{*d}}$, where $u$ is a Gaussian variable with unit variance. As noticed above, the jump of specific heat at $T_g$ is related to $B$ as: 
$\Delta C_p = \beta_g^2/B$, if we neglect the vibrational contribution (see discussion in section \ref{III-C}).

A TAP state $\alpha$ with a lower free energy ($u < 0$) is more stable against fragmentation, not only because the typical states that can `nucleate' now 
have a lower free-energy $f < f^*$, but also because the effective surface tension
is expected to increase with the stability of the state, and therefore with $f^*-f$ (see section \ref{III-E}). This means that the
size of the cavity beyond which this particular TAP state becomes unstable is increased. The balance between 
the free-energy gain and the surface energy loss now gives, to first order in $u$:
\be
\ell_\alpha \approx \ell^* \left[1 - \frac{Au}{\sqrt{\Delta C_p \ell^{*d}}} \frac{2\beta\beta_g}{\beta_K^2 -\beta^2}\right],
\ee
where $A>0$ is a numerical constant. Using $\ell^* = (\beta\Upsilon_0/\Sigma(T))^{1/d-\theta}$, the correction term is found to be small provided:
\be\label{bound}
\left(\frac{\beta_K^2 - \beta^2}{2 \beta \beta_g^2}\right)^{2\theta-d} \ll  \frac{(\beta \Upsilon_0)^d}{\Delta C_p^\theta},
\ee
which is valid arbitrary close to $T_K$ only if $\theta > d/2$, and marginally valid when $\theta=d/2$ which 
happens to be KTW's conjecture. If $\theta < d/2$, free-energy fluctuations in fact dominate the surface
energy cost, and quite interestingly, leads to an effective value of $\theta$ stuck at $d/2$. That $\theta=d/2$
plays a special role was in fact argued early on by KTW; it corresponds to $\ell^*$ diverging as $(T-T_K)^{-2/d}$ close to $T_K$, as expected for a disorder dominated 
phase transition \cite{KTW3}. If indeed $\theta=d/2$, the criterion Eq. (\ref{bound}) is never satisfied unless $\Delta C_p \ll 1$, which is not the case for fragile glass formers 
(see Fig. 3-b and below). We expect in this case fluctuation effects to be strong, as one may have anticipated from the ``wetting'' argument used by KTW to obtain $\theta=d/2$,
which also predicts that the width of the domain walls is itself $\ell^*$, i.e. that domain walls are in fact everywhere!

When $\ell^*$ is finite, there is an exponentially small probability to even observe the lowest possible free-energy
states, with $f=f_{\min}(T)$. These states can be called ``locally preferred structures'' (LPS) by analogy with theory of glasses
based on the existence of locally preferred packings (for a review, see \cite{Tarjus-review} and section \ref{VI-B}, and for recent interesting ideas about the nature of these LPS, see
\cite{Kurchan-New2}). These maximally stable states are robust against fragmentation up to size:
\be
\ell_{\max}^*(T) = \ell^*(T) \left(\frac{T+T_K}{T-T_K}\right)^{\frac{1}{d-\theta}},
\ee
which becomes much larger than $\ell^*$ close to $T_K$. The probability to observe such a large droplet is however
exponentially small, and given by:
\be
P(\ell_{\max}^*) \propto \exp[-\ell_{\max}^{*d} \Sigma(T)]
\ee
These fluctuations effects should have important consequences for the dynamics, in particular for fragile systems (because the quantity in the 
exponential behaves, for a fixed $T/T_K$, as $\Delta C_p^{-\theta/(d-\theta)}$). We will come back to this point in section \ref{IV-C}.

\subsection{Summary}

The point of the above section was to establish, on the basis of purely thermodynamic arguments, that the notion
of stable TAP states with a finite configurational entropy $\Sigma$ makes sense physically only if one zooms on a region of
space not larger than a certain length $\ell^*$, that diverges in mean-field or when $\Sigma \to 0$. Above $\ell^*$,
the system is a liquid, i.e. thermodynamically micro-phase separated on the scale $\ell^*$ into TAP states. Below $\ell^*$, on the other hand, the
system is an ideal glass even when $T > T_K$.

In the course of the argument, one has to introduce the somewhat hazy notion of a surface energy between two different TAP states, and assume that it grows as $\Upsilon_0 R^\theta$. 
There is little agreement at this stage about the value of $\Upsilon_0$ and $\theta$. Mean-field calculations and numerical simulations seem to favour $\theta=d-1$, possibly with strong 
finite size corrections \cite{Cavagna-surface0}, whereas Wolynes et al. recommend $\theta=d/2$. We have seen that fluctuation effects, absent in mean field, could renormalize the value of $\theta$, 
and that $\theta=d/2$ plays a special role in that respect.

\section{Dynamics in the mosaic state}

\subsection{Decorrelation and activation}
\label{IV-A}

Let us re-interpret the above results from a dynamical point of view. We have seen that if the configuration 
outside the cavity is frozen in state $\alpha$, the particles inside the cavity have an overwhelming probability
to stay in state $\alpha$ whenever $R < \ell^*$. Dynamically, some low energy excitations may sometimes be explored, 
but the system always reverts back to the $\alpha$ configuration. The dynamical structure factor $C(t)$ does not
decay to zero. The relaxation modes of length $R < \ell^*$ cannot be used to restore ergodicity in the system, 
and it is therefore self-consistent to assume that for these length scales, the environment is frozen, or more
precisely fluctuates around a well defined configuration. The motion on these length scales corresponds to a 
generalised `cage' effect and contributes to the so-called $\beta$-relaxation, i.e. the approach to the 
plateau value $C(t) \approx q^*$. Interestingly, the back and forth motion of groups of molecules between two or more 
configurations has been clearly observed using mesoscopic investigation techniques, see \cite{Israeloff}. 

When $R > \ell^*$, on the other hand, the configurational entropy becomes so large that the system ends up in a
completely different state $\gamma$ with only an exponentially small probability to come back to state 
$\alpha$, {\it even if the outside of the cavity is, by fiat, frozen}. Therefore, the 
relaxation time of the system is the time needed for a cavity of size $\sim \ell^*$ to leave state  $\alpha$ -- once this is 
achieved, the configuration inside the cavity will most probably never come back. 
The dynamical structure factor now decays to zero, and the assumption of a frozen environment is no longer self-consistent \cite{BB}.

A precise characterisation of the dynamical processes that allow the particles in the cavity to leave state $\alpha$ is an open problem. 
These processes must however satisfy some constraints. First, they have to be compatible with time reversal symmetry. Second,
the transition matrix should be such that many arrival states $\gamma$ can be reached once the system has left $\alpha$, otherwise the system would not gain
configurational entropy. Third, the dynamical process has to be compatible with the fact that TAP states are locally stable. As discussed above,
a crucial point is that on scales $R$ less than $\ell^*$ the system is like an ideal glass, and therefore the barrier to escape from the state 
$\alpha$ should grow with $R$.  On the basis of these remarks, a reasonable guess (inspired from known results in the context of pinned systems \cite{HH,YS,PLD}
and spin-glasses \cite{FHdyn,YT,BBSG,SGexp}) is that the arrival state 
will be reached through barrier crossing and the lowest energy barrier to significantly alter the configuration in a cavity of size $R$ grows with $R$ as a power-law: 
\be\label{DeltaR}
\Delta(R) = \Delta_0 R^\psi,
\ee
where $\Delta_0$ is an energy scale that presumably depends on the departure state (again, lower free-energy states
are more stable and should have a larger $\Delta_0$) and $\psi$ another exponent. 
A positive entropic correction is possible. This would be due to the fact that in order to leave 
the state $\alpha$ the system not only has to jump a barrier but also to find the target states between
all possible configurations. The higher the temperature the larger  the number of the target 
states relative to all available states. As a consequence this entropic contribution to the barrier should decrease at fixed $R$ when increasing $T$.

Naively, one may think that the reconfiguration events proceed by nucleating a droplet of another phase, leading to $\psi=\theta$
and $\Delta_0=\Upsilon_0$. This was KTW's original surmise, and is still advocated in Wolynes' recent papers (see
\cite{LW} for a review). More generally, one expects that the barrier should
be at least equal to the excess energy of the final state, leading to $\psi \geq \theta$, or $\psi=\theta$ and
$\Delta_0 \geq \Upsilon_0$ \cite{FHdyn}. Needless to say, a precise analytical or numerical determination of $\Delta_0$ and 
$\psi$ is currently beyond anyone's ability. Recent numerical results \cite{Cavagna-surface,Cavagna-surface0} 
suggest $\psi \approx 1$ and $\theta \approx 2$. This is actually quite puzzling in view of the expected bound $\theta \le \psi$.
At this stage it is unclear whether the violation of the bound $\theta \le \psi$ is related to the way $\theta$ 
is measured in \cite{Cavagna-surface,Cavagna-surface0} or signals that something is missing (or wrong) in the theoretical description.

Even the nature of the activated events that determines the
order of magnitude of $\Delta_0$ is quite obscure. Is $\Delta_0$ proportional to $T$, as assumed by Wolynes and associates, or
is $\Delta_0$ more akin to an elastic energy, and therefore proportional to $G_\infty(T)$? These uncertainties are related to our lack
of understanding of the physical nature of the activated events. Should one picture them as the activated fluctuations of the
grain-boundaries between glassites that sweep larger and larger regions of the bulk until $\ell^*$ is reached? Or is it the nucleation of 
small droplets within the bulk that grow and percolate? Or are these two pictures in fact equivalent because, as implied by the ``wetting''
argument of KTW that leads to $\theta=d/2$, the width of the interfaces are themselves of order $\ell^*$? What is the role played by ``excitation chains'',
of the type proposed by Glotzer et al. \cite{Glotzer-chain} and Langer \cite{Langer-chain}?

In any case, the typical time needed to decorrelate a cavity of size $\ell^*$ is obtained from Eq. (\ref{DeltaR}) as:
\be\label{taustar}
\tau_\alpha \sim \tau(\ell^*) \equiv \tau_0 \exp\left(\frac{\Delta_0 \ell^{*\psi}}{T}\right),
\ee
where we neglect all fluctuations at this stage (see section \ref{IV-C}), as well as a possible power-law prefactor $\ell^{*z}$ in the Arrhenius formula. 
From the above arguments, it is clear that the $\alpha$-relaxation time of the system is
$\tau(\ell^*)$: smaller length scales are faster (energy barriers are lower) 
but unable to decorrelate, whereas larger scales are orders of magnitude
slower so that the evolution on these scales will be short-circuited by a relaxation in parallel 
of smaller blobs of size $\ell^*$. The divergence of $\tau_\alpha$ when $\ell^* \to \infty$ is completely in line with the analysis of Montanari \& Semerjian \cite{MS}.

It is interesting to reformulate the above argument as a discussion of finite size effects. A cavity of size $R < \ell^*$ should relax to equilibrium in a time $\tau(R) \ll \tau(\ell^*)$, 
although this relaxation is incomplete in the sense that $C(t \gg \tau(R)) \to q^*(R) > 0$. As the size of the cavity increases beyond $\ell^*$, the relaxation time should 
increase and then saturate around $\tau(\ell^*)$, whilst $q^*(R) \to 0$. This scenario is expected for temperatures $T < T_d$, and preliminary numerical results seem to confirm it 
\cite{Cavagna-private}. Certainly, this is a major point that needs to be carefully tested in the future numerical investigations. 
When $T \geq T_d$, on the other hand, more complicated size effects should be present. One expects that the cavity equilibrates through non-activated relaxation channels corresponding to collective 
unstable modes of size $\ell_d \sim (T - T_d)^{-\nu}$ (see discussion in section \ref{III-D} and \cite{FM}). If $R < \ell_d$ these channels are not available and relaxation must be {\it slower},
not faster (see \cite{Kob,Karmakar}). Thus one expects that the convergence towards the bulk relaxation time $\tau_\alpha$ is from above for $T > T_d$ and from below when $T < T_d$. However, the situation might be
more complicated still because of the presence of activated events above $T_d$ (see below and section \ref{V-C}).

Before moving on, it is worth addressing a worry expressed by Langer: is it consistent to take into account the 
exponentially large number of TAP states to compute a static $\ell^*$ and then to claim that after a time $\tau(\ell^*)$ needed to
visit only a few of them the whole structure has evolved? Is the full configurational entropy $\Sigma(T)$ meaningful physically, or only a small fraction of it?
This is however a standard problem in statistical mechanics: although no physical system ever visits the exponential number $\exp(N \Sigma)$ of 
accessible states, we know that after a relaxation time much smaller than $\tau_0 \exp(N \Sigma)$, thermodynamical equilibrium is reached and
the use of the full entropy is warranted. What is important is not the number of states that are effectively visited, but rather the number
of states that {\it can} be visited in the course of dynamics. This discussion would however be relevant to determine the correct power-law 
prefactor $\ell^{*z}$ in the above Arrhenius formula for the relaxation time.

\subsection{Consequences: Adam-Gibbs and Vogel-Fulcher}
\label{IV-B}

Inserting the expression of $\ell^*$ given in Eq. (\ref{ellstar}) into Eq. (\ref{taustar}), we finally obtain:
\be
\ln \frac{\tau_\alpha}{\tau_0} =  \frac{\Delta_0}{T} \left(C_d \frac{\Upsilon_0}{T \Sigma(T)}\right)^{\frac{\psi}{d-\theta}},
\ee
where $C_d$ is a numerical constant which is difficult to estimate based on the above hand-waving arguments. 
The important point, though, is that the Adam-Gibbs inverse relation between relaxation time and configurational
entropy appears very naturally. Physically, it reflects that configurational entropy is the factor that limits 
the growth of `hard' glassy regions (glassites) which resist shear. Since these regions must evolve for the system to flow, 
and since the energy barrier must grow with the size of the glassites, the Adam-Gibbs correlation follows. 

The KTW conjecture $\psi=\theta=d/2$ allows one to recover precisely the Adam-Gibbs relation, whereas the naive
values $\psi=\theta=d-1$ lead to a stronger dependence, as $\ln \tau_\alpha \sim \Sigma^{-2}$ in $d=3$.
The values of the prefactors $\Delta_0, \Upsilon_0$ (and their possible fluctuations) are obviously relevant for a more quantitative comparison with experimental data. 
According to Wolynes et al., $\Delta_0=\Upsilon_0=\kappa T$, with
$\kappa$ a numerical constant nearly independent of all molecular details. \footnote{This assumption, however, looks difficult to reconcile with the recent results 
of Berthier and Tarjus \cite{Berthier-Tarjus}, who find that changing the attractive part of the Lennard-Jones potential does not change the thermodynamics (hence $\ell^*$) but changes 
a lot the dynamics.} Using the REM like model above for the
configurational entropy then leads to a modified Vogel-Fulcher form for the relaxation time: 
\be
\ln \frac{\tau_\alpha}{\tau_0} = \frac{2C_d \kappa^2\beta_g^2}{\Delta C_p (\beta_K^2-\beta^2)} \approx 
\frac{C_d \kappa^2 T_K}{\Delta C_p (T-T_K)} \quad {\mbox{when}} \quad \beta \approx \beta_K \approx \beta_g  
\ee

The calculation of the fragility parameter $m$, defined in Eq. (\ref{m-def}) is quite interesting. Within the Wolynes choice of parameters one finds, independently of the precise shape of $\Sigma(T)$:
\be
m= -T \left.\frac{\partial \log_{10} {\tau_\alpha}}{\partial T}\right|_{T_g} = m_0 \frac{\Delta C_p}{\Sigma(T_g)},
\ee
where $m_0=16$ is the number of decades separating the microscopic time scale and the conventional relaxation time defining a glass. 
But within the same Wolynes framework $\Sigma(T_g)$ is nearly universal, since by definition of $T_g$, $\widetilde{m_0} = C_d \kappa^2/\Sigma(T_g)$ (with $\widetilde{m_0}= \ln 10 \, m_0$)
The final result is therefore a simple relation between the fragility $m$ and the jump of specific heat $\Delta C_p$, first derived by Xia and Wolynes \cite{XW1}:
\be
m =  \frac{\widetilde{m_0} m_0}{C_d \kappa^2} \Delta C_p.
\ee
Experimentally, the slope between $m$ and $\Delta C_p$ is of order $7$ (see Fig. 3-b), leading to $\Sigma(T_g) \approx 2 k_B$ per particle, independently of the values of $\kappa$ and $C_d$.
\footnote{Lubchenko \& Wolynes \cite{LW,LW2} quote $\Sigma(T_g) \approx 1 k_B$ per ``bead''.}
This result is quite reasonable indeed: as mentioned in the introduction, molecular glasses are characterised
by an excess entropy of a few $k_B$ at $T_g$ (see Fig. 1). 

Another possible choice would be $\Delta_0 = \Upsilon_0 = \kappa T_g$, independently of temperature. The fragility parameter is now found to be:
\be
m = 2 m_0 + \frac{\widetilde{m_0} m_0}{C_d \kappa^2} \Delta C_p,
\ee
i.e. a non zero intercept in the relation between $m$ and $\Delta C_p$ can be accommodated by relaxing the strict proportionality between $\Delta_0, \Upsilon_0$ and temperature.

Within both hypothesis, the relevant mosaic length scale at $T_g$ is also universal, and given by:
\be
\ell^*(T_g) = \left(\frac{C_d \kappa}{\Sigma(T_g)}\right)^{2/3} = 
\left(\frac{\widetilde{m_0}}{\kappa}\right)^{2/3} \approx (7 \ln 10 C_d)^{1/3}
\ee
corresponding to a number of particles inside the `critical' cavity $N^*(T_g)=(4\pi/3) \ell^{*3} \approx 70 C_3$ in $d=3$. Taking $C_3 \sim 1$ for lack of other
natural choices, one finds $N^* \sim 70$; less than a hundred particles have to move together to make the system flow at the glass temperature. 
This is not inconsistent with various experimental estimates of this number in molecular glasses \cite{Ediger,Science,Dalle,Zamponi,chi3}. 
Note however that the associated length scale at $T_g$ is still very modest, $\ell^*(T_g) \sim 3$. 

Both the above choice for $\Delta_0$ and $\Upsilon_0$ is however subject to quibble and is in our opinion an open
and pressing problem. That the interface 
energy between two TAP states, and the energy barrier between them, decrease when $T$ is reduced, is not 
intuitive at all, in particular in view of the temperature dependence of the shear modulus. \footnote{Note however that Wolynes' 
interpretation of $\Upsilon_0$, in terms of the interface energy between the 
``liquid'' and a TAP state, and not between two typical TAP states, is a little different from ours above.}
Why should the microscopic mismatch between two TAP states decrease at low temperature whereas
the shear modulus (a proxy for a local barrier when multiplied by a microscopic volume $a^3$) does just the opposite? \cite{Dyre-exp,Nelson} 
This discussion is in fact related to Dyre's ``shoving model'' \cite{Dyre,Dyre-RMP}, where the
super-Arrhenius behaviour of the $\alpha$-relaxation time is entirely attributed to the growth of the 
high frequency shear modulus, through: $\ln \tau_\alpha/\tau_0 \propto G_\infty(T)/T$, with no contribution of
the configurational entropy. We will come back to this point in section \ref{VI-A}.

\subsection{Dynamic fluctuations: stretched exponentials and facilitation}
\label{IV-C}

As we explained in section \ref{III-F}, we expect strong local fluctuations to be present in glassy, amorphous systems. These fluctuations affect
both the size of the glassites $\ell^*$ and the energy barrier parameter $\Delta_0$. The relative fluctuations of the local relaxation time 
are given by:
\be
\frac{\delta \tau_\alpha}{\tau_\alpha} = \ln \frac{\tau_\alpha}{\tau_0} \times \left[\frac{\delta \Delta_0}{\Delta_0} +\psi \frac{\delta \ell^*}{\ell^*}\right].
\ee
Particularly low free energy states have a large $\ell^*$ and presumably a large $\Delta_0$; however $\Upsilon_0$ and $\Delta_0$ also fluctuate 
on their own, with relative fluctuations of order $\ell^{*-\omega}$, where $\omega$ is another unknown exponent. The free-energy induced
fluctuations, on the other hand, are of order $T_g \sqrt{\Delta C_p}\ell^{*-d/2}$. The total variance of the local fluctuations of 
relaxation time is therefore given by:
\be
\left\langle \left(\frac{\delta \tau_\alpha}{\tau_\alpha}\right)^2 \right\rangle = \ln^2 \frac{\tau_\alpha}{\tau_0}\times 
\left[A_\Upsilon \ell^{*-2\omega} + A_f \frac{T_g^2 \Delta C_p}{T^2 \Sigma(T)^2} \ell^{*-d}\right],
\ee
where $A_\Upsilon$ and $A_f$ are numerical constants with subscripts tagging their physical origin. As is well known, local fluctuations in the 
relaxation time induce both a stretching of the relaxation function, and a decoupling between viscosity and self diffusion, often called
SER violations \cite{Ediger,Tarjus2,GCSER}. A rough, but useful interpolation formula relates the exponent 
$\beta$ of the stretched exponential relaxation to the width of the relaxation time distribution through \cite{MB}:
\be
\beta(T) \simeq \left(1 + \left\langle \left(\frac{\delta \tau_\alpha}{\tau_\alpha}\right)^2 \right\rangle \right)^{-1/2}.
\ee
This relation is interesting. If we believe that $\ell^*(T_g)$ and $\Sigma(T_g)$ are indeed universal, 
then very fragile liquids with $\Delta C_p \gg \Sigma(T_g)$ should be such that, in $d=3$:
\be
\beta(T_g) \approx  \Sigma(T_g) \sqrt{\frac{\ell^{*3}}{A_f m_0^2 \Delta C_p}} \approx \sqrt{\frac{C_3}{A_f\,m }},
\ee
where we have used the linear relation between $m$ and $\Delta C_p$ obtained above. 
This $1/\sqrt{m}$ dependence appears to be compatible with the compilation of results provided in \cite{XW2}, where the proportionality
constant is found to be $\approx 3$, leading to a $A_f \approx C_3/9$. An inverse dependence between $m$ and $\beta$ is also reported in \cite{Alba}.

In the same limit of fragile liquids where the free-energy fluctuations should dominate, one finds that the
stretching exponent $\beta$ decreases when temperature is decreased, as seen experimentally. More precisely 
the relation reads $\beta(T) \propto T \ell^{*\theta/2-\psi}$ in the limit $\beta(T) \ll 1$. In particular, $\beta(T\to T_K^+) \to 0$
whenever $\psi \geq \theta/2$, in particular if $\psi = \theta$.

The typical fluctuations of the relaxation time are adequate to understand the shape of the correlation function 
around its inflexion point. But the long time behaviour of the relaxation will be dominated by very rare, 
but extremely stable states. One can come up with a droplet argument, inspired from spin-glasses \cite{FH}, 
for the relaxation function that illustrates this point. 
Let us consider droplets of all sizes $\ell$ around a given point in space. Each droplet can have a free-energy
$f$ not necessarily equal to $f^*$. For a given $\ell$ and $f$, the probability to be still in the initial 
configuration after time $t$ is:
\be
p_\ell(t|f) = \frac{1+z_\ell(f)e^{-\frac{t}{\tau(\ell)}}}{1+z_\ell(f)}, 
\qquad z_\ell(f) \approx \exp\left[(\Sigma(T)-\beta(f^*-f))\ell^d - \beta\Upsilon_0 \ell^\theta\right],
\ee
where $\tau(\ell) = \tau_0 \exp(\beta \Delta_0 \ell^\psi)$, and we neglect the fluctuations of $\Upsilon_0$ and 
$\Delta_0$ for simplicity, which is a better approximation for fragile materials according to the discussion above. The above equation 
tells us that after a time $\sim \tau(\ell)$, the probability to be still in the initial state reaches its 
equilibrium value $(1+z_\ell)^{-1}$, where $z_\ell$ is the weight of all other droplets of size $\ell$ that can nucleate. In principle, this
quantity should read:
\be
z_\ell(f) = \int {\rm d}f^\prime e^{\sigma(f^\prime,T)\ell^d -\beta(f^\prime-f)\ell^d - \beta\Upsilon_0 \ell^{\theta}},
\ee
and a saddle point approximation leads to the above expression with $f^\prime \approx f^*$ and $\sigma(f^\prime,T) \approx \Sigma(T)$.

The correlation function is the probability that no droplet, whatever its size, was able to change state. Therefore:
\be
C(t) \approx \prod_\ell \left[\int {\rm d}f \frac{e^{(\sigma(f,T)-\beta f)\ell^d}}{Z(\ell)} p_\ell(t|f)\right].
\ee
This expression is complicated to analyse in general but is quite interesting since it should in principle provide an approximate
description of $C(t)$ in the $\beta$ regime $C(t) \approx q^*$ in terms of droplet excitations. 
For very large $\ln t$, a saddle point estimation leads to the following result:
\be\label{relax-droplets}
\ln C(t) \sim_{\ln t \gg 1} -\Delta C_p T_g^2 (\beta_K-\beta)^2 \left(\frac{T \ln t}{\Delta_0}\right)^{d/\psi},
\ee
i.e. a relaxation slower than any stretched exponential, that is dominated by the extremely low free-energy, 
``preferred structure'' states with $f=f_{\min}$. Note that:
\begin{itemize}
\item This asymptotic decay law allows one to define a characteristic relaxation time $\tau_\infty$. From $C(t>\tau_\infty) < \varepsilon$, one 
finds:
\be
\ln \tau_\infty \propto \left(T-T_K\right)^{-2 \psi/d},
\ee
which diverges with the same exponent as $\ln \tau_\alpha$ whenever $\theta=d/2$, independently of the value of $\psi$.
\item This slow relaxation tail is expected to be present
in the range $T_d < T < T_0$. Above $T_0$, even the deepest states become unstable (see also section \ref{V-C} below).
\end{itemize}

Note that when an activated event takes place within a droplet of size $\ell^*$, the boundary conditions of the
nearby droplet changes. There is a substantial probability that this triggers, or facilitates, and activated
event there as well, possibly inducing an ``avalanche'' process that extends over the dynamic correlation length scale $\xi_d > \ell^*$. 
The dynamics on length scales $< \ell^*$ is, within RFOT, inherently cooperative, but one 
may attempt to construct a coarse-grained description of the dynamics beyond $\ell^*$. What is the appropriate model? 
It seems to us that even if some facilitation mechanism is indeed highly plausible, the coarse-grained model should be built 
as kind of contagion model of activity with Poisson like activated initiators, rather that a Kinetically
Constrained Model with strictly conserved mobility defects \cite{GC,GCPNAS,GC-review} 
(see \cite{Cand-Dauchot-Biroli,Dauchot-new} 
for a recent discussion of this point, and section \ref{VI-C} below). In any case, the relation between the dynamic correlation length $\xi_d$, defined for example through 
three- or four-point point correlation functions \cite{JCP1,IMCT} and the mosaic length $\ell^*$ is at this stage an important open problem (see \cite{Dalle,Zamponi}). 

\subsection{Two new predictions of RFOT}
\label{IV-D}
\subsubsection{Energy relaxation}

The mosaic picture of RFOT theory suggests an interesting scenario for the relaxation of the energy per particle 
after a sudden quench from high temperatures, $T \gg T_d$, which in turn may provide a test for the theory and 
a way to measure (numerically and experimentally) important quantities such as $\Upsilon_0(T)$ and the exponents
$\theta$ and $\psi$. Although some predictions of RFOT in such an out of equilibrium aging regime were made in \cite{LW}, we
believe that the discussion below is new.

The idea is that after a sudden quench, TAP states appear locally but initially have a very small extension $\ell$.
After a rather short stage after the quench where the initial excess energy disappears, a slower regime where the 
TAP droplets have to grow to reach their equilibrium size $\ell^*$. The residual excess energy is primarily 
concentrated in the interface regions. If the typical size of the droplets at time $t_w$ after the quench is 
$\ell(t_w) \ll \ell^*$, it is reasonable to expect that the excess energy $e$ per unit volume is given by:\cite{YT}
\be
e(t_w,T) - e^*(T) \propto \frac{\Upsilon_0(T) [\ell(t_w)]^\theta}{[\ell(t_w)]^d}
\ee
where $e^*(T)$ is the equilibrium energy at temperature $T$. If we assume that the droplets have to cross the same kind of energy 
barriers to grow as the ones that are dynamically relevant in equilibrium, the time dependent droplet size should
be given by:
\be
\ell(t_w) \propto \left(\frac{T \ln \widetilde t_w}{\Delta_0(T)}\right)^{1/\psi}, \qquad \widetilde t_w = \frac{t_w}{\tau_0}
\ee
When $\ell(t_w)$ gets close to $\ell^*$, the final relaxation stages correspond to smaller scale rearrangements and are expected to be much faster. 
Therefore, a natural conjecture for the energy relaxation of the mosaic state is:
\be\label{scaling}
e(t_w,T) - e^*(T) = \Upsilon_0\left(\frac{T \ln \widetilde t_w}{\Delta_0}\right)^{\frac{\theta-d}{\psi}}
{\cal F}\left(\frac{T \ln \widetilde t_w}{\Delta_0 \ell^{*\psi}}\right),
\ee
with ${\cal F}(x \to 0)$ equal to a numerical constant, whereas ${\cal F}(x \to \infty)$ decays quickly to zero. 
The decay of the excess energy should therefore be an inverse power of the log of time, at variance with models 
of conserved mobility defects where the decay should be faster: $\sim t_w^{-d/z}$. 
The claim is that the scaling form Eq. (\ref{scaling}) should account for different temperatures and allow one, 
if correct, to extract useful information on $\Upsilon_0(T)$ and $\Delta_0(T)$. 

Preliminary numerical simulations seem to be compatible with Eq. (\ref{scaling}) \cite{Chiara}. Careful 
calorimetric experiments might allow one to test this prediction for real supercooled liquids. However, a major
technical obstacle is to be able to reach very fast cooling rates, or else one should generalise Eq. (\ref{scaling}) to an arbitrary 
temperature cooling scheme $T(t_w)$, as was done in \cite{RFIM} for a similar problem. For example, one could study regular
quenches with constant cooling rate $\Gamma$. Up to logarithmic accuracy, the system is equilibrated up to a length $\ell(\Gamma)$
given by:
\be
\ell(\Gamma) \propto \left(\frac{T^* \ln \frac{1}{\Gamma \tau_0}}{\Delta_0(T^*)}\right)^{1/\psi}, 
\ee
where $T^*$ is the crossover temperature at which the energy barriers start growing (in fact, a slightly more precise formula would 
replace $T^*/\Delta_0(T^*)$ by $T_{int}/\Delta_0(T_{int})$ where $T_{int}(\Gamma)$ is some intermediate $\Gamma$-dependent temperature between $T^*$ and the
final temperature $T$.) As a consequence the residual energy should scale as:
\be\label{scaling2}
e(\Gamma,T)-e^*(T) \propto \Upsilon_0(T)\left(\frac{T^*  \ln \frac{1}{\Gamma \tau_0}}{\Delta_0(T^*)}\right)^{\frac{\theta-d}{\psi}}
\ee
If one believes that most of the excess contribution to the age dependent volume or dielectric constant also
comes from the interfaces, then similar scaling laws should hold for these quantities as well, with $\theta$ 
replaced by $d_s$, the fractal dimension of the interfaces. If $d_s=d-1$, the decay of these quantities 
as $(\ln \widetilde t_w)^{-1/\psi}$ should  give direct access to the value of $\psi$. These ideas could be used to interpret the volume
experiments of Kovacs \cite{Kovacs} or the dielectric experiments of Leheny \& Nagel \cite{Leheny} or of Luckenheimer et al. \cite{Lucken}

\subsubsection{Non linear flow curves}

\begin{figure}
\begin{center}
\centerline{\epsfig{file=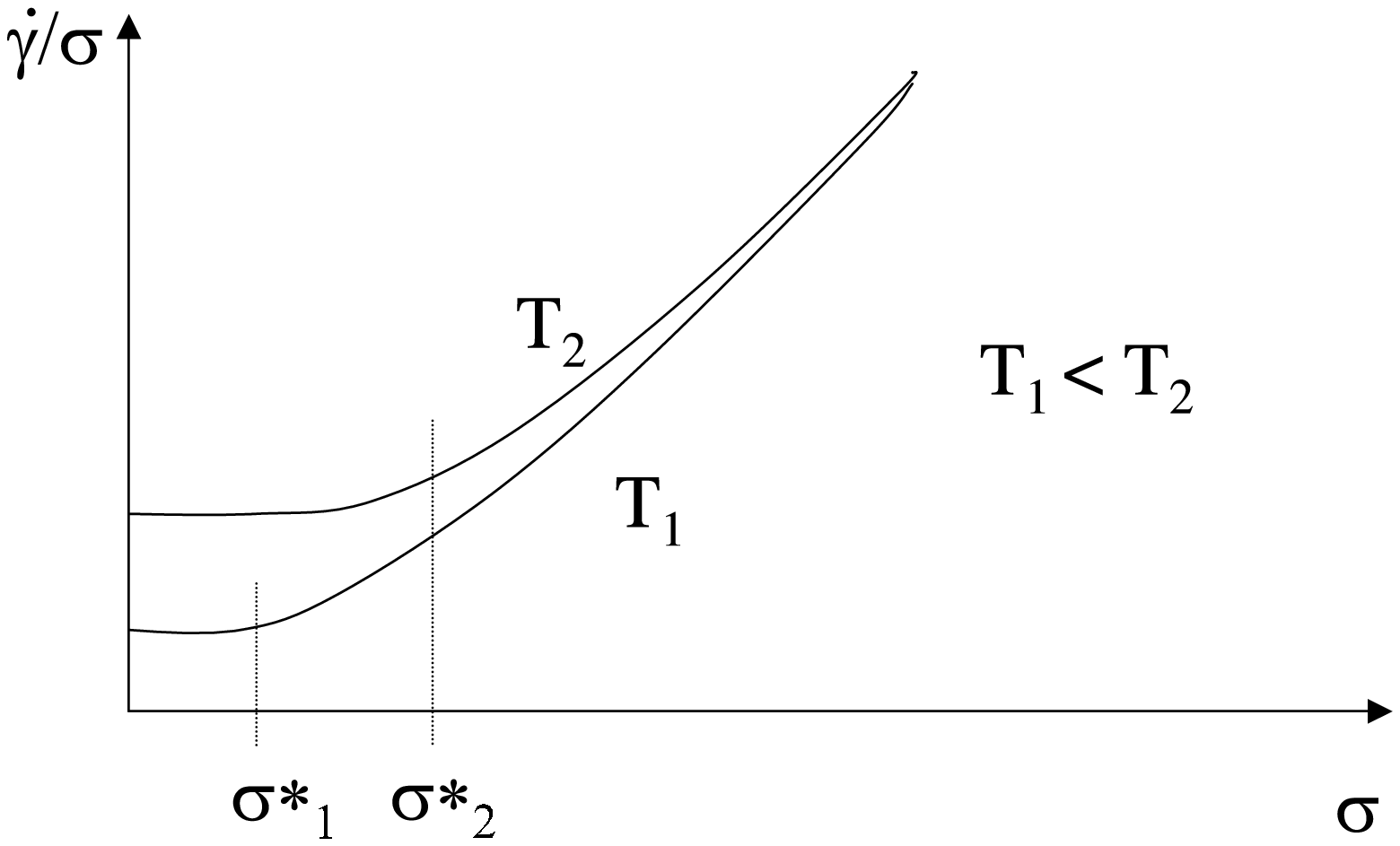,height=0.3\textheight}}
\end{center}
\caption{Sketch of the shear dependent inverse viscosity $\eta^{-1} \equiv \dot \gamma/\sigma$ as a function of $\sigma$. For $\sigma < \sigma^*$, $\dot \gamma/\sigma \approx 1/\eta_l$, 
the zero-shear (linear)
velocity. For $\sigma > \sigma^*$, some significant shear thinning is expected. If the RFOT scenario is correct, the curves for different temperatures $T_1 < T_2$ 
should look as drawn (at least for low enough temperatures when $G_\infty(T)$ is approximately constant), 
with lower temperatures corresponding to larger viscosities {\it and} lower shear thinning stress. Other theories lead to the opposite 
behaviour, where $\sigma^*$ increases upon lowering the temperature.}
\label{flow}
\end{figure}

Another interesting consequence of the putative mosaic structure of the liquid is the non-linear response to
an external shear. \footnote{This problem was very recently addressed within the context of RFOT by Lubchenko \cite{Lubchenko09}, with somewhat different 
conclusions.} The elastic energy stored in a volume $R^d$ due to a shear $\sigma$ is given by $\sigma^2 R^d/2G_\infty$. 
If the elastic energy of each glassite is smaller than the typical energy barriers $\Delta_0 \ell^{*\psi}$ that the liquid has to spontaneously cross to flow, 
the rheology will be Newtonian, with a shear rate $\dot \gamma = \sigma/\eta_l$, where $\eta_l = G_\infty \tau_\alpha$ is the linear viscosity. 
But if this elastic
energy becomes larger than $\Delta_0 \ell^{*\psi}$, the external shear lowers the barriers to flow and speeds up 
the flow. In this regime, the highest barriers correspond to a smaller droplet length scale $\ell_\sigma$ such that
both energies balance:
\be
\Delta_0 \ell_\sigma^{\psi} \sim \sigma^2 \ell_\sigma^d\sigma^2/2G_\infty \to 
\ell_\sigma = \left(\frac{G_\infty \Delta_0}{\sigma^2}\right)^{\frac{1}{d-\psi}},
\ee
such that $\eta$ becomes strongly reduced by $\sigma$ (shear-thinning), as:
\be
\eta(\sigma < \sigma^*) \approx \eta_l; \qquad \eta(\sigma > \sigma^*)\propto 
\eta_l^{Y(\sigma)}, \quad {\mbox {with}} \quad Y(\sigma)=\left(\frac{\sigma^*}{\sigma}\right)^{\frac{\psi}{d-\psi}} < 1,
\ee
where $\sigma^*$ is the crossover shear stress where the rheology becomes non-linear, corresponding to 
$\ell_\sigma=\ell^*$. Assuming for simplicity $\psi=\theta$, and $\Upsilon_0 \sim \Delta_0 \sim G_\infty$, one finds:
\be
\sigma^*(T) \sim \sqrt{T \Sigma(T) G_\infty}.
\ee
This predicts, rather non-trivially, that the cross-over shear stress where the viscosity starts plummeting 
decreases as the temperature decreases and the viscosity itself increases, at least if $G_\infty$ is approximately 
constant. As $T \to T_K$, $\sigma^*(T) \sim \sqrt{T-T_K}$. The shape of the flow curve should schematically look as in Fig. \ref{flow}, 
at least at low enough temperatures where the variation of $G_\infty(T)$ with temperature can be neglected. 
But since in many fragile systems $G_\infty(T)$ increases quite substantially between $T^*$ and $T_g$. The behaviour of $\sigma^*$ with temperature and the corresponding 
flow curves could then be quite different. 

Intuitively, the reduction of $\sigma^*$ as temperature is decreased is intimately linked to the growth of the
amorphous order correlation length $\ell^*$, below which the system behaves as a true thermodynamic glass. 
Experimental confirmation of such an evolution with temperature would be a very valuable confirmation of the
premises of RFOT, in particular because alternative pictures lead to the opposite evolution with temperature.

\subsection{Summary}

RFOT asserts that the fast growth of the $\alpha$ relaxation time is primarily due to activated rearrangements of 
glassy droplets of ever increasing size $\ell^*$, in turn induced by the reduction of configurational entropy. This 
framework allows to account naturally for the Adam-Gibbs correlation between dynamics and excess entropy. It 
allows one to rationalise the relation between fragility, relaxation time broadening and specific heat jump. 
The existence of a non-trivial glassy length scale $\ell^*$ should show up in the non-linear rheological 
properties of supercooled liquids, that should become more pronounced as $\ell^*$ increases, i.e. at lower temperatures. 
We also expect that the growth of the mosaic length $\ell(t_w)$ in out-of-equilibrium situations should have interesting observable consequences. 

\section{The mysterious MCT/RFOT crossover}

As we repeatedly argued above, one of the very strong selling point of RFOT is that Mode-Coupling Theory is naturally embedded within the theory and describe the initial
stages of slowdown of the  dynamics, at temperatures higher than the dynamical (MCT) transition $T_d$. MCT offers several precise predictions for the scaling of the relaxation function, 
and for the divergence of the relaxation time (see \cite{Gotze1,Gotze2,Das} and section \ref{II-B.3} for a short summary). Within an energy landscape interpretation, 
at high temperature the system is typically near unstable saddle points.
The slowdown is due to the fact that these saddles become less and less unstable as $T$ approaches $T_d$ from above, see \cite{Cavagna1,Cavagna-saddles}. As shown in  \cite{IMCT}, relaxation through 
these unstable modes involves a growing number of particles: the dynamic correlation length grows like $\ell_{d} \sim (T-T_c)^{-1/4}$ in mean-field. 

In finite dimensions, however, the relaxation time does not diverge because at some point activated events become more efficient a relaxation channel than unstable modes. This
is precisely in line with Goldstein's early insight about the existence of a crossover temperature $T^*$ separating ``free flow'' for $T > T^*$ from activated dynamics for 
$T < T^*$. But the details of the crossover are still poorly understood theoretically, and potentially very constraining for RFOT. In spite of several attempts, only very qualitative 
RFOT predictions for this dynamical crossover are available \cite{LW2,Wolynes-nature,Wolynes-Bagchi,Cates-Ramaswamy}, see also \cite{Das}.  In the following we will review the evidence, or lack thereof, 
of a clear dynamical crossover around $T^*$ of the kind envisaged by RFOT-MCT. We will also propose some very crude ideas to rationalise a somewhat puzzling situation. 

\subsection{Encouraging items}
\label{V-A}
\begin{figure}
\begin{center}
\centerline{\epsfig{file=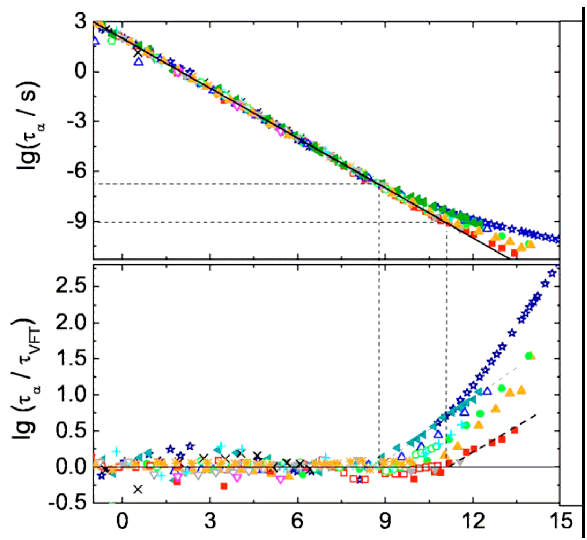,height=0.3\textheight}}
\end{center}
\caption{Evidence of a crossover temperature $T^*$ such that $\tau_{\alpha} \sim 10^{-7}$, in a representation where Vogel-Fulcher law becomes a straight line. Lower panel: same data, but
divided by the Vogel-Fulcher fit adequate for $T < T^*$. Here the crossover is most conspicuous. From \cite{Blochowitz}.}
\label{crossover}
\end{figure}

Many hints of a crossover in the dynamics around a temperature such that $\tau_{\alpha} \sim 10^{-7}$ sec. have indeed been reported in the literature \cite{Novikov-magic}. 
We show for example in Fig. \ref{crossover} the relaxation time $\tau_\alpha$ as a function of temperature in a representation where the Vogel-Fulcher law becomes a straight line \cite{Blochowitz}. This clearly reveals a change of 
behaviour around $\tau_{\alpha} \sim 10^{-7}$ sec, which is even more apparent in the bottom panel of Fig. \ref{crossover}, where the ratio of $\tau_\alpha$ to its low temperature Vogel-Fulcher
fit is represented. Several authors have insisted that the effective energy barrier $\Delta(T)=T \ln \tau_\alpha/\tau_0$ starts growing significantly below a certain temperature, 
that we identify with $T^*$, as:
\be\label{deltaT*}
\Delta(T) = \Delta_>  + \Delta_0 \left(1 - \frac{T}{T^*}\right)^\varphi,\qquad (T < T^*),
\ee
where $\Delta_>$ is the temperature independent barrier above $T^*$, and $\varphi \approx 8/3$ \cite{Tarjus,Tarjus2} or $\varphi = 2$ \cite{GCU} -- see section \ref{VI-B}.

The literature lending experimental or numerical support to MCT is abundant and reviewing it is far beyond the scope of this paper; we refer to \cite{Gotze1,Gotze2,Cummins,Das} for 
overviews. We note that MCT seems to account well for some aspects of the initial slowing down of the dynamics, in particular the appearance of a two step relaxation decay 
with a non trivial $\beta$ regime that abides Time-Temperature superposition (see Eq. (\ref{tts}). 
The phenomenon of dynamical heterogeneity, that experimentally or numerically emerges already within the MCT regime, is also correctly captured 
by the theory \cite{BBMCT,IMCT,JCP1} -- at least qualitatively. This is actually a very stringent test for MCT; as a matter of fact the founding fathers of MCT never believed 
that the theory could predict the growth of such a dynamic length scale! But had that not been the case, it would have been
hard to argue that MCT is a physically sound theory for supercooled liquids. 

However, quite unfortunately, none of the {\it quantitative} MCT predictions can be tested beyond any quibble. This is because
all these predictions only become exact (a) in mean field and (b) extremely close to the dynamic singularity $T_d$ (see ref. \cite{SBBB}). Not only critical fluctuations are expected
to renormalize all MCT predictions whenever the dimension of space is less than $d_c = 8$ \cite{IMCT,JCP1,BBSER}, but the dynamic singularity itself in fact disappears and 
becomes a mere crossover.
Therefore MCT cannot claim victory before we have a full understanding of these two issues. The details of the MCT-RFOT crossover is clearly part of this predicament.

\begin{figure}
\begin{center}
\centerline{\epsfig{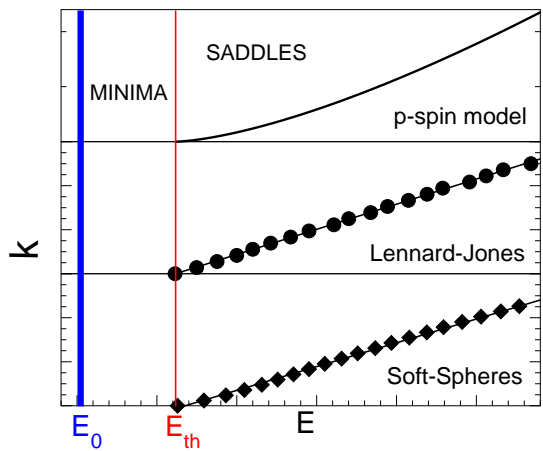}}
\end{center}
\caption{Index $k$ (fraction of unstable directions) of the most probable saddles as a function of their energy, for the $p$-model (analytical result) and 
for two models of glass formers (numerical 
results). In agreement with the mean-field scenario, the most probable saddles seem to become stable below a certain energy threshold $e_{th}$. 
For the $p$-spin model, $k \propto (e-e_{th})^{3/2}$. Courtesy of A. Cavagna and I. Giardina.}
\label{saddles}
\end{figure}

There has also been some numerical efforts to investigate directly the nature of the saddle points of the energy landscape that are probed by the dynamics \cite{Cavagna-saddles}. 
The results are compatible with the idea of an energy threshold (or more probably an energy crossover) above which saddles are indeed mostly unstable. The average index 
(i.e. the fraction of negative eigenvalues of the Hessian matrix) of the saddles as a function of the energy per particle is shown in Fig. \ref{saddles}, both for realistic models of glass formers 
and for the p-spin model. As we mentioned above, the vanishing of the saddles' index is associated to the divergence of a correlation length, which in turn should be 
associated to a delocalisation of the corresponding eigenvector of the Hessian matrix. Although direct evidence for this is lacking (to say the least -- see \cite{Coslovich}), 
numerical investigations of dynamic 
correlations within the
expected MCT region indeed reveal the existence of a growing dynamic length $\xi_d=\ell_d(T)$ \cite{TWBBB,JCP2,Dalle,Stein,Karmakar}. The quantitative agreement with the
predictions of MCT is however still under debate. From that perspective, one important qualitative result is that dynamical correlation length appears 
to increase quite fast (as a power law of relaxation time) in the MCT region, crossing over to a much slower,logarithmic growth at low temperatures \cite{JCP2,Dalle}, as expected if activated dynamics 
sets in. Thus one indeed finds some indication of a cross-over in the structure of 
dynamical heterogeneities compatible with the MCT-RFOT scenario.

In summary, MCT certainly reproduces qualitatively many of the physical phenomena pertaining to  the dynamics
of moderately supercooled molecular glass formers. However, it cannot claim victory on the basis of its quantitative predictions: some of them 
are indeed remarkable but there are also notable failures that will be discussed in the following section. One could argue that this is inherent to the fact 
that in finite dimensions the MCT transition must necessarily morph into a cross-over. This is indeed reasonable, 
but it also implies that validating the theory by testing its quantitative predictions is a hopeless task.
We believe that one should instead devise demanding tests to ascertain that the physical phenomena taking place around Goldstein's cross-over temperature $T^*$ 
are indeed the ones encoded, even in a oversimplified way, by the MCT formalism.  A truly smoking gun evidence should result from the study of finite size effects. 
These should be strongly anomalous, in the sense that above $T^*$, smaller system should relax {\it more slowly} than larger systems, in particular when $R < \ell_d$, 
since unstable modes should be stabilised by the boundary conditions (see section \ref{III-D}). Instead, below $T^*$, the relaxation mechanism should consist in activated dynamics and within RFOT one expects that smaller systems should relax faster, at least until $R \approx \ell^*$. Some evidence for such an anomalous size dependence has indeed
been reported in \cite{Kob,Karmakar,Cavagna-private}, but more work in that direction is clearly needed.

\subsection{Worrying items}
\label{V-B}
The most striking feature that appears to be in blunt disagreement with the MCT predictions is that activated events are in fact already present well above $T_d$: 
\begin{itemize}
\item the average energy of the inherent structures seem to dip at an onset temperature $T_0 > T_d$ \cite{Sastry-onset,Kob-onset};
\item numerical work clearly reveal activation between `meta-basins' that dominate the long time dynamics above the estimated MCT temperature \cite{DRB,Heuer};
\item Vogel-Fulcher fits and Adam-Gibbs correlations with the configurational entropy have been reported {\it above} $T_d$ both for experimental and numerical data \cite{Stickel,Karmakar}; 
the power-law increase $\tau_\alpha \propto (T-T_d)^{-\gamma}$ predicted by MCT holds at best over three decades in $\tau_\alpha$ and even this has been questioned;
\item The shape of the relaxation function $C(t)$ in the $\alpha$ region, or of the $\alpha$ peak in dielectric spectra, are suspiciously similar above and below $T_d$ (see e.g. \cite{Blochowitz}).
\end{itemize}
All these elements suggest that the cross-over at $T^* \sim T_d$ -- if it indeed exists -- is not very sharp, 
whereas the RFOT-MCT scenario suggests, at least naively, that the relaxation channels in the $\alpha$-regime should be completely different: 
unstable modes at high temperatures crossing over to activation at lower temperatures. 

A related worry is the fact that all estimates (numerical and experimental) of the dynamic correlation length $\xi_d$ are remarkably smooth around $T_d$. This is 
puzzling because the simplest (and probably too naive) interpretation of the analytical results of Franz \& Montanari \cite{FM} suggests that $\xi_d$ should grow 
as $\ell_d \propto (T-T_d)^{-\nu}$ until the MCT relaxation time becomes larger than the activated relaxation time related to the mosaic length 
scale. This should occur when:
\be
\ell_d^z \sim e^{\Delta_0 \ell^{*\psi}/T},
\ee
where $z$ is the MCT dynamical exponent, given by $z=4 \gamma$ in mean-field (see Eq. (\ref{MCT-gamma})). But because of the exponential dependence of the activated time scale, this
crossing is expected to occur whilst $\ell^{*} \ll \ell_d$, suggesting that the dynamic correlation length $\xi_d$ may in fact be non-monotonous with temperature, with a (curious) drop
around $T_d$. Of course, $\xi_d$ and $\ell^*$ may not necessarily be the same length at all, but it still a challenge to explain how the two regimes patch together seamlessly.

Let us finally note that the value of $T_d$ calculated within the MCT approximation is usually substantially higher than the value needed to fit MCT predictions to experimental or 
numerical data \cite{Reich-onset}. This could be considered as worrying for the MCT approach altogether. However, as shown in \cite{ABB}, MCT should in fact be understood as a kind of Landau 
theory for the
glass transition, that predicts generic scaling form for the relaxation function. In particular, all the predictions quoted in section \ref{II-B.3} above are valid beyond the MCT 
{\it approximation}, which might indeed be quite imprecise as far as the value of $T_d$ is concerned. We believe that a more accurate prediction of $T_d$ should be obtained using
static replica theory instead (see \cite{MP} and Appendix A). The only case where this has been done is hard spheres, where the critical dynamic volume fraction $\phi_d$ was computed to
be $\phi_d=0.5159$ using the MCT approximation and $\phi_d \approx 0.62$ using replicated HNC, indeed closer to the numerically determined value $\phi_d=0.592$ \cite{Ludo-Witten}.

\subsection{Ways out of the conundrum: a Ginzburg criterion for MCT?}
\label{V-C}

The presence of activated events above $T_d$ are in fact expected for systems of finite size and/or finite dimensions. Take for example a mean-field model such as the Random Orthogonal 
Model (ROM) for which unstable saddles and stable minima are strictly ``de-mixed'' for infinite system size $N \to \infty$. For finite size $N$, there is always a probability 
(exponentially small in $N$) to find the system stuck in a minimum {\it above} the (sample dependent) transition temperature $T_d$. Even if these events are rare, they clearly dominate 
the long-time equilibrium relaxation $C(t)$. These rare events are even more important when one considers the disorder averaged correlation function $\overline{C}(t)$, 
because of the order $N^{-1/2}$ sample to sample fluctuations  of the critical temperature $T_d$: at a given temperature, some samples are `hot' and dominated by the MCT unstable mode scenario, 
whereas others are `cold' and dominated by activation over
finite barriers. The physics of these sample-to-sample fluctuations has been studied in great details for the ROM in \cite{SBBB} recently. Numerical simulations of the dynamics of ROM clearly
reveal long time relaxation tails above $T_d$ that are not accounted for by MCT.

Similar fluctuation effects are also expected in finite dimensions and may affect the MCT scenario in a profound way. 
One can classify spatial fluctuations in three different categories: 
\begin{itemize}
\item the usual critical fluctuations that renormalize the MCT exponents below the upper critical dimension $d_c=8$ \cite{IMCT,JCP1,BBSER};  
\item the `nucleation' fluctuations that lead to activated dynamics;
\item the finite dimensional counterpart of the sample to sample fluctuations found in the ROM, and expected in any disordered system.
\end{itemize}
Let us focus on the last two. In order to understand their relevance,  we first derive a 
Ginzburg-Harris criterion for the MCT transition in finite dimensions. Neglecting for a while activated processes, 
we recall that the liquid above $T_{d}$ is formed by dynamically correlated regions of size $\ell_d$. 
This length measures the typical extension of the unstable modes and diverges as $\ell_d \propto (T-T_d)^{-1/4}$.
The free-energy of a correlated region, computed within a time scale less than $\tau_\alpha$, fluctuates from region to region. 
Within the simple REM (parabolic) description of the configurational entropy used throughout this paper, these fluctuations 
are of order $\delta f(\ell_d) \sim T_g \sqrt{\Delta C_p} \ell_d^{-d/2}$, which can be interpreted as local temperature fluctuations, of order:
\be
\delta T(\ell_d) = \frac{\delta f(\ell_d)}{\left|\frac{\partial f^*}{\partial T}\right|} \sim \frac{T^2}{T_g \sqrt{\Delta C_p \ell_d^d}}.
\ee
Assuming mean-field is correct, $\ell_d \propto (T-T_d)^{-\nu_{mf}}$ with $\nu_{mf}=1/4$ one finds that the temperature fluctuations within a region of size $\ell_d$ are of order:
\be
\delta T(\ell_d) \sim \frac{T_d^{2-d/8}}{T_g \sqrt{\Delta C_p}} (T-T_d)^{d/8}.
\ee
Clearly, $\delta T(\ell_d)$ should be much smaller than $T-T_d$ itself, otherwise the nature of the slowing down would be totally changed. 
Whenever $d > 8$, this criterion is always satisfied in the critical region $T \to T_d$ \footnote{Here there is a subtlety related to the existence of conserved quantities in the dynamics. 
If this is the case, as it for real liquids, then the upper critical dimension is $d_c=8$. Instead, without any conserved quantities, as it is the case for the p-spin models with
Langevin dynamics for example, the upper critical dimension is $6$. The validity of the previous argument, which assume scaling, could therefore
be questionable. Nonetheless a more careful treatment \cite{SBBB} shows that the final result is correct.}. 
In $d < 8$, however, this is only valid outside a Ginzburg-Harris region defined by:
\be
\frac{T_{GH} - T_d}{T_d} \gg  \left(\frac{T_d}{T_g}\right)^{\frac{8}{8-d}} \frac{1}{\Delta C_p^{\frac{4}{8-d}}}.
\ee
For MCT predictions to hold, on the other hand, one must have $T - T_d \ll T_d$. Both criteria can be simultaneously satisfied only if $\Delta C_p \gg 1$, i.e. for 
very fragile systems. The dynamic length corresponding to the boundary of the Ginzburg-Harris region reads:
\be
\ell_{GH} \sim \left(\frac{T_d}{T_g}\right)^{\frac{2}{d-8}} \Delta C_p^{\frac{1}{8-d}} 
\ee
which for $d=3$, $T_d/T_g \sim 1.3$ and $\Delta C_p \sim 20$ (corresponding to very fragile systems, see Fig. 3-b) is still very modest: 
$\ell_{GH} \sim 1.65$! The corresponding relaxation time 
\be
\log_{10} \frac{\tau_\alpha}{\tau_0} = 4 \gamma \log_{10} \ell_{GH} \approx 2
\ee
for $\gamma = 2.5$. This could explain why MCT predictions at best only explain the first 3 decades of increase of the relaxation time, before crossing over to a new regime. 

Within the Ginzburg-Harris region, fluctuations are important and the exponent governing the divergence of $\ell_d$ must be
renormalised from $\nu_{mf}=1/4$ to a value $\nu \geq 2/d$, such that the above criterion is asymptotically satisfied ($\nu > 2/d$), or marginally satisfied ($\nu=2/d$), 
see \cite{Spencer} for a proof that this inequality must be satisfied
generically for disordered systems. 

Let us now consider the fact that there are other types of fluctuations as well. The usual critical
fluctuations should only change the value of the exponents close enough to $T_d$. In this case the above arguments above still hold 
but with a value of $\nu$ possibly different from the mean field exponent $\nu_{mf}=1/4$ if the standard Ginzburg region turns out to be
wider that the above Ginzburg-Harris region, which is unlikely unless $\Delta C_p$ is really large. Fluctuations related to activated events, on the
other hand, are in competition with the ones above, induced by the self-consistent disorder. These are the only ones remaining in $d > 8$ and they also blur the 
MCT transition. Understanding the interplay between these two types of fluctuations is an open problem. 
We sorely lack deeper analytical investigations of the MCT transition in finite dimensions, and we can only venture to formulate some uncontrolled conjectures. 

A possible scenario is that $\nu$ is precisely equal to the lower bound $2/d$; which means that the probability that $T - \delta T(\ell_d) < T_d$ is always of order unity, 
such that $T_d$ cannot be sharply defined. 
In this case the MCT transition is blurred in the following way: a finite fraction $1 -\phi$ of regions of size $\ell_d$ are governed by the MCT relaxation mechanism and ``fast'', 
while the complementary fraction $\phi$ are effectively {\it below} 
the dynamic transition. These latter regions have lost all unstable relaxation modes and should be governed by the slow mosaic relaxation mechanism with $\ell^*(T_d) \sim 1$. 
If this is the case, the short time part of the relaxation function $C(t) > \phi q^*$ is dominated by the faster, MCT channel whereas the long time 
part, corresponding to $C(t) < \phi q^*$ is dominated by activation processes. In fact, the asymptotic relaxation regime based on the droplet argument above 
[see Eq. (\ref{relax-droplets})] should still hold above $T_d$, because of the unavoidable presence of rare, but deeply stable glassites up to the onset temperature $T_0$.
This naturally explains why activated effects are indeed observed above $T_d$ and why $\alpha$ peaks look rather similar above and below $T_d$.

\begin{figure}
\begin{center}
\centerline{\epsfig{file=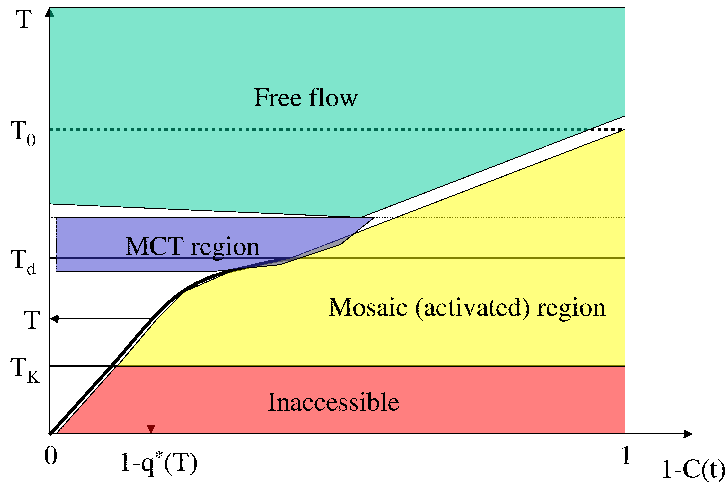,height=0.35\textheight}}
\end{center}
\caption{Cartoon of the conjectured MCT-RFOT crossover in the time-temperature plane for $d < 8$. We parameterise time between $0$ and $1$ as $1-C(t)$ (x-axis). Below the hypothetical $T_K$, 
$C(t \to \infty)=q^* > 0$, and the red region is inaccessible. The yellow region corresponds to activated dominated dynamics, that extends up to the onset temperature $T_0 > T_d$ if $t$ is large enough. 
The blue MCT 
``sliver'' region is confined to the immediate vicinity of $T_d$ and for short times, mostly in the $\beta$ regime. The green region at high temperatures corresponds to free flow, when barriers are 
irrelevant because typical saddle points are highly unstable.}
\label{cartoon}
\end{figure}

Conversely, there should still be occasional unstable soft modes below $T_d$, that dominate the short time behaviour of the correlation function.
It would be very important to formulate these arguments more precisely, even at a phenomenological level. In any case, the following ideas seem to us crucial to understand 
the MCT/RFOT crossover and avoid contradictions with empirical data:
\begin{itemize}
\item Fluctuation effects dominate the physics around $T_d$ and lead to a {\it coexistence} of MCT and activated dynamics, which are hard to disentangle on intermediate time 
scales. This does not happen in mean-field situations, nor in the Kac limit, and requires analytical tools unavailable at this stage.
\item In the temperature-time plane, there should only be a sliver region that is well described by MCT predictions (see Fig. \ref{cartoon}).
\item The crossover between the MCT relaxation mechanism and the activated mechanism in practice takes place whilst all correlation lengths $\xi_d, \ell_d$ and $\ell^*$ are still very small 
and close to each other such that no `kink' or non monotonicity of $\xi_d(T)$, that could indeed occur in mean-field, will ever be detected experimentally. 
\end{itemize} 
The above arguments also suggest an interesting lead to study the MCT transition in finite dimensions: the important parameter that controls the width of the Ginzburg-Harris region
appears to the specific heat jump $\Delta C_p$; at least one kind of fluctuations can be tuned down  for models systems such that $\Delta C_p \to \infty$. This may allow one to understand 
the MCT cross-over in more details. 

\section{Comparison with other theoretical approaches}

In order to assess the achievements of RFOT and understand its strengths and weaknesses, it is useful to compare the basic premises of the theory, as well as its most salient predictions, 
to those of other approaches. There is in fact a bevy of different theoretical pictures for the glass transition. 
We certainly do not want to review all of them but will pick of few proposals that seem relevant to us, either because they represent viable alternatives to RFOT, 
or because they have been actively discussed in the recent years (or both !). \footnote{We refer the reader to \cite{Langer-chain,Donth,Ken,Freed,Procaccia} for other interesting ideas about the
glass transition.}

\subsection{Elastic models}
\label{VI-A}

Dyre's ``shoving model'' \cite{Dyre,Dyre-RMP} postulates that elementary events that allow supercooled liquids to flow are nucleation of ``voids'' that allow particles to move around 
and unjam. The energy cost $\Delta$ of such voids comes from the elastic deformation of the surroundings during the short instant during which the void is created. 
This deformation only involves shear, and Dyre argues that the corresponding energy barrier is:
\be\label{Dyre1}
\Delta(T) \propto G_\infty(T) R^{*3}
\ee
where $R^*$ is related to the size of the critical void, which is assumed to be temperature independent, and $G_\infty$ the high frequency shear modulus, relevant for these rare, but supposedly quick, 
activated events. 

The non-Arrhenius behaviour of $\tau_\alpha$ is therefore {\it entirely ascribed} to $G_\infty$; the crossover temperature $T^*$ is associated to a noticeable stiffening of the
liquid at high frequencies. Since the thermal fluctuations of the particles around their amorphous equilibrium position is also dominated by shear modes, the plateau 
value $u^2_p$ of the mean-squared displacement are also given in term of $G_\infty$ as:  
\be
u^2_p \equiv \langle u^2(\tau_0 \ll t \ll \tau_\alpha) \rangle \approx \frac{2T}{G_\infty(T)}.
\ee
Interestingly, the two equations above relate the short time displacements of the particles to the slow relaxation time of the system. Eliminating $\Delta(T)$ leads to:
\be\label{Dyre2}
\ln \frac{\tau_\alpha}{\tau_0} = A \frac{a^2}{u^2_p},
\ee
where $a$ is the average inter-atomic distance and $A$ is a temperature independent numerical constant. This relation was actually first proposed by Hall \& Wolynes in 1987 \cite{HW}, 
but within a framework quite different from RFOT -- in fact, this relation is very unnatural within the context of Wolynes' ``standard'' version of RFOT, see below for more. 

Now, there are a number of very good points about Dyre's theory. First it is extremely simple and the assumptions are straightforward and transparent. Second, the model fares quite well
at accounting for the non-Arrhenius dependence of $\tau_\alpha$ with temperature. Plotting $\ln \tau_\alpha$ as a function of $G_\infty/T$ leads to much better (but not perfect) straight lines 
than when plotted against $1/T$, as indeed predicted by Eq. (\ref{Dyre1}) \cite{Dyre-RMP,Dyre-exp,Nelson}. Third, the correlation between $\ln \tau_\alpha$ and short time motion, 
Eq. (\ref{Dyre2}), is well supported by recent experiments \cite{Niss}. Actually, recent simulations \cite{Harrowell} have shown that this relationship holds even locally in model 
supercooled liquids: `soft' regions where $u^2_p$ is large indeed relax faster than `hard' regions. 

In this purely local picture, where the super-Arrhenius behaviour is totally determined by $G_\infty$, the Adam-Gibbs correlations is deemed not fundamental but merely fortuitous. 
Despite this fact, the model, if correct, has at least to be compatible with these Adam-Gibbs correlations: between energy barriers and excess entropy on the one hand, between
jump of specific heat and fragility on the other hand. A way out is suggested below. Another correlation that should be explained is why more fragile materials seem to have a lower stretching 
exponent $\beta$ \cite{XW2,Alba}, and a stronger violation of SER \cite{Ediger}, suggesting more fluctuations. Of course, one can always argue that these correlations are, again, nonexistent 
or artefacts. 

In summary, the correlation between the effective energy barrier and the high-frequency shear modulus, that explains at least part of the super-Arrhenius behaviour, is certainly 
striking and suggestive. The most serious issue with this approach is, however, related to its core assumption: that the glassy slowing down is a purely local phenomenon, and that 
the size $R^*$ of the critical void is temperature independent. This is at odds with our current understanding of dynamical heterogeneities, and with all 
recent theoretical and numerical results on the growth of a characteristic length as temperature is decreased (see e.g. \cite{JCP1,JCP2}).  
Experimental results on dynamical correlations also clearly favour a growing length scale \cite{Ediger,Science,Dalle,Marty,Lechenault}. 
In particular, the very recent measurements of the non-linear dielectric constant of glycerol unambiguously reveal that coherent amorphous order indeed propagates over larger 
and larger distances as the temperature is reduced \cite{chi3}. Although one could still argue that these facts are mere consequences and not primary causes of the slowing
down, recent numerical simulations \cite{Nphys,Cavagna-private} seem to rule out that the growth of the effective barrier is due to a purely local process. 
In particular, as we mentioned in the discussion of finite size effects above (see section \ref{IV-A}), the relaxation time $\tau(R)$ of a cavity of size $R$ grows with $R$ up to the
thermodynamic mosaic length $\ell^*$ and saturates for $R > \ell^*$ \cite{Cavagna-private}. If the explanation of the slowing down was local in space and only due to the growth of 
$G_\infty(T)$, one would expect no change in $\tau(R)$ as soon as $R > R^*$, where $R^*$ is temperature independent, contrary to numerical observation. \footnote{It would in fact be interesting to 
measure $u_p^2$ in the same conditions. A breakdown of the relation between this observable and the relaxation time as $R$ is varied may indicate that Eq. (\ref{Dyre2}) is less 
fundamental than anticipated.} Note that since $G_\infty(T)$ is expected to saturate at low enough temperatures, the energy barrier should stop increasing if
$R^*$ is indeed temperature independent. In the Dyre model, $\tau_\alpha$ should revert to a purely Arrhenius behaviour at low temperature, 
as indeed expected if no growing order of any kind is present in the system. Finally, a possible problematic point of this model is the value of $R^*$, found to be in the range
$3-4$ Angstroms using the data of \cite{Nelson}. It would be very important to rationalise this number in terms of a realistic microscopic rearrangement process. 

Dyre's model and RFOT might in fact be more akin than may appear at first sight. Even if one accepts that an important contribution to the super-Arrhenius behaviour is indeed 
the anomalous growth of $G_\infty(T)$ when $T$ is less than the Goldstein temperature $T^*$, then we should explain the microscopic mechanisms for such a sudden change of behaviour of the system
around $T^*$. Here, the Goldstein argument of a crossover between unstable saddles and stable minima seems unavoidable, because any growth of $G_\infty$ is nearly tautologically related to the 
stability increase of the local configurations. Therefore, an MCT-like mechanism should be at play, at least to understand the behaviour of $G_\infty(T)$. Within an 1-RSB like 
scenario, the increase of $G_\infty(T)$ is associated to a change of dominant metastable states: as $T$ is decreased, the system probes deeper and more stable minima of the energy
landscape. Interestingly, as noted in section \ref{III-E} above (see also Fig. \ref{ginfty}), the rapid change of $G_\infty(T)$ is in fact associated to an important change in the vibrational entropy. 
This could provide a natural explanation of the Adam-Gibbs correlation, without any {\it direct} link between the configurational 
entropy and energy barriers, but only a co-variation of these quantities, as shown in Fig. \ref{ginfty} and discussed in \cite{Dyre-RMP,Wyart-New}. 

But if this 1-RSB mechanism is indeed the explanation for the increase of $G_\infty(T)$, 
the mosaic argument should be valid too, re-introducing the increase of the mosaic length $\ell^*$ as a source of super-Arrhenius behaviour. A possibility is that in the experimental
temperature range, the energy barrier $\Delta(T) = \Delta_0 \ell^{*\psi}$ increases both because $\Delta_0 \propto G_\infty$ increases, as in Dyre's shoving model, and because $\ell^*$ 
increases, although in a very modest fashion. This would allow one to account for the clear curvature in the plots of $\ln \tau_\alpha$ as a function of $G_\infty/T$ (see e.g. 
\cite{Nelson}, Figs. 5 and 6), and reconcile Dyre's mechanism with the existence of a growing length scale. 
On the other hand, all the quantitative estimates based on Wolynes's choice of RFOT parameters (see section \ref{IV-B}) would need a complete overhaul. 
Another possibility is that the barrier crossing leading to relaxation on the scale $\ell^*$ involve smaller and smaller displacements 
per particle, $\delta$, upon lowering the temperature. This would lead to an effective barrier that is of the order $G_\infty \delta^3 \ell^{*\psi}$, where $\delta^3$ decreases 
approaching $T_K$ and $\ell^{*\psi}$ increases, providing an interpretation of the correlation between $\Delta(T)$ and $G_\infty$, and 
leaving room for $\ell^*$ to grow. In any case, a convincing reconciliation between the shoving model and RFOT is needed for both theories; this means in particular elucidating the relation 
between $\ell^*$ and $R^*$ (if any).
This seems to us a very important point to clear up in future research. 

Finally, coming back to the issue of a growing length scale, a clear-cut distinction between the shoving model with a fixed $R^*$ and the mosaic 
picture with a growing $\ell^*(T)$ is the behaviour of the cross-over stress $\sigma^*$ beyond which shear-thinning is expected. In the 
mosaic picture, $\sigma^*$ should scale as $\sqrt{T G_\infty \Sigma(T)}$. In the shoving picture, on the other hand, $\sigma^*$ is simply given by $G_\infty$. 
This should lead to testable differences: a relatively fast growth of $\sigma^*$ when $T$ decreases for the shoving model and a roughly constant or even a mild decrease for RFOT. 

\subsection{Frustration Limited Domains}
\label{VI-B}
This class of models has been advocated by Kivelson, Tarjus \& coworkers since the mid-nineties \cite{Tarjus,Tarjus2,Tarjus-review}, building upon older ideas dues to 
Frank, Sadoc et al., Nelson and others \cite{Frank,Sadoc,DNelson}. The basic tenets of this approach have been very clearly articulated in the review \cite{Tarjus-review}, 
which we are satisfied to reproduce almost identically here: 
{\it

1) A liquid is characterised by a locally preferred structure (LPS) which is different than that of the crystalline phases.

2) Because of geometric frustration, the LPS characteristic of a given liquid cannot tile the whole space.

3) It is possible to construct an abstract reference system in which the effect of frustration is turned off.

Frustration naturally leads to collective behaviour on a mesoscopic scale (...). The collective property comes from the growth of some
ordered phase (in which the liquid LPS spreads in space) induced by the proximity of an (avoided) critical point.  The limitation of 
the scale over such a growth can take place results from frustration that aborts the phase transition and leads to ``avoided criticality''.}

There are some analogies, but also deep differences, between the ideas of Frustration Limited Domains (FLD) and RFOT. As in RFOT, there is a crossover
temperature $T^* \sim T_d$ where some kind of local thermodynamical order set in: amorphous TAP states in one case, PLS in the other. This gives rise to high energy barriers because 
collective motions are needed for any rearrangement to take place. 
As in RFOT, the size $\ell^*$ of the glassites are limited for any given temperature $T < T^*$, and
grows when $T$ decreases. As in RFOT, the energy barrier is postulated to grow as a power of $\ell^*$. 

However, the size $\ell^*$ is limited by configurational entropy in the case of RFOT, 
and by long range elastic energy that frustrates the local order in the case of FLD. More precisely, the proponents of
FLD argue that $\ell^*$ grows as $(1-T/T^*)^\nu/K^{1/2}$ where $K$ is an a-dimensional parameter measuring the strength of frustration, and $\nu$ is the exponent governing the growth 
of the correlation length of the unfrustrated transition ($K=0$). As expected, $\ell^*$ diverges if frustration is absent ($K \to 0$). 
The energy barrier for the relaxation of the glassites is assumed to be given by $\Delta = A\Gamma \ell^{*2}$, where
$\Gamma(T)$ is the surface tension between two different LPS. Since the (avoided) phase transition towards LPS takes place at $T^*$, 
it is reasonable to assume that $\Gamma(T)$ itself vanishes at $T^*$ as $\Gamma \propto T^{*1-2\nu}(T^*-T)^{2\nu}$. This leads to an energy barrier that behaves as \cite{Tarjus,Tarjus2,Tarjus-review}:
\be\label{deltaT**}
\Delta(T < T^*) = \Delta_> + \frac{A T^*}{K} \left(1-\frac{T}{T^*}\right)^{4\nu},
\ee
where we have added a high temperature energy barrier that governs the frequency of elementary moves. This has the shape alluded to earlier, see
Eq. (\ref{deltaT*}) and was used to fit the data shown in Fig. 2 with $\nu$ in the range $[1/2,3/4]$ \cite{Tarjus,GCU}. The value $\nu=2/3$ in fact corresponds
to an Ising like transition of the unfrustrated model in $d=3$. The model was extended to account for fluctuations of $\ell^*$ and predict the broadened
dielectric spectrum as a function of temperature and frequency, with remarkable success \cite{Tarjus-review}. 

However, we note that below a certain temperature, sufficiently far from $T^*$, both $\ell^*$ and $\Gamma$ should saturate and the barrier $\Delta(T)$
should become constant. As within the shoving model, FLD predicts reversion to a purely Arrhenius behaviour at low temperature, whereas RFOT 
suggests that $\Delta(T)$ continues to grow as the configurational entropy goes to zero.

Recently, this scenario was carefully tested \cite{Tarjus-hyperbolic} by simulating a mono-atomic liquid on the hyperbolic plane, where crystallisation is avoided 
because the non-zero curvature frustrates the crystalline order. As a consequence, this setting provides, {\it mutatis mutandis}, a benchmark to study the FLD theory. 
The results that have been found are in overall agreement with the above scaling arguments and assumptions. Signatures of LPS in glass-forming model systems have also
been reported, see \cite{Coslovich-LPS}.

There are two main issues that we would like to address concerning FLD theory. The first one is its compatibility with 
the Adam-Gibbs like kinetic/thermodynamic correlation. Kivelson and Tarjus \cite{Tarjus-entropy} have argued that FLD could cope reasonably well with the correlation 
between excess entropy and relaxation time. What is unclear to us is whether the correlation between fragility and specific heat is compatible with FLD. 
Simple -- but arguably not very reliable -- arguments suggest that there might indeed be a problem. The main difficulty is to obtain a good estimate of $S_{xs}$ within FLD. 

Kivelson and Tarjus argued in \cite{Tarjus-entropy} that the proximity of the unfrustrated phase transition leads to a singular contribution to the excess entropy given by 
$S_{xs} \approx S(T^*)-A\left(1-T/T^*\right)^{3\nu-1}$, where $A$ is a numerical constant independent of $K$, and hence of the fragility. \footnote{Note that for $\nu=2/3$, 
this predicts a simple linear behaviour of $S_{xs}(T)$ 
below $T^*$.} Although this expression indeed predicts a decreasing $S_{xs}$ below $T^*$, it implies an incorrect evolution with fragility, since it leads to a specific heat jump at $T_g$
totally independent of fragility!
 
Another physically transparent (but still moot) estimation the configurational entropy is the following.
Let $S_0$ be the total zero temperature entropy of the unfrustrated model (i.e. a measure of the degenerescence of the LPS). Since each glassite of size $\ell^*$ can independently be in $e^{S_0}$ 
different LPS, the configurational entropy per particle at temperature $T$ is given by:
\be
\Sigma(T)= \frac{S_0}{\ell^{*3}} \propto S_0 K^{3/2} (1-T/T^*)^{-3\nu}.
\ee
Comparing with Eq. (\ref{deltaT**}), this leads to an Adam-Gibbs relation for a given material (i.e. a given $K, S_0$):
\be
\log \frac{\tau_\alpha}{\tau_0} = \frac{\Delta_>}{T} +  A' \frac{T^*S_0^{4/3} K}{T \Sigma(T)^{4/3}},
\ee
with an exponent $-4/3$ instead of the usual Adam-Gibbs exponent $-1$. These values are probably hard to distinguish experimentally.

Now, taking the derivative of the above relation with respect to temperature allows one to compute the fragility $m$ as:
\be
\frac{m-m_0}{m_0} =  \frac{4\Delta C_p}{3\Sigma(T_g)} = \frac{4\nu T^*}{T^*-T_g}
\ee
where we neglect the contribution of $\Delta_>$ at $T_g$ for fragile materials, which turn out to be such that $K \to 0$ \cite{Tarjus}. 
The above direct relation between fragility and $T^*-T_g$ could be tested empirically. This would allow one to measure $\nu$ directly.
Within the same approximation, the very definition of $T_g$ leads to:
\be
\frac{AT^* (1-T_g/T^*)^{4\nu}}{T_g K} \approx m_0 \longrightarrow \frac{T^* - T_g}{T^*} \approx \left(\frac{m_0K}{A}\right)^{1/4\nu} \to_{K \to 0} 0, 
\ee
showing indeed that $m \to \infty$ when $K \to 0$. From this relation, we also find that the size of the domains at $T_g$ is 
$\ell^*(T_g) \propto K^{-1/4}$ and (weakly) grows as a function of fragility, whereas $\Sigma(T_g) \propto S_0 K^{3/4}$.
The last information we need is $\Delta C_p$ computed from the explicit form of $\Sigma(T)$:
\be
\Delta C_p \approx 3\nu \frac{S_0 K^{3/2}}{(1-T_g/T^*)^{1+3\nu}} \propto S_0 (1-T_g/T^*)^{3\nu-1} \propto S_0 K^{\frac{1-3\nu}{4\nu}},
\ee
which, combined with the above relation finally leads to:
\be
\frac{m-m_0}{m_0} \propto \left(\frac{\Delta C_p}{S_0}\right)^{\frac{1}{1-3\nu}}.
\ee
If we assume that $S_0$ does not vary much between different materials, while the change of $K$ is responsible for the change of fragility, 
we find that $m$ and $\Delta C_p$ should be {\it inversely correlated} whenever $\nu > 1/3$. For example $(m-m_0)/m_0 \propto \Delta C_p^{-1}$
for the favoured value $\nu=2/3$. Another way to see this is that $\Delta C_p$ goes to zero when $K \to 0$ when $\nu > 1/3$. 
The previous estimation of the configurational entropy does not take into account the entropy due to the wandering of 
domains between LPS. If $S_0$ is not a constant but scales with $\ell^*$ then $\Sigma(T)$ would also 
have a different scaling with $\ell^*$. Assuming $\Sigma(T)\propto 1/\ell^{*3-d_f}$, where $d_f$ is a certain fractal dimension, one can find the correct correlation 
between specific heat jump and fragility for some range of $d_f$. However, for these very same values of $d_f$ the Adam-Gibbs
relation would be completely altered in a way detectable in experiments. 

As a conclusion, it seems that obtaining the correct evolution with fragility of the configurational entropy and the specific
heat jump is a very non trivial constraint for FLD theories. Because of the absence of an analytically reliable way 
to estimate the configurational entropy it would be worth addressing this question using simulations, in particular on the hyperbolic plane where FLD ideas 
have been shown to be valid \cite{Tarjus-hyperbolic}.  

Another quite interesting aspect of FLD theory is that any reasonable analytical approximation used to deal with explicit models for which the FLD scenario 
is expected to hold (such as, for example, Brazovskii's frustrated Coulomb model) in fact all lead to RFOT-MCT equations \cite{Sch-Wol,KTW4,Grousson,KacW}. 
The crucial assumption that the number of 
LPS of the system is small, and not exponential in $N$, could therefore be unwarranted. An exponential number of amorphous metastable states emerge instead, induced by frustration. 
This extensive configuration entropy is in fact at the root of the positive correlation between fragility and jump of specific heat, captured by
RFOT.

Finally, the crossover stress in the FLD theory should be given by $\sigma^* \propto \sqrt{G_\infty \Gamma /\ell^*} \sim K^{1/4}(1-T/T^*)^{\nu/2}$, which now {\it increases} as $T$ is decreased. 
When $T=T_g$, $\sigma^*(T_g)/\sqrt{T_g G_\infty}$ should scale as $K^{3/8}$ for very fragile materials. Actually, the FLD behaviour of $\sigma^*$ when $T$ is lowered is intermediate 
between the shoving model, which corresponds to an increase proportional to $G_\infty$, and RFOT, which corresponds to a variation slower than $\sqrt{G_\infty}$. Non-linear 
rheology data for a range of fragile materials would certainly be very useful to distinguish between all these pictures.

\subsection{Others}
\label{VI-C}
Another active line of research is devoted to Kinetically Constrained Models (KCM, for reviews see e.g. \cite{Sollich-review,GC-review}). The idea behind this family of models is that 
one might be able to find a coarse-grained effective description of the dynamics of glass forming liquids where the complicated microscopic interactions that
slow down the dynamics are replaced by simple ``kinetic'' constraints. For example, one could replace the initial problem by a lattice model, where
each cell contains a certain number of particles, enough for the coarse-grained description to be valid. Each cell is then ``active'' if some
motion can take place, or ``inactive'' if jammed. The density of active cells is assumed to be given by a simple exponential: $\rho = \exp(-J/T)$,
where $J$ is an activation energy. The number of active cells is supposed to be {\it strictly conserved}: activity cannot spontaneously 
appear nor disappear, but only hop to nearest neighbours. This is where kinetic constraints come into play. A simple rule is that the hop 
is only allowed if the target cell is itself surrounded by a sufficiently large number of active cells. Several other 
rules have been considered, leading to different models, some exhibiting very interesting properties, such as the existence of a genuine dynamical 
transition at some critical density of active cells $\rho_c$ \cite{BFT}.

But the truly interesting outcome of these models is that kinetic constraints lead to cooperative dynamics: mobile regions must cluster together
in order for the system to evolve. Without constraints, the relaxation time $\tau_\alpha$ would be simply proportional to $\rho^{-1}$ in the dilute regime $\rho \to 0$, leading to 
an Arrhenius slow-down. By imposing cooperativity, kinetic constraints can deeply affect this behaviour, and give rise
to super-Arrhenius behaviour, for example $\tau_\alpha \sim \rho^{-J/T}$ as $T \to 0$, which reproduces Bassler's $1/T^2$ law mentioned in section \ref{I-B} \cite{Sollich-review,GC-review}. 
KCM also offer interesting ideas to rationalise dynamical heterogeneities and SER violations \cite{GCSER}. 

However, the basic assumption here is that kinetic constraints can lead to non-trivial and interesting effects, even if the thermodynamics 
is completely trivial. By construction, KCM has a hard time explaining correlations of the Adam-Gibbs type. In fact, the large specific heat jump $\Delta C_p$ 
at $T_g$ for fragile glasses (several $k_B$ per particle), cannot be accounted by the freezing of dilute mobility defects \cite{BBT}. The thermodynamics of
each cell in the coarse-grained model must be contributing to $\Delta C_p$ in order to cure
the problem of having too small a specific heat jump at $T_g$ \cite{GC-reply}. On the other hand, the correlation between
kinetic and thermodynamics, and its evolution with fragility, then requires that 
these correlations are encoded in the thermodynamics of the cells. But this appears to be self-contradictory, because by assumption nothing at all happens within inactive cells, 
while these represent the overwhelming majority of cells as $T$ approaches $T_g$ -- so nothing big should freeze at $T_g$ since it is already frozen above $T_g$!   

In any case, the crucial issue is to work out how a system of dense, interacting particles can {\it really} be mapped onto a KCM at sufficiently large length scales. 
The mechanism by which the system becomes locally rigid \footnote{Alternative words for this are: ``jammed'', ``immobile'' or ``inactive'', but 
all these ideas reflect the fact that the relevant configurations are at least locally stable.} in the first place must be accounted for. Most people, including the advocates
of KCM, now seem to agree that this local rigidity sets in around a rather well defined temperature $T^*$ (see e.g. \cite{GCU}); this is an essential ingredient of all
theories reviewed so far (RFOT, elastic models, FLD and KCM). In order to account for such an abrupt change in the local rigidity and the appearance of constrained 
dynamics below $T^*$, an MCT type of scenario seems to us difficult to avoid -- with stable, amorphous and/or locally preferred structures, becoming thermodynamically predominant below $T^*$. 

Once this local rigidity is installed, the scale beyond which an effective, simplified description is possible should correspond to
the size of an elementary cooperative volume, i.e. the number of particles involved in an elementary activated event. This would be $\ell^*(T)$ 
in the framework of RFOT or FLD, and $R^*$ in the shoving model. However, this issue has only been very cursorily discussed in the 
KCM literature \cite{GCPNAS}, although crucial to understand what regime of length and time scales KCM are supposed to describe. As mentioned 
in section \ref{IV-C}, one expects that some facilitation mechanism should be present also within RFOT on length scales larger than $\ell^*$: 
when an activated event takes place within a droplet of size $\ell^*$, the boundary conditions of the
nearby droplet changes, which can trigger a second activated event, possibly inducing an ``avalanche'' process 
that extends over the dynamic correlation length scale $\xi_d > \ell^*$. However, we fail to see why activity should be an almost conserved quantity, 
as assumed in the series of papers of Garrahan \& Chandler \cite{FA,FA2,KAgas,GC,GCPNAS,GC-review}. Actually, we believe that a more
likely situation is that activity becomes {\it less and
less} conserved upon lowering the temperature, as it happens in granular systems close to the jamming transition \cite{Cand-Dauchot-Biroli,Dauchot-new}. 
Spontaneous, thermally activated events occurring within immobile regions should be more and more important as the system slows down.
Clearly, this is a very important issue that should hopefully be settled by numerical simulations. 

Finally, we want to mention the work of M. Moore \& collaborators \cite{Moore1,Moore2,Moore-Yeo,Moore-Tarzia}, where the working hypothesis is that all models of glasses that are 1-RSB 
in mean field can 
in fact be mapped onto a {\it spin-glass in a magnetic field} in finite dimensions. This is indeed supported by some analytical and numerical calculations. 
Although some aspects of the physics of spin glasses look very similar to that of glasses, there are also striking differences,
such as the continuous vs. discontinuous nature of the order parameter at the transition. Nevertheless, some efforts to reproduce the phenomenology
of glasses within that framework have been reported \cite{Moore-Tarzia}.

If Moore's conclusion turned out to be valid, this would be quite a fatal blow for RFOT, except if the time and length scales
below which the mean-field 1-RSB physics survives are large enough to cover the experimentally accessible window.  However, some of the calculations bolstering this strong 
claim are quite complex and rely on subtle assumptions \cite{Moore-Yeo}, which in our opinion are not yet firmly established. 
Moore's conclusion is also at odds with the success of 1-RSB calculations for realistic models of liquids and hard-spheres reviewed in \cite{MP}. Furthermore, recent numerical 
simulations of some finite dimensional models clearly demonstrate that 1-RSB effects do seem to survive away from mean-field, at least over a large time window \cite{coniglio,SBBB2,Enzo}. 
This whole issue is  related to the fate of MCT in finite dimensions, that we have commented on in section \ref{V-C}. We lack firmly grounded theoretical work on the 
interplay between 1-RSB and finite dimensional fluctuations, without which RFOT will remain a beautiful, but fragile, phenomenological theory.

\section{Discussion and conclusions}

\subsection{Successes and Difficulties of RFOT}

Let us summarise the strong and the weak aspects of RFOT that are strewn in various parts of this paper. First, some
undeniable successes:

\begin{itemize}
\item From a theoretical point of view, RFOT is a natural finite dimensional interpretation of the mean-field theory of 
generic complex systems with many minima. Two characteristic temperatures appear naturally: (i) a dynamical temperature $T_d$ where the phase 
space gets fragmented into different minima, which in mean-field corresponds to a Mode-Coupling non-ergodic transition, and in finite
dimensions to the temperature where some high-frequency rigidity appears and leads to activated dynamics; (ii) a static
temperature $T_K$ where the configurational entropy vanishes  and where the system, if equilibrated, would enter an ideal glass state. These two
temperatures echo the well-known Goldstein and Kauzmann temperatures that describe the phenomenology of supercooled liquids. Item (i)
appears to be a crucial ingredient of all viable theories to date (RFOT, FLD, elastic models, and even KCM): glasses lose their rigidity as temperature increases because at some point 
local stability is lost.

\item Any reasonable analytic approximation that describes the thermodynamics or the dynamics of realistic models of supercooled liquids, using 
e.g. density functional techniques, replica theory, the Bethe-Peierls approximation, the projection operator formalism or 
self-consistent re-summation schemes, {\it all lead to a 1-step replica symmetry broken low temperature phase} (characteristic of systems with an extensive configurational entropy) 
and to a non-ergodic transition of the type predicted by Mode-Coupling Theory. Mode-Coupling Theory can actually be derived within a Landau approach of discontinuous glass 
transitions, as a generic expansion where the analogue of the `order parameter' is the (small) difference between the correlation function $C(t)$ and its plateau value $q^*(T_d) > 0$. 
From all this, it is hard to see how a truly first principle statistical mechanics approach to the glass problem can lead to
a very different theory.
It is always possible that strong fluctuations or non perturbative effects completely change the picture in three dimensions, as is the case for example of the Kosterliz-Thouless 
transition in two dimensions. However, because of the firm mean-field foundation of RFOT, one would have to come out with compelling physical arguments for such a scenario.  

\item RFOT suggests that the liquid between $T_K$ and $T_d$ can be thought of as a mosaic of glass nodules or ``glassites'' with a spatial extension
$\ell^*(T)$ limited by the configurational entropy. 
Regions of size smaller than $\ell^*(T)$ are glasses: they are effectively below $T_K$ and cannot relax, even on very long time scales. 
Regions of size greater than $\ell^*$ are liquid in the sense that they explore with time an exponentially large number of unrelated configurations, and
all correlation functions go to zero. The relaxation time of the whole liquid is the relaxation time of glassites of size $\ell^*$. 
The crucial assumption, that {\it thermodynamics alone fixes the value of  $\ell^*$}, 
is well borne out by static and dynamic simulations of a liquid inside a cavity. The existence of a thermodynamic ``point-to-set'' correlation
length $\ell^*$ that diverges as $T \to T_K^+$ is supported by exact analytical calculation in the Kac limit of weak, long range interactions \cite{FM}. This
scenario is also compatible with recent experiments that measure the non-linear (third harmonic) dielectric constant $\chi_3(\omega)$ of a standard molecular glass, glycerol. 
If some transient amorphous order sets in 
over a length scale $\ell^*$, $\chi_3(\omega)$ is expected to grow as a power of $\ell^{*}$ when $\omega \tau_\alpha \sim 1$ but to remain small for $\omega \to 0$, 
since no static transition takes place, at variance with spin-glasses where $\chi_3(\omega=0)$ diverges at the transition. Precisely such a behaviour was seen in the Saclay group \cite{chi3}.

\item Within RFOT, the large excess entropy of supercooled liquids at $T_g$ (several $k_B$ per molecules) is attributed to the existence of an exponential number of unrelated 
stable amorphous configurations. The essential role played by the configurational entropy allows one to account very naturally, in a way not too sensitive to 
additional assumptions, for a series of empirical correlations between thermodynamics and dynamics: Adam-Gibbs relation between the relaxation time and the configurational entropy, 
correlation between the fragility and the jump of specific heat 
$\Delta C_p$, and between the stretching of the relaxation function and $\Delta C_p$. 
At least some of these correlations are not addressed, or difficult to interpret within alternative descriptions, 
such as Dyre's shoving model, the Frustration Limited Domain theory, or Kinetically Constrained Models. 
Although one can of course take the view that all these correlations are entirely
accidental, we do believe that they impose strong constraints on the theory of glasses and have to be taken 
into account \footnote{Google scholar counts 1882 citations of the Adam-Gibbs paper on November 5th, 2009, since the beginning of the nineties.}.
\end{itemize}

There is also a worrying list of loose ends, caveats and difficulties:

\begin{itemize}
\item The existence of the mosaic length $\ell^*$ heavily relies on the hazy concept of ``surface tension'' or mismatch energy between different amorphous states. It is not easy to give a precise 
operational meaning to this idea, that would allow one to compute or measure this quantity (see however \cite{Cavagna-surface,Cavagna-surface0} for an attempt). Correspondingly, there is no consensus, neither 
on the value of exponent $\theta$ determining the size dependence of such a mismatch energy, nor on the dimensional prefactor $\Upsilon_0$ and its temperature dependence. The
situation is even worse for energy barriers. We have no idea whatsoever about the nature of the activated events allowing the system to jump between on amorphous state to another 
on scale $\ell^*$ -- are these made of compact droplets, fractal objects, excitation chains, void nucleation? The exponent $\psi$ and prefactor $\Delta_0$ fixing the height of the
barriers as $\Delta=\Delta_0 \ell^{*\psi}$ are even less understood. Wolynes' prescription corresponds to $\theta=\psi=d/2$ and $\Delta_0=\Upsilon_0=\kappa T$. Although based on 
suggestive arguments, it is by no means obvious and not supported at this stage by any first principle calculation. When available, these first principle calculations 
(analytical or numerical) suggest $\theta=d-1$, but fluctuations effects are probably neglected. We also find difficult to fathom why $\Delta_0$ should decrease with temperature.

\item Related to the previous point, shouldn't the energy barrier scale $\Delta_0$ in fact be related to the high frequency shear modulus $G_\infty$? This quantity is known to have a non 
trivial temperature dependence that may account for an important fraction of the increase of the energy barrier between $T^*$ and $T_g$. This is the claim made by Dyre and others, in
the context of elastic models where the energy barrier is directly proportional to $G_\infty$, and therefore also related to the short time fluctuations 
$\langle u^2 \rangle \propto T/G_\infty$ of the particles around their average position. But if most of the energy barrier increase is due to the behaviour of the shear modulus, there
is little space left for the essential prediction of RFOT, i.e. that the super-Arrhenius behaviour is due to the increase of the mosaic length $\ell^*$, itself driven by 
configurational entropy. We have discussed this issue in more depth in section \ref{VI-A}; understanding whether the empirical correlation between the energy barrier $\Delta$ 
and the high frequency shear modulus $G_\infty$ is fortuitous or fundamental seems to us one of the most perplexing quandary that needs to be resolved.

\item Finally, the crossover between a high-temperature MCT regime and a low temperature activated (mosaic) regime, supposed to be a strong selling point of RFOT, is very poorly understood
even at a phenomenological level. There is certainly no clean crossover between a well developed MCT scaling regime above $T^*$ and activated dynamics below $T^*$, but rather an
intricate situation, perhaps resembling the sketch in Fig. \ref{cartoon}, where activated events appear well above $T_d$. We believe that this crossover is dominated by fluctuations, due both to
the finite dimensional effects that affect the MCT regime and to a broad distribution of local energy barriers. 

\end{itemize}

In order to put the MCT crossover on a less shaky ground, two pieces of evidence are needed: First, a smoking gun experiment or a simulation that 
conclusively shows the existence of a MCT-like mechanism. This could be realized by studying finite size effects and equilibration inside small cavities both below and above $T_d$.     
Second, we need a theoretical approach able to lead to precise predictions in the crossover region. This will require some major technical breakthrough. 
Until then, this potentially informative crossover regime will remain somewhat mysterious and inconclusive.

\subsection{Topics not addressed here, open problems and conjectures}

There is a number of other properties of glasses that RFOT might be able to account for, but
that we have chosen to leave out, both for lack of space and because we feel that our
level of understanding is not satisfactory. These are in particular the very low temperature properties, dominated
by `two-level systems', where quantum effects become important. Lubchenko \& Wolynes argue that some of these 
low temperature properties are inherited from the mosaic structure as the liquid freezes at $T_g$ \cite{LW}. This
is a very interesting topic on which we hope to return in the future. The relation of the so-called ``Boson-peak'' with 
MCT-like soft modes has also spurred a spree of activity in recent years, see in particular \cite{Parisi-boson,Wyart,Nagel}.
We have only rapidly touched upon some aging properties when discussing energy relaxation after a quench in section \ref{IV-D}, see also \cite{LW}.
Much more has to be done, in particular concerning two-time quantities, fluctuation-dissipation theorem, effective temperatures, 
temperature jumps and temperature cycles. One exciting idea is that the notion of ``temperature chaos'' makes sense in systems
without quenched disorder. More precisely, one could expect that the dominant state $\alpha$ within a cavity with a fixed
boundary condition abruptly change as temperature is varied, as a result of level crossing (see \cite{Krzakala-Martin,Yoshino-Rizzo}).
This could induce some interesting rejuvenation effects.

Although RFOT appears to contain all the ingredients to understand the strong violations of SER in fragile liquids (see section \ref{IV-C}),
we lack a more detailed understanding of the so-called fractional Stokes-Einstein relation that approximately holds between $T_g$ and $T^*$, i.e. $D_s \eta/T \propto \tau_\alpha^k$, 
where $D_s$ is the self-diffusion constant and $k>0$ an exponent that appears to increase with the fragility of the liquid. 
The argument must be that some activated events of size $\ell < \ell^*$ contribute to the diffusion constant but not to the stress relaxation,
for example by allowing the permutation of particles without structural rearrangements, as in a perfect crystal for example. A related poorly understood problem is the dynamics of a 
tracer particle driven by an external force in the disordered, but time dependent environment created by the surrounding particles (see \cite{fuchs-tracer} for recent results in the MCT regime).

We have not dwelled on the subject of dynamical heterogeneities either. It is clear that while RFOT predicts strong spatial heterogeneities and temporal 
intermittency in the dynamics, it is not clear to us how one should compute the dynamical correlation length $\xi_d$ and compare it to various recent numerical and 
experimental determinations of these dynamical correlations. Whereas cooperative activation events take place over the mosaic length $\ell^*$, it might be
that facilitation effects leads to avalanches of activity that spreads over much longer length scales, in which case $\xi_d \gg \ell^*$. Cooperativity and
correlation do not mean the same thing, although it is by no means easy to define cooperativity unambiguously, i.e. in a such way that one devise a precise 
protocol to measure it (see \cite{Cand-Dauchot-Biroli} for a recent attempt in granular media).

What do we need at this stage? First, some major analytical progress on a realistic, finite dimensional system where RFOT/MCT is expected to hold, to understand in detail the 
interplay between spatial fluctuations and activated events. We need to understand whether the cartoon sketched in Fig. \ref{cartoon}, that delineates the different 
regions of applicability of MCT and 
RFOT, makes sense or not. The Ginzburg-Harris criterion we discussed above tells us that the limit of an infinitely large jump of 
specific heat should ease some of the problems.  A toy-model where non perturbative activated effects can be studied thoroughly would be extremely valuable. Some of the ideas 
sketched in the Appendices might turn out to be useful.

Short of theoretical ideas, one should at least try to find a finite dimensional version of a 1-RSB mean-field model  
that can be convincingly simulated both in the MCT regime and in the activated/mosaic regime. This would allow one to test directly some of the building 
blocks of RFOT. Unfortunately, the only models in that category can only be simulated above the MCT temperature $T_d$, and cannot be equilibrated below (at least up to now).
Although the study of the high temperature side of $T_d$ brings a decent amount of information (in particular on the role of activated events in that region) the situation
is frustrating. It might be the case that models that are `hard' enough (in the sense that $q^*=C(\tau_0 \ll t \ll \tau_\alpha)$ is close to unity) so that 1-RSB 
effects survive in low dimensions, are also, for the very same reason, extremely difficult to equilibrate numerically \cite{SBBB}. Intuitively, these are systems with a ``golf-course'' energy 
landscape and very narrow canyons. Is this a necessary prerequisite to form a glass?

One could also hope for a smoking gun experiment of some sort, that either definitely rules out or strongly support the RFOT scenario. We have seen, for example, that the role
of shear on viscosity is quite different in RFOT and in alternative theories, in particular the behaviour of shear thinning as a function of temperature. Energy or volume relaxation after 
a deep quench and other aging/rejuvenation experiments could also be very informative. Any direct proof of the existence of a configurational entropy dominated point-to-set length 
would be decisive. One could surely do with more ``cavity'' numerical simulations and/or real experiments on the statics and dynamics of a liquid trapped within walls made 
of the same frozen liquid. Certainly, the idea that one can force a small enough system of size $< \ell^*$ into an ideal glass phase even when $T > T_K$ is worth exploring experimentally.

Finally, if the relaxation time $\tau_\alpha$ really varies exponentially with the size of the glassites $\ell^*$, we must come to terms with the fact that cooperative regions are 
doomed to remain small, maybe $10^3$ molecules if we are lucky, or $\ell^* \sim 10$ at most. Scaling relations are at best guiding lights to understand general trends, but can never be tested
accurately. Sub-leading corrections will always soil the asymptotic values of the exponents that toil and sweat might be able to produce once quite formidable theoretical 
challenges unravel. The glass transition problem will probably only be solved like in the game of Go, by a slow, patient siege. Because our plea for a smoking gun experiment will
probably never be granted, we should at least remember Feynman and put down all the facts that disagree with our pet theory, as well as those that agree with it.

\section*{Acknowledgements} We want to warmly thank V. Lubchenko and P. G. Wolynes for asking us to write up our thoughts on RFOT, and for their patience. 
We thank L. Berthier, A. Cavagna and G. Tarjus for a careful reading and feedback on the manuscript.  
Finally, we also want to thank all our collaborators
on these subjects, namely: C. Alba-Simonesco, A. Andreanov, L. Berthier, E. Bertin, A. Billoire, R. Candelier, A. Cavagna, L. Cipelletti, 
L. Cugliandolo, O. Dauchot, D. S. Fisher, Y. Fyodorov, T. Grigera, W. Kob,
J. Kurchan, F. Ladieu, F. Lechenault, A. Lef\`evre, D. L'H\^ote, 
M. M\'ezard,  K. Miyazaki, C. Monthus, D. Reichman, T. Sarlat, G. Tarjus, M. Tarzia, C. Thieberge, C. Toninelli, P. Verrocchio and M. Wyart.  
We have benefited from enlightening conversations over the years with C. Brito, C. Cammarota, M. E. Cates, D. Chandler, C. Dasgupta, J. C. Dyre,  
S. Franz, J. P. Garrahan,  P. Goldbart, A. Heuer, R. Jack, J. Langer, A. Montanari, M. Moore, S. Nagel, G. Parisi, S. Sastry, P. Sollich, T. Witten,  P. G. Wolynes, H. Yoshino and F. Zamponi.\\
We were partially supported by ANR DYNHET 07-BLAN-0157-01.

\appendix

\section{Analytical approaches to metastable states and configurational entropies}

\subsection{Density functional approaches}

As we discussed in section II-B, the solution of some mean-field spin-glasses unveils a very rich and 
interesting scenario that bears some resemblance with the physics of glass forming liquids. Still, 
there seems to be a long way before making a quantitative connection with the behaviour of realistic model of supercooled liquids.
The aim of this Appendix is show how this goal can achieved, at least to some extent. The specific issues we address are:
\begin{itemize}
\item How can one identify metastable states theoretically?
\item How can one compute their physical properties? 
\item How can one check that a given system displays indeed many metastable states (without resorting to a dynamical or 
numerical analysis)?
\end{itemize}
For simple systems, like a ferromagnet, the standard procedure consists in identifying the minima of the free energy 
functional. In the case of liquids, this would correspond to the density functional defined as 
\begin{equation}\label{dft}
{\mathcal F}(\{\rho(\mathbf x)\})=-\frac1 \beta\log \frac{1}{N!}\int \prod_{i=1}^N d\mathbf x_i \exp\left(-\beta {\cal H}+
\int d\mathbf x \mu(\mathbf x)\left[\sum_{i=1}^N\delta(\mathbf x-\mathbf x_i)-\rho(\mathbf x) \right]\right)
\end{equation}
where ${\cal H}$ is the Hamiltonian of the system and $\mu(\mathbf x)$ is a chemical energy fixed by requiring that 
$\langle \sum_{i=1}^N\delta(\mathbf x-\mathbf x_i) \rangle=\rho(\mathbf x)$. The average in the previous expression 
is obtained using the Boltzmann weight on the RHS of Eq. (\ref{dft}). 

By analysing the minima of the density functional one can scan the free energy landscape and find whether there exist metastable
states or not. Such a plan has been indeed followed using a simple approximation for ${\mathcal F}(\{\rho(\mathbf x)\})$,
called the Ramakrishnan-Youssouf density functional. It consists in only retaining, in the diagrammatic 
part of  ${\mathcal F}(\{\rho(\mathbf x)\})$, the linear and quadratic terms in $\rho(\mathbf x)-\rho_0$ where 
$\rho_0$ is the average density of the system. It reads:
\begin{equation}\label{dftry}
{\mathcal F}_{RY}(\{\rho(\mathbf x)\})={\mathcal F}(\rho_0)+\frac 1 \beta \int d\mathbf x \rho(\mathbf x) \log(\rho(\mathbf x)/\rho_0)
-\frac 1 2 \int d\mathbf x d\mathbf y  (\rho(\mathbf x)-\rho_0) c(\mathbf x-\mathbf y)(\rho(\mathbf y)-\rho_0)
\end{equation}
where $c(\mathbf x-\mathbf y)$ is the directed correlation of the liquid, defined in Fourier space as 
$c(k)\equiv\frac{1}{\rho_0}\left(1-\frac{1}{S(k)}\right)$, where $S(k)$ is the structure function of the liquid, i.e.
$S(k)=\langle \rho(k) \rho(-k)\rangle$. 

The functional ${\mathcal F}_{RY}$ has been studied in great detail. It allows one to develop a quantitative theory of crystallisation by restricting 
to periodic solutions \cite{Oxtoby} and, concerning liquids at low temperatures, to reveal the existence of metastable amorphous solutions, i.e. glassy metastable states.
This route was pioneered by Wolynes and Stoessel \cite{StoesselW}. This first work has been followed up recently by many others \cite{Dasgupta}. 
It is an approach that has the advantage of being very direct and concrete. Its drawbacks are that quantitative computations are difficult and 
computing properties like the configurational entropy is daunting, both numerically and analytically. 

\subsection{Boundary Pinning Field and Replicas} 

In the following we shall describe an approach based on the replica method which, although more 
abstract has the clear advantage that analytical computations for finite 
dimensional systems, even realistic ones such as hard spheres or binary Lennard-Jones mixtures, 
becomes feasible (although quite involved) \cite{MP}. 

To understand how replicas come about in a model without disorder, let assume that the system is 
in a regime of temperatures where there are indeed many very long-lived metastable states and that the 
Gibbs-Boltzmann measure is distributed over all of them, with the weight corresponding to the 
corresponding free energy, like in mean field models (see section \ref{II-B.1}) . 
In order to study the statistical property of a typical metastable state, we focus on a very large cavity of radius $R$, carved in an otherwise 
infinite (or much larger) system. As explained in intuitive terms in section \ref{III-B}, the basic idea  is to apply a suitable boundary external field in an 
attempt to pin the system in one of the possible metastable states \footnote{Our presentation is different from that of the original papers \cite{Monasson,Franz-Parisi}
but based on the same ideas}. Contrary to simple cases, e.g. the ferromagnetic transition for which a positive or negative magnetic field selects states,
the external ``field" (or analogously the boundary condition) one has to impose to select a given amorphous state, is as 
unpredictable as the state one wants to select. To overcome this difficulty one can take an equilibrium configuration $\alpha$, 
freeze the position of all particles outside the cavity, and use this as a boundary condition \cite{BB}. If the system is a thermodynamic glass characterised by many metastable states 
then this boundary condition should force the system inside the cavity to be into the same metastable state as the equilibrium configuration $\alpha$. 
An illustration of this technique to the simple case of the Ising model helps to clarify and make concrete this discussion. We invite the reader to look at Appendix B for this
purpose. 

Concretely, the procedure consists in computing the cavity partition function, for a fixed  $\alpha$:
\begin{equation}
Z_{\alpha}(R)=\sum_{{\cal C}}\exp(-\beta H({\cal C}))\delta(q^{out}({\cal C},{\cal C}_{\alpha})-1)\quad,
 \end{equation}
 where $q^{out}({\cal C},{\cal C}_{\alpha})$ is a suitably defined overlap that measures the similarity between 
 density configuration ${\cal C}$ and that of the $\alpha$ state in the space outside the cavity. 
 When the overlap equals unity the two configurations are the same outside the cavity. In the large $R$ limit, the intensive free energy of the metastable state, 
 obtained by taking the logarithm of the partition function, is expected to be self-averaging and independent of ${\cal C}_{\alpha}$.
 Physically, this means that the overwhelming majority of the metastable states sampled by the equilibrium Boltzmann measure are characterised by the same intensive free energy.  
 
Although we started from a problem without quenched disorder, we find that the analysis of the metastable states leads us to 
a problem where the configuration ${\cal C}_{\alpha}$ plays the role of a (self-induced) quenched disorder. 
In order to proceed further and compute the intensive free energy of a typical metastable states we have 
therefore to average over  ${\cal C}_{\alpha}$. As usual for quenched disorder problem, one can make use of the 
replica trick:  
\begin{equation}
 \langle\ln Z_{\alpha}\rangle_{\alpha}=\lim_{m\rightarrow 1}  \frac{\ln \langle Z_{\alpha}^{m-1}\rangle_{\alpha}}{m-1}
 \end{equation}
 In order to compute the average $\langle Z_{\alpha}^{m-1}\rangle_{\alpha}$ one can introduce
 replicated configurations  and write:
 \begin{equation}
\langle Z_{\alpha}^{m-1}\rangle_{\alpha}=\frac{\sum_{{\cal C}_{\alpha};{\cal C}_{1}\cdots{\cal C}_{m-1}}
\exp(-\beta H({\cal C}_{\alpha}))\prod_{a=1}^{m-1}[\exp(-\beta H({\cal C}_{a}))\delta(q^{out}({\cal C}_{\alpha},{\cal C}_{a})-1)]}{\sum_{{\cal C}_{\alpha}}\exp(-\beta H({\cal C}_{\alpha}))}
 \end{equation}
As usual with replicas one computes the above sum for integer and positive values of $m-1$ and then makes an analytical continuation to make the $m\rightarrow 1$ limit.   
It is important to notice that the average in the numerator of the previous expression can be rewritten as 
the partition function of $m$ replicas constrained to be identical outside the cavity but free to fluctuate 
inside since in the above expression ${\cal C}_{\alpha}$ is no longer different from the other replicas.

Let us denote the logarithm of the partition function of the $m$ constrained replicas as $-\beta F_{m}$. Once this quantity is known, one can compute the
partition function of the large cavity as:
\begin{equation}\label{def-F}
 {\cal F} \equiv - T \langle\ln Z_{\alpha}\rangle_{\alpha}= \lim_{m\rightarrow 1}  \frac{ (F_{m}-F_{1})}{m-1}=
  \left. \frac{\partial F_{m}}{\partial m} \right |_{m=1}
 \end{equation}
This gives the free energy of one typical state inside the cavity. Since we are interested in thermodynamic quantities,
henceforth we will consider $R$ to be very large.  If there are many states, i.e. an exponential number in the 
size of the system, then the free energy of the cavity ${\cal F}$ may be different from the unconstrained one $F_{1}$. 
This can be seen by rewriting the replicated partition function as a sum over 
all states with their Boltzmann weight. If the constraint is strong enough to force the replicas $a=1, \dots, m-1$ to
fall into the same state as $\alpha$ itself, then:
\footnote{The alert reader should be very suspicious about this assumption, and rightly so.  This will be critically reconsidered later in this Appendix.}
\begin{equation}  
\frac{F_{m}}{N}=-T \frac{\ln \sum_{\alpha}e^{-\beta f_{\alpha}mN}}{N}=-T \frac{\ln \int df \exp(N[-\beta f m+\sigma(f,T)])}{N}=f^* m - T \Sigma(T),
\end{equation}
where $f^*$ is the free energy density that maximises the argument of the exponential, $\Sigma(T)=\sigma(f^*,T)$ and $N$ is the number
of particles inside the cavity. Using Eq. (\ref{def-F}), one immediately finds that $f^* = {\cal F}/N$ is the intensive free energy of a typical 
metastable state. But the free energy of the system without constraint is $F_1 = f^*-T \Sigma(T)$ that contains the configurational entropy contribution. 
The replica method allows one to obtain both quantities, which then yields the configurational entropy: 
\begin{equation}\label{configentr-replica}
\Sigma(T)=-\beta F_{1}+\beta \left. \frac{\partial F_{m}}{\partial m} \right |_{m=1}=\beta \left. \frac{\partial }{\partial m}\left[\frac{F_m}{m} \right]\right |_{m=1}
\end{equation}
Hence, we have found that computing the statistical properties of metastable states 
reduces to the computation of the thermodynamics of $m\rightarrow 1$ replicas with 
the constraint that the overlap outside a spherical cavity of radius $R$ is equal to one. 
In practice one has to do a computation for $m$ replicas and take
the space dependent overlap $q_{a,b}(r)$ between them in the bulk of the cavity as an order parameter. 
Hence, one has to compute as accurately as possible the free energy as a function of the 
overlap $q_{a,b}(r)$, and then find the stationary points. 

One always finds a trivial solution with uncoupled replicas, i.e. $q_{a,b}(r)=0$ for $a \neq b$. 
This is expected since if it was not for the boundary condition the replicas would be indeed completely uncoupled. 
If this is the only solution then $F_m=mF_1$, and $\Sigma=0$ as it should be.
One has therefore to inspect whether another solution exists. 
If it is the case then the constraint outside the cavity plays the role of a boundary condition 
for $q_{a,b}(r)$ and selects this coupled replica solution.    

The approach outlined above is the starting point for several investigations of the glass transition that explicitly encapsulates the tenets of RFOT.
It provides a microscopic basis to the arguments of section \ref{III-B} on entropy driven cavity melting leading to the mosaic state. 
We have explained that one has to compute as accurately as possible the m-replica free energy as a function 
of the overlap $q_{a,b}$ and to look for a coupled replica solution. When the system is translation invariant one 
naturally looks for homogeneous solutions and this is indeed what has been done to compute quantitatively
the configurational entropy and the properties of the metastable states and of the glass phase, see e.g. \cite{franzcardenas} and \cite{MP} for a review. 
But the study of inhomogeneous solutions is also very interesting. In \cite{Dzero,Franz} it has been 
shown that between $T_d$ and $T_K$ (as computed by using the homogeneous solution), one finds 
an inhomogeneous solution with an overlap equal to one on the boundary that vanishes away from it on distances greater than a certain $\ell^*$.  
This shows that the coupled replica solution actually is unstable and the metastable order does not propagate further than the length $\ell^*$, 
which can be therefore identified with the mosaic length-scale discussed heuristically in section \ref{III-B}.  This provides a microscopic way to compute the exponent 
$\theta$. Up to now, all computations have lead to the $\theta=d-1$, which is different from the value $d/2$ coming from KTW's wetting argument, see  \cite{Dzero,Franz}
for more details. 

\subsection{Application: the ``Weak Glass'' expansion and the nature of the glass transition}
 
In order to apply the procedure described in the previous section to a real glass-former, one has to be able  compute the thermodynamic properties of $m$ replicas with the condition 
that their overlap is one outside a very large region ${\cal S}$ of linear size $R$. This leads to study a multi-component system characterised 
by an infinitely strong attraction between replicas outside ${\cal S}$ and no inter-replica interaction inside ${\cal S}$. Same replica particles
interact everywhere via the physical potential $V$. Using the diagrammatic formalism of liquid state theory \cite{moritahiroike} for this multi-component system 
one can write the Helmholtz free energy, which we will denote with a slight abuse of notations by $F_m$, as a functional of the densities $\rho^a(\mathbf x)$ 
and correlation functions $C^{ab}(\mathbf x, \mathbf x')=\langle \sum_i\delta(\mathbf x-\mathbf x_i^a) \sum_j\delta(\mathbf x'-\mathbf x_j^b)\rangle$. 
As usual, it is more practical to use the following representation for the correlation function: 
 \be
 C^{ab}(\mathbf x, \mathbf x')=\rho_0^2(1+h^{ab}(\mathbf x,\mathbf x'))+\rho_0\delta_{ab}\delta(\mathbf x-\mathbf x')
 \ee
where $\rho_0$ is the density of the liquid we are focusing on. 
The general expression of $F_m$ is quite complicated  \cite{moritahiroike,mezardparisi}. In 
our case it can be simplified using explicitly that there is no interaction between replica inside ${\cal S}$
and assuming translation invariance. This last assumption is certainly true far from the boundaries and simplifies the discussion without losing too much generality. 

As discussed above, the absence of metastable states translates into a solution
where the replicas are completely uncoupled: $\rho_0^a=\rho_0$ and $h^{ab}=h\delta_{ab}$. 
Instead, the coupled solution corresponds to $\rho_0^a=\rho_0$ but $h^{ab}=h_\ell\delta_{ab}+h_g(1-\delta_{ab})$ and allows one to unveil the existence of metastable states. 
In the $m\rightarrow 1$ limit, $h_\ell$ corresponds to the liquid correlation function and $h_g$ to the so-called called non-ergodicity factor $q^*$: 
the plateau value of the density-density time correlation function, which measures the fraction of frozen-in density fluctuations.   
Note that contrarily to the case of spin glasses where replica symmetry breaking is spontaneous \cite{Parisi}, the boundary conditions break explicitly this symmetry in 
the present case.

In the large $R$ limit we find: 
\bea\label{functional}
\beta\frac{F_m}{mN}&=&\beta\rho_0\mu
+\rho_0\ln (\lambda^3\rho_0)-1)+\frac{1}{m}\sum_{ab} \int \frac{d\mathbf k}{(2\pi)^3}L(\rho_0 \widehat h^{ab}(\mathbf k))
\\
&+&\frac{1}{2m} \sum_{ab}\rho_0^2\int d\mathbf x(1+h^{ab}(\mathbf x))[\ln (1+h^{ab}(\mathbf x))
+\delta_{ab}\beta V(\mathbf x))]-\frac{1}{2m} \sum_{ab}\rho_0^2\int d\mathbf x h^{ab}(\mathbf x)+ {\cal T}
\eea
where $\mu$ is the chemical potential, the hat on $h^{ab}(\mathbf k)$ denotes the Fourier transform, $\lambda$
the thermal De-Broglie wavelength, $L(u)$ the function $-\log(1+u)+u-u^2/2$
and $\cal T$ the sum of all the more-than-doubly connected diagrams which are composed of $\rho_0$ circles 
and $h^{ab}$ lines.  

It is clear that the replica uncoupled ansatz leads to a RHS which is not a function of $m$ and, hence, to a zero
configurational entropy via Eq. (\ref{configentr-replica}). The replica coupled ansatz instead leads to a 
non zero  configurational entropy. However, in order to obtain the correct result one has to minimise $F_m$ with respect to 
$\rho_0$ and $h^{ab}$, plug back the solution into the expression of $F_m$ and finally take the derivative
with respect to $m$, which means performing at some previous stage the analytic continuation in $m$. 
Of course this is impossible to perform exactly and one must resort to approximations. This 
is what has been done in \cite{mezardparisi,franzcardenas}, see also \cite{MP}. 
In the following we shall present what we call a ``weak glass'' expansion with the purpose of showing that the glass transition is necessarily first order in 
$h_g$ despite the fact that it is second order thermodynamically.

The weak glass expansion assumes that the boundary pinning field induces only a very small off-diagonal correlation, i.e.
that the frozen-in density fluctuations $h_g$ are very small. We do not discuss the physical conditions that make this assumption self-consistent but 
surmise that it is possible to find a (presumably long-ranged) potential $V$ such that this is the case. One can collect all terms which are of lowest order in $h_g$ keeping at the same 
time the $h_\ell$ contributions to all order. The zeroth order in $h_g$ is by construction independent of $m$. It coincides with the 
free energy for the one-component liquid we are focusing on and corresponds to  
the usual free energy of diagrammatic liquid theory as a function of $h_\ell$. Analysing the expansion of $\cal T$ in $h_g$, one finds that
the first contribution containing $h_g$ is necessarily at least third order.  
Since all the other terms contain linear and quadratic terms in $h_g$, we can drop the $\cal T$ contribution, which is sub-leading
within the weak glass expansion. The other important observation, that can be checked straightforwardly, is that the terms containing $h_g$
are $O(m-1)$ in the limit $m\rightarrow 1$. This has two important consequences. First, one recovers the usual liquid theory 
diagrammatics for $m\rightarrow 1$, as one should. Second, it reveals that a non zero $h_g$ lead to a non trivial dependence on $m$
of the RHS of (\ref{functional}) and, hence, to a non zero configurational entropy via Eq. (\ref{configentr-replica}).

As a summary, we find that the minimisation of $F_m$ in the  limit $m\rightarrow 1$ leads to two {\it independent} 
variational problems. First, the usual one of liquid theory, from which one obtains the exact solution for the liquid $h(\mathbf x)$ and $\rho_0$.
The second one gives access to the properties of metastable states in which the liquid can remain trapped for a long time. 
It corresponds to the minimisation of the $m-1$ part of $F_m$. This is a functional 
of $h_\ell$, $h_g$,$\rho_0$. One has therefore to plug the values of $h_\ell$ and $\rho_0$ obtained from the solution of the first
variational problem and minimise the result with respect to $h_g$. 
The minimisation equation determining $h_g$ reads, to lowest order:
\[
\hat h_g(\mathbf k)=\left(1-\frac{1}{S(\mathbf k)} \right)\hat h_g(\mathbf k)+O(\hat h_g^2),
\]
where $S(\mathbf k)$ is the exact structure factor of the liquid.
This shows that a continuous transition leading to a non zero $h_g$ is impossible except if the
structure factor diverges for some vector(s) $\mathbf k$, but this is not observed in experiments, and would correspond to a quite different physical situation.

Therefore, one concludes that when a non zero $\hat h_g(\mathbf k)$ appears it necessarily does so in a discontinuous fashion. 
Thus, we find one of the very remarkable properties of glass-forming liquids which distinguish them
from spin-glasses: when the dynamics starts to slow down above the glass transition, i.e. when
metastable states emerge, the plateau in density-density correlation function 
appears discontinuously. This happens at a temperature which is identified with the dynamical temperature $T_d$ as discussed in sect \ref{II-B.3} .
In order to understand quantitative the behaviour of $h_g$ as a function of temperature one should go beyond the weak glass expansion and re-sum in some way
an infinite subclass of diagrams. It would be interesting to pursue further this weak glass expansion. This approach would be the counterpart
of density functional theory for crystallisation where one makes an expansion of the exact functional in powers of the $\rho(\mathbf x)-\rho_0$.
It may shed light on many important aspects of the dynamical transition that are very poorly known in finite dimensions.

  \section{A toy model of entropy driven cavity melting}
  
  In the following we shall focus on a toy model of entropy driven cavity melting. The aim of this section is to 
   make more concrete several ideas explained previously and to shed some new light on procedures and 
   intuitive ideas introduced in the study of the mosaic state. The model we focus on is 
  the one dimensional Ising spin glass model with quenched random couplings $J_i$, equal to $+J$
  and $-J$ with probabilities $1/2$. 
  A gauge transformation $S_i \rightarrow \prod_{j=0}^i (J_i/J)S_i$ makes the model equivalent 
  to the pure ferromagnetic Ising model. Thus, at zero temperature it is characterised by two degenerate ground states. 
  The zero temperature order does not survive to any finite temperature. The mechanism that destroys the order 
  provides a cartoon version of the entropy driven cavity melting discussed in section \ref{III-B}. 
  In the following we will discuss this model quite in detail from this perspective. 
    
  \subsection{A recap on the low temperature physics of the one dimensional spin glass Ising model}
  The Hamiltonian of the one dimensional Ising model reads:
  \[
  H=-\sum_i J_i S_i S_{i+1} \qquad S_i=\pm 1
  \]
  At $T=0$ and with open boundary conditions the model presents two degenerate ground states corresponding respectively to all spins up and all spins down after the gauge transformation.  
  Reversing all the spins on an interval of size $\ell$ costs 
  an energy $4J$. This is independent of the length $\ell$ of the interval. In fact one can interpret this process as the
  creation of two defects. These are called kinks or domain wall and each one of them separates a region with all spins up from one with all spins down. 
  One can considers the low temperature phase of the 1D Ising model as a dilute gas formed by these excitations. Since each excitation costs an energy $2J$ the density of 
  kinks is $\rho \propto e^{-2J/T}$. This implies that a typical equilibrium configuration of the spin glass Ising model at low temperature consists -- after gauge transformation -- of 
  up and down regions which alternate along the 1D lattice on the scale $L_K=\rho^{-1}$, see Fig \ref{fig:Ising}. 
Note, however, that in the original gauge, the spin configurations look completely amorphous and without any sign of growing long range order. The spatial average of the connected correlation 
functions between two spins at distance $x$ reads: 
\[
\frac 1 N\sum_{i,j; j-i=x} \langle S_i S_j \rangle_c =\frac 1 N\sum_{i,j; |i-j|=x} \prod_{k=i}^{j-1} \tanh \beta J_k
  \] 
In the large $N$ limit it becomes zero except for $x=0$, i.e. the configuration seems to lack any kind of long range order. 
Moreover, the Metropolis dynamics of this system become slow at zero temperature since it is governed by the diffusion of these rare defects.  
For instance, the time between two consecutive flips of a spin is of the order of the square of the distance
between defects, i.e. $\tau_0 e^{4J/T}$ where $\tau_0$
is the time on which each spin attempts to flip with the Monte Carlo dynamics.  

\subsection{A mosaic with few tiles}
\begin{figure}
\center
\includegraphics[width=0.69\columnwidth]{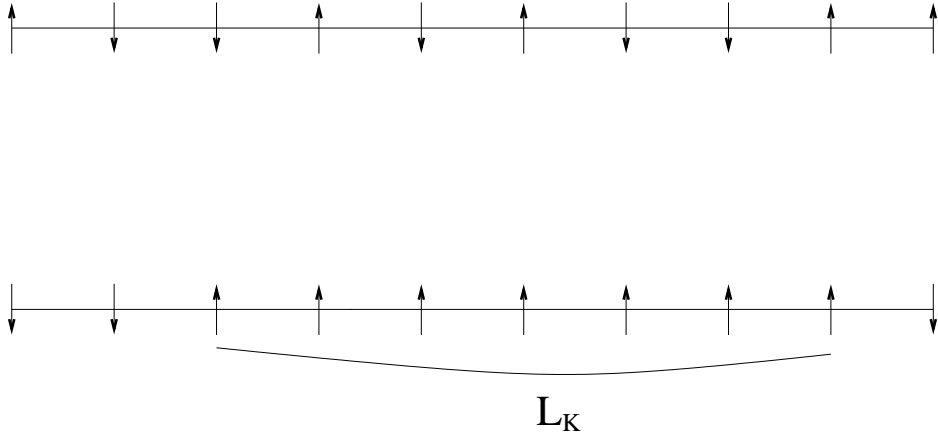}
\caption{Top: typical low temperature configuration for the Ising spin glass at low temperature. Bottom: corresponding
configuration after gauge transformation. $L_K$ is the typical distance between kinks.}
\label{fig:Ising}
\vspace{-0.5cm}
\end{figure}

The low temperature configuration of the 1D Ising model shown in Fig. \ref{fig:Ising} can be interpreted as a mosaic state composed
by just few tiles. There is indeed a strong similarity with the glassy mosaic. Finding back the physics described above from 
the mosaic point of view is an interesting exercise as we shall show in the following. 

Let us follow the procedure outlined for RFOT in section \ref{III-B}. For simplicity we will work always directly on the pure Ising model, i.e. 
after having performed the gauge transformation. The mean field analysis of the problem is often 
related to the Curie-Weiss mean field equations: $m=\tanh 2\beta Jm$, which would predict a phase transition 
at temperature $T_c=2J$. However, one can also write down the equivalent of the TAP free energy \cite{TAPISING}, 
as one would naturally do in a RFOT analysis. In this 1D case the free energy as a function of the local magnetization 
can be computed exactly \cite{TAPISING}. For simplicity, we just reproduce its continuum approximation which reads: 
\[
F[\{m(x)\}]=\int dx \left [D(\nabla m(x))^2 + g(m^2(x)-1)^2\right]
\]
where $D$ and $g$ are two microscopic constants. 
The analysis of the low free energy minima reveals that there are two absolute minima corresponding to uniform plus or 
minus magnetisations. Moreover, there are many other minima that corresponds to have $L$ domain walls inside the system. Their free energy
(or energy since they coincide in the low temperature limit we are focusing on) is higher than the ground state one of an amount $\Delta F=2JL$. 
One can easily compute the number of such states: it corresponds to the different ways to place the kinks on the lattice and reads $\frac{N!}{(N-L)!L!}$. 
This leads to an intensive 'configurational entropy' equal to 
$S_c=-\rho \log \rho -(1-\rho)\log(1-\rho)$, where $\rho=L/N$ is the density of the kinks. Note that for the Ising case, the minima are not separated by 
barriers contrary to what one expects for the glassy minima of super-cooled liquids.\\
Following the same procedure described in the previous section for glasses, we compute the partition function 
of the system by summing over all states and weighting each term by its Boltzmann weight:
\[
Z=\sum_\alpha e^{-\beta F_\alpha}=e^{-\beta F_{GS}}\int d\rho e^{N\left[-\rho \log \rho -(1-\rho)\log(1-\rho) -\beta 2J \rho \right] }
\]
where $F_{GS}$ is the free energy for the state without kinks.   
The integral can be performed by the saddle point method. The value of $\rho$ that dominates the integral at low temperature is given by $\rho^*\simeq \exp \left( -\frac{2J}{T}\right)$. 
So we find that the 'configurational entropy', $S_c(\rho^*)$, is small and goes to zero at $T=0$. Note, however, that real space configurations still looks completely disordered. 
Only applying the gauge transformation one can discover a growing order. Thus, at low temperature we are in a situation very similar to glasses: the spatial average of 
the two point correlation function is zero except at zero distance, indicating apparently no growing order, and at the same time the dynamics slows down very fast.
Using the argument developed in section \ref{III-B}, we would expect that the competition between configurational entropy and 
surface tension leads to a mosaic state with a length $\left(\frac{\Upsilon}{TS_c}\right)^{1/(d-\theta)}$. In this simple 
1D case $d=1, \theta=d-1=0$ and it is natural to take $\Upsilon=2J$ (or more generally a constant times $J$). This leads to an estimate of the mosaic length $e^{\frac{2J}{T}}$.   

This result can substantiated by computing explicitly the average overlap inside a frozen cavity, as described in section \ref{III-D} and Appendix A. Let's consider an equilibrated configuration $\{S_i^{eq}\}$ 
and freeze all the spins at a distance larger than $R$ from a given spin $S_i$. We analyse the thermodynamics  of the cavity with this boundary condition, in particular we compute 
the average value of $S_i$.  It is easy to check that this
is given by: 
\[
\langle S_i \rangle_{\{S_i^{eq}\}}=\frac{S_{-R+i}^{eq}\prod_{k=-R+i}^{i}\tanh \beta J_k + S_{i+R}^{eq}\prod_{k=i}^{R+i-1}\tanh \beta J_k
}{1+S_{-R+i}^{eq}S_{i+R}^{eq}\prod_{k=-R+i}^{R+i-1}\tanh \beta J_k }
\]
As expected, this average value depends on the equilibrated configuration only through its boundary spins. As in Appendix A,
we shall now compute the average overlap between $\langle S_i \rangle_{\{S_i^{eq}\}}$ and the value of the spin $S_i^{eq}$ in the original equilibrated configuration:
\[
q(R)=\frac 1 N \sum _i \sum_{\{S_i^{eq}\}} \frac{e^{-\beta H(\{S_i^{eq}\})}}{Z} S_i^{eq}\langle S_i \rangle_{\{S_i^{eq}\}} 
\]
Note that we also average over the site $i$. The results will be therefore equivalent to averaging over the quenched 
disorder. The computation of $q(R)$ is rather straightforward and will not be detailed here. The final result is:
\[
q(R)=\frac{2 (\tanh \beta J)^{2R}}{1+(\tanh \beta J)^{2R}}
\]  
The function $q(R)$ decreases monotonically on a length-scale $\xi=\frac{1}{2\ln \tanh \beta J}$. 
In the low temperature limit we are interested $\xi\simeq  e^{2J/T}/4$. This  
coincides with $L_K$ up to a constant factor. 
Thus we find the same results than in the previous heuristic argument
and we conclude that the system is indeed in a mosaic state of length $L_K$. Furthermore,
we find that the configurational entropy of the system is equal to $S_c(\rho^*)$. Since
the intrastate entropy is zero this is also the entropy of the system and indeed it coincides with the correct result for the 
1D Ising spin glass. 
\subsection{Further analogies and differences}
We have found that the procedure outlined in Appendix A to study the mosaic state successfully applies to the toy model
considered here. Of course there are important differences between the 1D Ising spin glass and the behaviour conjectured
by RFOT for a glass-forming liquid. The most important ones are that (i) metastable states are not separated by barriers that grow with size, 
actually they are connected by zero modes; (ii) the configurational entropy on the mosaic length scale $\xi$
grows only logarithmically with $\xi$, contrary to three dimensional glasses where it is expected to scale as $\xi^\theta$
with $3/2 \le \theta\le 2$. This means that the mosaic of the Ising spin glass has much less tiles than the one conjectured
for 3D supercooled liquids. (iii) Although linear responses and two point correlation functions do not show any long range 
order, three point responses and the square of the connected correlation function do so, as in spin glasses. In glasses
this is expected to take place at finite frequency only.  

Despite these differences the analysis of the toy model is interesting because it allows one to make concrete many concepts
and procedures. It could also be used to test new ideas, as we do in the following.

Clearly, RFOT is rooted in 
the analysis of mean field glassy systems. However, as we have seen, its finite dimensional extension departs in very important ways.
In order to develop a complete and self-consistent theory, it seems necessary to develop an approach which 
does not make too much reference to mean field systems. In many instances, there are observables and concepts 
that are well defined within the mean-field approximation only. Their extension to finite dimensional is somehow fuzzy
and make the whole RFOT construction shaky.
A very good candidate would be a renormalisation group description of the ideal glass transition. Let us use the toy model 
to have a first hint of what the outcome would be. In the 1D spin glass Ising model the real space renormalisation group can 
be performed exactly. Using the Migdal-Kadanoff prescription, one integrates out the spins on the, say, odd sites to get 
a new effective coupling between even spins. The exact RG relation reads:
\[
\beta J^{eff}_{i,i+2}=\frac12 \ln \frac{1+\tanh \beta J_i \tanh \beta J_{i+1}}{1-\tanh \beta J_i \tanh \beta J_{i+1}}
\] 
By iterating the RG transformation one finds that $\beta J_{eff}$ becomes of the order of one on scales of the order $L_K$. 
This means that on this scale (and above) the system is effectively at high temperature: it has almost no stiffness and 
the entropy is of the order of one. This suggests an RG way to characterise the mosaic state in liquids. 
Integrating out degrees of freedom up to scale $\ell$, one should find that the entropy increases, 
the surface tension between metastable states decreases, and the relaxation timescale increases. When $\ell$ equals the 
point to set length one should find a liquid characterised by a vanishing surface tension between states. 
Thus, metastable states should become unstable and the related configurational entropy disappears. 
Clearly, the above scenario is highly speculative but we do hope that future works will be performed along these lines.

\end{document}